\documentclass[manuscript]{acmart}
\usepackage{amsmath,amsfonts}
\usepackage{algorithmic}
\usepackage{algorithm}
\usepackage{array}
\usepackage{textcomp}
\usepackage{stfloats}
\usepackage{url}
\usepackage{verbatim}
\usepackage{graphicx}

\usepackage{xcolor}
\usepackage{hyperref}

\usepackage{multirow}
\usepackage{graphicx}
\usepackage{booktabs}
\usepackage{float}

\usepackage[normalem]{ulem}
\useunder{\uline}{\ul}{}
\usepackage{enumitem}

\usepackage{threeparttable}

\usepackage{subfigure}
\usepackage{pgfplots}
\usepackage{tikz}
\usepackage{tikz-qtree}

\usetikzlibrary{matrix}
\usetikzlibrary{patterns}
\usepgfplotslibrary{groupplots}

\definecolor{1}{RGB}{102,102,255}
\definecolor{2}{RGB}{102,206,245}
\definecolor{3}{RGB}{255,247,102}
\definecolor{4}{RGB}{255,178,102}
\definecolor{11}{RGB}{146,209,79}
\definecolor{5}{RGB}{194,209,163}
\definecolor{6}{RGB}{102,163,255}
\definecolor{7}{RGB}{163,223,196}
\definecolor{8}{RGB}{194,163,163}
\definecolor{9}{RGB}{194,102,163}
\definecolor{10}{RGB}{255,102,102}
\definecolor{red1}{RGB}{199,21,133}
\definecolor{red2}{RGB}{255,102,102}
\definecolor{red3}{RGB}{255,182,193}
\definecolor{blue1}{RGB}{25,25,112}
\definecolor{blue2}{RGB}{30,144,255}
\definecolor{blue3}{RGB}{135,206,250}

\AtBeginDocument{%
  }

\setcopyright{acmlicensed}
\copyrightyear{2018}
\acmYear{2018}
\acmDOI{XXXXXXX.XXXXXXX}

\acmJournal{JACM}
\acmVolume{37}
\acmNumber{4}
\acmArticle{111}
\acmMonth{8}

\begin{document}

\title{Causal Deconfounding via Confounder Disentanglement for Dual-Target Cross-Domain Recommendation}

\author{Jiajie Zhu}
\email{jiajie.zhu1@students.mq.edu.au}
\orcid{0000-0001-8673-1477}
\author{Yan Wang}
\authornote{Yan Wang is the corresponding author.}
\email{yan.wang@mq.edu.au}
\orcid{0000-0002-5344-1884}
\affiliation{%
  \institution{Macquarie University}
  \city{Sydney}
  \state{NSW}
  \country{Australia}
}

\author{Feng Zhu}
\affiliation{
  \institution{Ant Group}
  \city{Hangzhou}
  \country{China}}
\email{zhufeng.zhu@antgroup.com}
\orcid{0000-0003-4200-0423}

\author{Zhu Sun}
\affiliation{%
  \institution{Singapore University of Technology and Design, Singapore}
  \city{Singapore}
  \country{Singapore}}
\email{sunzhuntu@gmail.com}
\orcid{0000-0002-3350-7022}

\renewcommand{\shortauthors}{Zhu et al.}

\begin{abstract}
  In recent years, dual-target Cross-Domain Recommendation (CDR) has been proposed to capture comprehensive user preferences in order to ultimately enhance the recommendation accuracy in both data-richer and data-sparser domains simultaneously. However, in addition to users' true preferences, the user-item interactions might also be affected by confounders (e.g., free shipping, sales promotion). As a result, dual-target CDR has to meet two challenges: (1) how to effectively decouple observed confounders, including single-domain confounders and cross-domain confounders, and (2) how to preserve the positive effects of observed confounders on predicted interactions, while eliminating their negative effects on capturing comprehensive user preferences. To address the above two challenges, we propose a \textbf{C}ausal \textbf{D}econfounding framework via \textbf{C}onfounder \textbf{D}isentanglement for dual-target \textbf{C}ross-\textbf{D}omain \textbf{R}ecommendation, called CD2CDR. In CD2CDR, we first propose a confounder disentanglement module to effectively decouple observed single-domain and cross-domain confounders. We then propose a causal deconfounding module to preserve the positive effects of such observed confounders and eliminate their negative effects via backdoor adjustment, thereby enhancing the recommendation accuracy in each domain. Extensive experiments conducted on seven real-world datasets demonstrate that CD2CDR significantly outperforms the state-of-the-art methods.
\end{abstract}

\ccsdesc[500]{Information systems~Recommender systems}
\ccsdesc[500]{Computing methodologies~Neural networks}

\keywords{Cross-Domain Recommendation, Confounder Disentanglement, Causal Deconfounding}

\maketitle

\section{Introduction}

\begin{figure}[ht]
\centering
\includegraphics[scale=0.36]{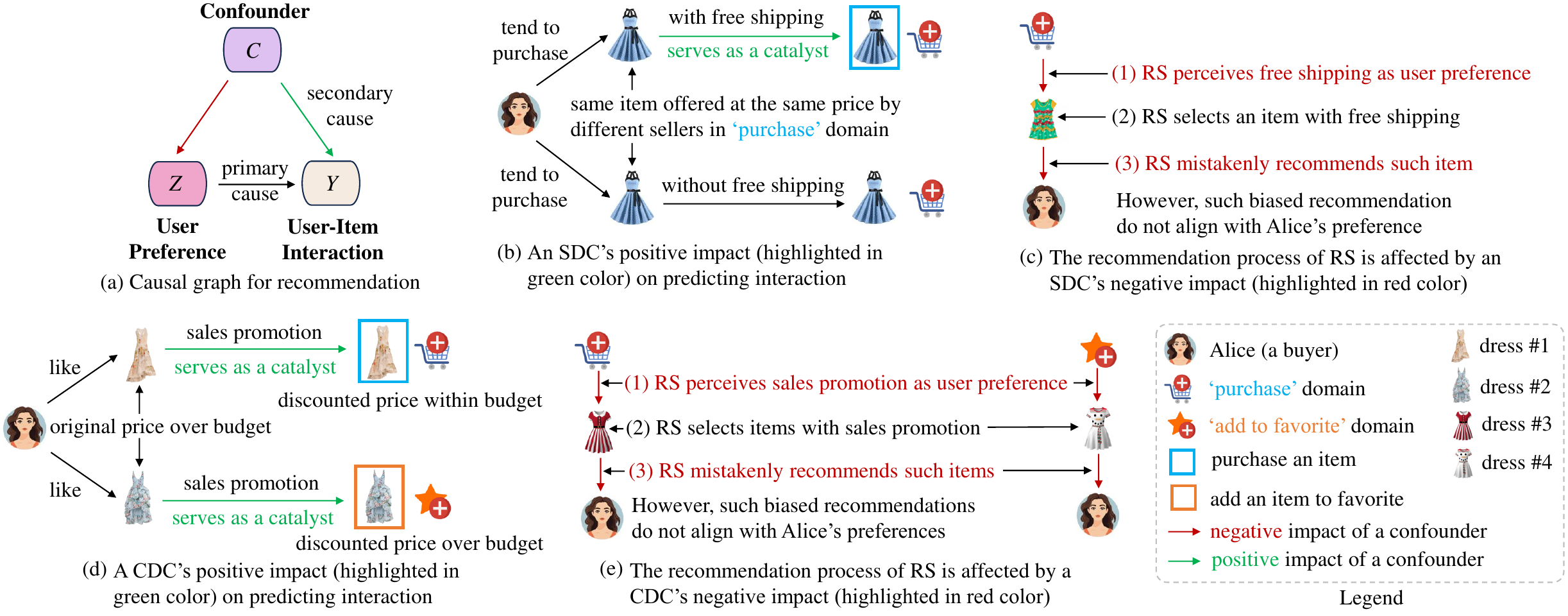} 
\vspace{-0.15in}
\caption{Examples to depict single-domain confounder (SDC) and cross-domain confounder (CDC) \protect\cite{du2022invariant}.}
\label{confounder_example}
\vspace{-0.2in}
\end{figure}

Cross-Domain Recommendation (CDR) aims to transfer valuable information from a relatively data-richer source domain to a relatively data-sparser target domain to improve recommendation performance, forming \emph{single-target CDR} \cite{zhu2019dtcdr}. Effective CDR requires the source and target domains to share certain relatedness while maintaining distinctions in user intents, user behaviors, or item categories \cite{luo2023mamdr}. For instance, on e-commerce platforms like Taobao\footnote{Taobao is a Chinese Customer-to-Customer (C2C) platform that facilitates transactions between individual sellers and buyers, similar to eBay.}, different purchase scenarios, such as ‘what to take when travelling’ and ‘how to dress up for a party’, share overlapping user interests (e.g., finding suitable clothing or accessories), while maintaining distinctions in user intents, such as practicality for travel and aesthetics for parties \cite{du2019sequential,zhang2023hierarchical}. Similarly, on Tmall\footnote{Tmall is a Chinese Business-to-Customer (B2C) platform designed for brand merchants and flagship stores, similar to Amazon.}, behaviors such as ‘add to favorite’ and ‘purchase’ can be regarded as business domains, both reflecting user interest but differing in their focus as exploration- and purchase-oriented actions, respectively \cite{li2021debiasing,zhang2023connecting}. In contrast, on Amazon, domains often refer to different item categories (e.g., ‘movie’ and ‘book’) that share user interests in certain features (e.g., genres or styles), while differing in domain-specific item features and user preferences \cite{chen2023toward,song2024mitigating}. These relatedness and distinctions together define the principle of domains, thereby ensuring CDR's adaptability to diverse recommendation scenarios.

In addition, on top of the same above-mentioned principle of domains, Dual-Target CDR \cite{ijcai2021p639} has been proposed to capture comprehensive user preferences, and thus enhance the recommendation accuracy in both data-richer and data-sparser domains simultaneously, which are source domains and target domains as well. However, in addition to users' true preferences, the user-item interactions might also be affected by confounding factors. A confounding factor, termed as \emph{confounder} in causal inference \cite{gao2022causal,zhu2023causal}, affects both the treatment and the outcome \cite{wu2025instrumental,luo2024survey}, which can be broadly interpreted as user preference and user-item interaction respectively (see Fig. \ref{confounder_example}(a)) in the context of recommender systems (RSs) \cite{gowda2021pulling,zhu2024mitigating}. 

In the dual-target CDR scenario, observed confounders can be divided into two types, i.e., \emph{single-domain confounder} (SDC) and \emph{cross-domain confounder} (CDC). SDC only affects user preference and user-item interaction in one specific domain and has been widely studied in the existing literature \cite{zhang2021causal,wang2021deconfounded}. By contrast, CDC affects both domains, which, however, has been overlooked in existing dual-target CDR methods. Essentially, SDC is a simplified version of CDC. Below we first briefly review SDCs and then provide an in-depth analysis of CDCs, both illustrated with examples from Tmall, where ‘purchase’ and ‘add to favorite’ are regarded as two domains, as they align with the principle of domains.

SDCs have both positive and negative impacts on predicting user-item interactions in their corresponding domain. For instance, as shown in Figs. \ref{confounder_example}(b)-\ref{confounder_example}(c), ‘free shipping’ is an SDC in the ‘purchase’ domain. Consider a scenario where the same item is offered at the same price by different sellers. One seller provides free shipping, while the other provides shipping with an additional cost. Thus, the offer with free shipping positively influences Alice’s decision to purchase the item with free shipping. As for negative impact, a data-driven RS improperly perceives ‘free shipping’ (i.e., an SDC shown in Fig. \ref{confounder_example}(c)) as Alice’s preference in the ‘purchase’ domain. As a result, the data-driven RS mistakenly recommends an item with free shipping that Alice does not actually like to her. This misalignment, referred to as \emph{confounding bias} \cite{li2023causal}, results in biased recommendations.

In fact, the confounding biases also exist in cross-domain scenarios. More importantly, CDCs have both positive and negative impacts on predicting user-item interactions in both domains. For example, as illustrated in Figs. \ref{confounder_example}(d)-\ref{confounder_example}(e), ‘sales promotion’ is a CDC, because it simultaneously affects ‘purchase’ and ‘add to favorite’ domains. On the one hand, this ‘sales promotion’ CDC has a positive impact. In fact, while Alice's true preference is the primary cause of her behaviors in both domains, ‘sales promotion’ is a secondary cause that serves as a catalyst. With a new sales promotion on dresses \#1 and \#2, both of which Alice likes but previously found over her budget, she immediately purchases dress \#1 that has become affordable within her budget. By contrast, since the discounted price of dress \#2 is still over her budget, she adds it to favorite for future consideration, looking forward to a further price reduction. On the other hand, this ‘sales promotion’ CDC has a negative impact too. As depicted in Fig. \ref{confounder_example}(e), a data-driven RS improperly perceives ‘sales promotion’ (i.e., a CDC) as Alice’s preference in both domains. As a result, the data-driven RS mistakenly recommends dresses \#3 and \#4 with sales promotion to Alice, but Alice actually does not like them.

Based on the above discussion, an effective dual-target CDR should \emph{deconfound both observed single-domain and cross-domain confounders}, which includes three tasks, namely, (1) identify and decouple such observed confounders, (2) preserve their positive impacts on predicted interactions, and (3) eliminate their negative impacts on user preferences \cite{yu2023deconfounded}. However, existing dual-target CDR approaches overlook the above observations. Thus, a novel dual-target CDR model should be proposed to incorporate such insights for comprehensively understanding user-item interactions.

To effectively advance dual-target CDR, the following two key challenges need to be addressed.

\noindent \textbf{CH1.} \emph{How to effectively decouple observed cross-domain confounders in addition to single-domain confounders to comprehensively understand user-item interactions in dual-target CDR?} The existing dual-target CDR methods either employ graph clustering strategy \cite{li2022debiasing} and variational information bottleneck \cite{cao2022cross}, or identify unobserved domain-specific confounders first, and then utilize causal techniques, e.g., inverse propensity score (IPS) estimators \cite{li2021debiasing} and invariant learning \cite{zhang2023connecting}, to obtain debiased representations (it is worth mentioning that domain-specific confounders in existing works are different from SDCs in our work, see Section \ref{sec:II-B} for elaboration). However, none of them explicitly decouples observed CDCs, and thus it is hard to obtain a comprehensive understanding of user-item interactions in each domain.

\noindent \textbf{CH2.} \emph{How to preserve the positive impacts of observed confounders on predicted interactions, while eliminating their negative impacts on capturing comprehensive user preferences, thereby enhancing the recommendation accuracy in both domains?} Most existing causal methods \cite{wang2022unbiased,xu2023deconfounded} tend to eliminate the confounders' negative impacts, in order to obtain the debiased comprehensive user preferences for recommendation. However, most of them overlook the confounders' positive impacts, and thus limit their efficacy in enhancing the recommendation accuracy \cite{zhan2022deconfounding}.

\noindent \textbf{Our Approach and Contributions.} To address the above two challenges, we propose a novel causal deconfounding framework via confounder disentanglement for dual-target CDR. To the best of our knowledge, this is the first work in the literature that explicitly decouples observed CDCs, and incorporates observed confounders' positive impacts into debiased comprehensive user preferences for dual-target CDR. The characteristics and contributions of our framework can be summarized as follows:

\begin{itemize}
    \item We first propose a \textbf{C}ausal \textbf{D}econfounding framework via \textbf{C}onfounder \textbf{D}isentanglement for dual-target \textbf{C}ross-\textbf{D}omain \textbf{R}ecommendation, called CD2CDR, which can disentangle two types of observed confounders (i.e., SDCs and CDCs), eliminate their negative impacts to obtain debiased preferences, and preserve such confounders' positive impacts, thereby enhancing the recommendation accuracy in both domains;
    \item To address \textbf{CH1}, we propose a confounder disentanglement module to effectively disentangle observed SDCs and CDCs. In this module, we devise a dual adversarial structure to disentangle SDCs in each domain and apply half-sibling regression to decouple CDCs, thus obtaining a comprehensive understanding of user-item interactions in each of both domains;
    \item To address \textbf{CH2}, we propose a causal deconfounding module to deconfound disentangled observed SDCs and CDCs via backdoor adjustment. Specifically, we design a confounder selection function to mitigate such observed confounders' negative effects, thereby recovering debiased comprehensive user preferences. We then incorporate the observed confounders' positive effects into such debiased user preferences to enhance the recommendation accuracy in both domains;
    \item Extensive experiments conducted on seven real-world datasets demonstrate that our CD2CDR outperforms the best-performing state-of-the-art baseline with an average increase of 6.17\% and 8.23\% w.r.t. HR@10 and NDCG@10, respectively.
\end{itemize}

\section{Related Work}
\subsection{Single-Target and Dual-Target CDR}
\noindent \textbf{Single-Target CDR.} Single-Target CDR \cite{ijcai2021p639} focuses on addressing the data sparsity problem by utilizing the abundant information available in a data-richer domain to improve the recommendation performance in a data-sparser domain. The existing single-target CDR approaches can be divided into two categories: (1) content-based transfer, and (2) feature-based transfer \cite{ijcai2021p639}. Content-based transfer \cite{kanagawa2019cross} leverages user/item attributes and textual information to establish links across domains. By contrast, feature-based transfer \cite{hu2019transfer,fu2019deeply} employs machine learning techniques to extract user/item embeddings or rating patterns \cite{ijcai2019p587} for cross-domain transfer.

\noindent \textbf{Dual-Target CDR.} Unlike single-target CDR, dual-target CDR \cite{zhu2019dtcdr} aims to enhance the recommendation accuracy in both data-richer and data-sparser domains by bidirectionally sharing knowledge, providing a basis for its expansion into Multi-Target CDR \cite{zhu2021unified,guo2023disentangled}. The existing dual-target CDR methods can be roughly classified into two classes: (1) conventional methods, and (2) disentanglement-based methods. Conventional methods utilize two base encoders to transform each domain’s interaction data into embeddings, which are then symmetrically incorporated through various transfer layers \cite{li2020ddtcdr,liu2020cross}. By contrast, disentanglement-based methods employ variational autoencoder (VAE) \cite{cao2022disencdr} or other decoupling approaches \cite{zhang2023disentangled} to disentangle domain-shared user preferences for common knowledge transfer, and then incorporate such preferences with decoupled domain-specific or domain-independent user preferences to capture comprehensive user preferences. However, these methods overlook that, in addition to users’ true preferences, users’ final decisions are also influenced by confounders, which limits their ability to accurately predict user-item interactions, leading to suboptimal recommendation results.

\subsection{Deconfounded Recommendation}
\label{sec:II-B}
In recent years, causal learning \cite{ijcai2022p787,xu2025causal} has been introduced into RSs due to its ability to effectively tackle confounding problems arising from confounders, which can be classified into two types: observed confounders and unobserved confounders. For observed confounders, the existing deconfounded RSs adopt inverse propensity weighting (IPW) \cite{sato2020unbiased} or backdoor adjustment \cite{wang2021deconfounded} to address the observed specific confounders, such as item popularity \cite{zhang2021causal} and video duration \cite{zhan2022deconfounding,he2023addressing}. For unobserved confounders, the existing deconfounded RSs either add additional assumptions \cite{liu2021mitigating,liu2023debiased} or infer substitutes for confounders \cite{wang2020causal,zhang2023debiasing} to alleviate the confounding bias.

Moreover, recent research efforts have extended confounder debiasing into CDR scenarios, focusing mainly on unobserved confounders. The unobserved confounders can be further categorized into two classes: domain-specific confounders (e.g., purchase-guided domain setting) and general confounders (e.g., the display position of items) \cite{li2021debiasing}. Both classes cannot be captured from the datasets, and thus they are different from the observed SDCs and CDCs. Most of existing approaches tend to remove the negative influences of domain-specific confounders \cite{li2021debiasing,zhang2023connecting} or general confounders \cite{zhao2023sequential,xu2024rethinking}, but overlook the positive influences of such confounders, leading to an incomplete understanding of user-item interactions. In contrast to unobserved confounders \cite{liang2024deconfounding,huang2024multi}, which are hidden and often difficult to estimate, observed confounders can be explicitly decoupled. As long as observed confounders are accurately disentangled, they can facilitate the design of effective deconfounding module for more precise deconfounding. However, none of existing approaches explicitly decouples observed CDCs, and preserves the positive influences of observed confounders on predicted interactions, and thus it is hard to achieve a comprehensive understanding of user-item interactions.

\subsection{Disentangled Recommendation}
Disentangled representation learning (DRL) has recently gained increasing interest in RSs, aiming to decouple users’ true preferences from confounding factors for more robust recommendation. For example, MacridVAE \cite{ma2019learning} models disentangled embeddings of user intentions from user-item interactions at both macro- and micro-level to reduce the impact of confounding factors. Moreover, DICE \cite{zheng2021disentangling} tends to decouple users’ interests and conformity to extract the desired causes of user-item interactions for robust recommendation.

In addition, DRL has also been effectively applied to multi-interest recommendation \cite{chen2021multi}, where the goal is to identify and separate users' diverse preference facets. One key challenge in this area is the collapse issue, where initially distinct interest embeddings become increasingly similar during training, resulting in a loss of diversity and a failure to capture users' multifaceted preferences. To address this challenge, researchers have proposed various disentanglement-based strategies, which can be broadly classified into two groups \cite{du2024disentangled}.

The first group of approaches primarily tackles the collapse issue through statistical regularization, aiming to enforce diversity among interest embeddings. Rather than modeling semantic differences directly, these approaches impose various mathematical constraints on the learning process to discourage homogeneity in the representation space. For instance, REMI \cite{xie2023rethinking} enhances multi-interest representation learning by introducing an interest-aware hard negative mining strategy alongside routing regularization, effectively preventing the routing collapse in capsule networks. Furthermore, VALID \cite{tran2023multi} achieves disentanglement within the regularization framework by iteratively refining personalized item clusters via latent attention.

The second group of approaches mainly handle the collapse issue through representation-guided refinement, focusing on optimizing the bidirectional relationships between interest embeddings and their corresponding items. For example, DisMIR \cite{ma2020disentangled} leverages disentanglement techniques to separate user intentions within sequential behavior patterns, applying self-supervised learning in the latent space to maintain interest diversity and prevent representation collapse over time. Moreover, Re4 \cite{zhang2022re4} disentangles user interests by implementing backward flows from interests to items.

In addition to multi-interest recommendation, DRL has also been applied to causal recommendation. Existing disentanglement-based methods \cite{wang2022causal} first decouple the semantic-aware intent embeddings, and then employ causal intervention \cite{yu2023deconfounded,liu2024interact} to alleviate the confounding bias. Unlike prior works \cite{du2024towards}, our work focuses on decoupling both observed SDCs and CDCs.

\begin{table}[ht]
\setlength{\abovecaptionskip}{0cm}
\setlength{\belowcaptionskip}{0cm}
\caption{Important notations.}
\label{tab:important_notation}
\centering
\begin{tabular}{cc}
\toprule[0.7pt]
\textbf{Symbol}                      & \textbf{Definition}                    \\ \midrule[0.5pt]
$d$              & the embedding dimension      \\
$m$         & the number of users          \\
$n$         & the number of items          \\
$\mathcal{U}$         & the set of users  \\ 
$\mathcal{V}$         & the set of items          \\
${\mathbf{R}} \in {\{ 0,1\} ^{m \times {n}}}$         & the interaction matrix  \\
$y_{ij}$         & the interaction of user ${u}_i$ on item ${v}_j$          \\
$\hat y_{ij}, \hat y_{ik}$         & the predicted user-item interactions          \\
${*^A}$, ${*^B}$  & the notation for domain $A$ and $B$, respectively\\
$\mathbf{Z}_{sha}$          & the domain-shared user preferences \\
$\mathbf{Z}_{spe}$   & the domain-specific user preferences \\ 
$\mathbf{Z}_{ind}$   & the domain-independent user preferences\\ 
$\mathbf{E}_u^*$  & the comprehensive user preferences \\
$\mathbf{E}_v$  & the item embeddings \\
$\mathbf{C}_{sd}$   & the single-domain confounders\\ 
$\mathbf{C}_{cd}$   & the cross-domain confounders\\
$S(\cdot)$, $T(\cdot)$  & the generator in domain $A$ and $B$, respectively\\
${H}( \cdot )$  & the discriminator \\
$J$         & the number of cluster centroids\\ 
$p(c)$     & the uniform distribution for prior probability\\
$\lambda$  & the weight of cycle consistency loss\\
$\alpha $ & the regularization parameter\\
$\mathbf{W}$  & the weight matrix\\
\bottomrule[0.7pt]
\end{tabular}
\vspace{-0.15in}
\end{table}

\subsection{Deconfounded Domain Generalization}
Domain generalization \cite{zhou2022domain,wang2022generalizing} aims to train models on labeled data from source domains to enhance their generalization ability across unseen target domains by learning domain-invariant feature representations. However, confounders influencing both features and labels can undermine such representations, preventing models from capturing the true causal effects. In recent years, causal inference techniques \cite{sheth2022domain,zhang2023video} have been employed to address these confounding problems in domain generalization, thereby enhancing the model’s ability to generalize accurately across varied domains. For instance, a line of existing works \cite{liu2022causal,zhang2022learning} simply adopts the average value of all domain features in each domain as the confounder, and employs backdoor adjustment to capture the true causality. Another line of works incorporates interventional pseudo-correlation augmentation \cite{ouyang2022causality} or adversarial training \cite{wang2023deconfounding} to remove the confounders to better generalize to the unseen domain. There is also another line of works that exploits the instrumental variables \cite{yuan2023instrumental} or learns substitutes \cite{jin2024unbiased} to eliminate the unobserved confounders and capture the invariant features for domain generalization. However, most of existing works either neglect domain-variant features or use off-the-shelf features as confounders instead of explicitly decoupling such confounders, leading to degraded deconfounding performance.

\section{Preliminaries}

\subsection{Problem Definition}
For improved readability, in Table \ref{tab:important_notation}, we present the important notations used in this paper. The paper explores a fully overlapping dual-target CDR scenario in the domains ${D^A}$ and ${D^B}$, with a common user set $\mathcal{U}$, the size of which is denoted as $m = |\mathcal{U}|$. Let ${\mathcal{V}^A}$ (of size ${n^A} = |{\mathcal{V}^A}|$) and ${\mathcal{V}^B}$ (of size ${n^B} = |{\mathcal{V}^B}|$) denote the item sets in the domains ${D^A}$ and ${D^B}$, respectively. The raw feature vector of each item in ${D^A}$ (or ${D^B}$) is defined as $\mathbf{E}_{vr}^A \in \mathbb{R}^{d^A}$ (or $\mathbf{E}_{vr}^B \in \mathbb{R}^{d^B}$), where ${d^A}$ (or ${d^B}$) is the dimensionality of features. The interaction matrices are denoted as ${\mathbf{R}^A} \in {\{ 0,1\} ^{m \times {n^A}}}$ and ${\mathbf{R}^B} \in {\{ 0,1\} ^{m \times {n^B}}}$ in ${D^A}$ and ${D^B}$, respectively.

To improve the performance of dual-target CDR, it is crucial to explicitly consider the impacts of observed confounders. These confounders include single-domain confounders ${\mathbf{C}_{sd}}$ and cross-domain confounders ${\mathbf{C}_{cd}}$, both of which simultaneously influence user preferences and user-item interactions. Addressing the impacts of these confounders necessitates significant adjustments to existing dual-target CDR models. Hence, it would be beneficial to propose a novel deconfounding framework that is highly extendable and compatible with most off-the-shelf dual-target CDR models. For this purpose, since DIDA-CDR \cite{zhu2023domain} is a representative and state-of-the-art dual-target CDR model, we choose it as the foundation for our problem definition.

DIDA-CDR has effectively decoupled three essential components of user preferences for modeling comprehensive user preferences $\mathbf{E}_u^*$, thus achieving good recommendation results in each domain. These three components include:

\begin{figure}[!t]
\centering
\includegraphics[scale=0.4]{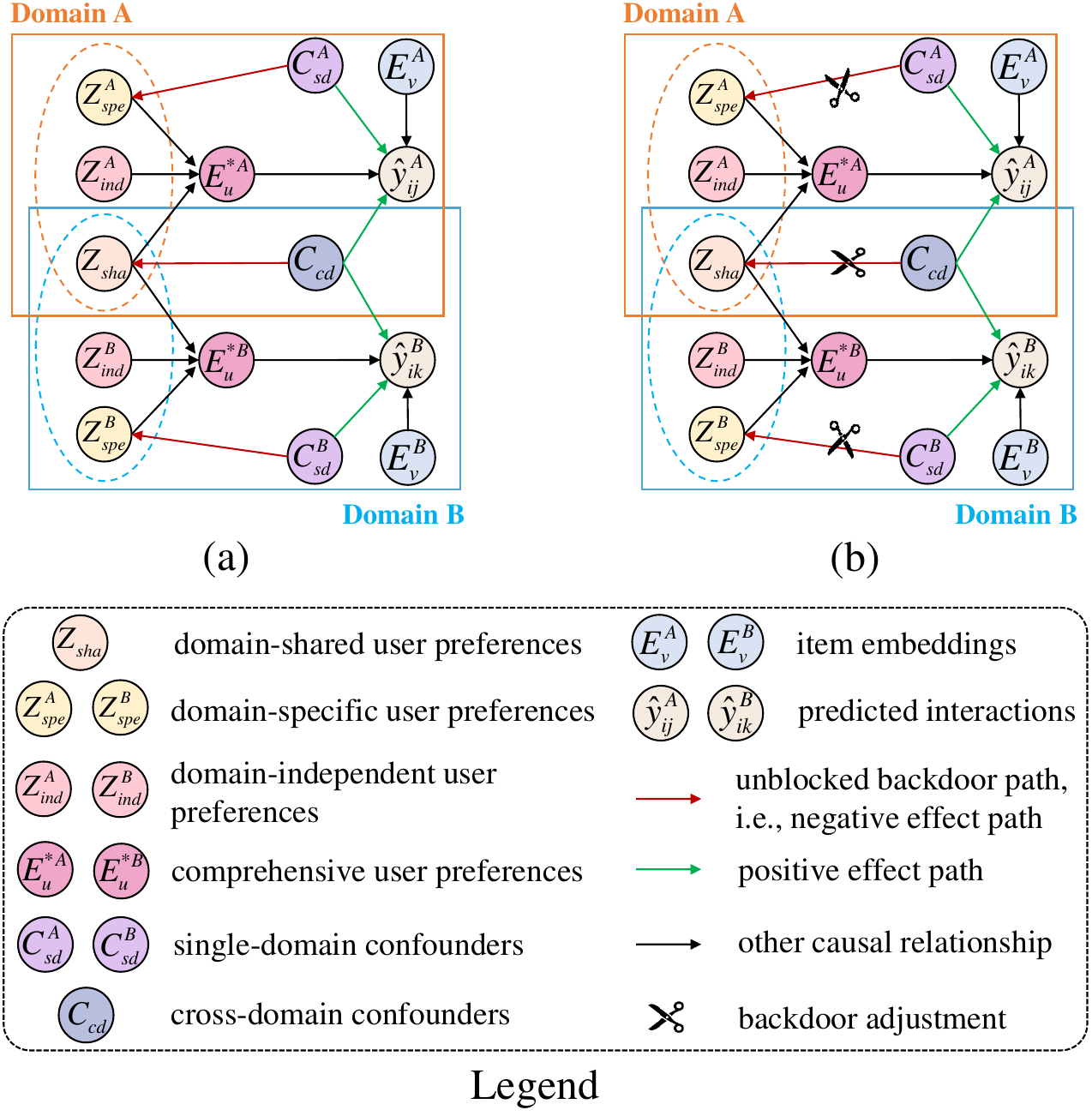} 
\vspace{-0.15in}
\caption{Causal graphs of dual-target CDR for deconfounding observed SDCs and CDCs. (a) Original causal graph. (b) Deconfounded causal graph after eliminating such observed confounders' negative effects by blocking backdoor paths via backdoor adjustment, as indicated by scissors \protect\cite{wang2022causal}.}
\label{causal_graph}
\vspace{-0.32in}
\end{figure}

\begin{itemize}[leftmargin=*]
\item[(1)] \emph{domain-shared user preferences} ${\mathbf{Z}_{sha}}$, which have the same meaning in each of both domains. For instance, users might prefer items in the sports ‘category’, which is the domain-shared preference covering both the ‘purchase’ and ‘add to favorite’ domains, reflecting consistent preferences across both domains.
\item[(2)] \emph{domain-specific user preferences} ${\mathbf{Z}_{spe}}$, which are unique to one domain. For example, in the ‘add to favorite’ domain, users might prefer ‘luxurious’ items that they cannot afford to purchase but still wish to add them to favorite, while in the ‘purchase’ domain, users might prefer ‘practical’ items that offer good value for money.
\item[(3)] \emph{domain-independent user preferences} ${\mathbf{Z}_{ind}}$, which are seemingly common in both domains but have different meanings in each domain \cite{zhu2023domain}. For instance, in the ‘purchase’ domain, a preference for ‘professional features’ refers to choosing items that are specialized and match the user’s current skill level or needs. Specifically, a beginner photography enthusiast might purchase an entry-level professional digital camera, which can mount different lenses for learning photography, emphasizing practicality and suitability for immediate use. By contrast, in the ‘add to favorite’ domain, a preference for ‘professional features’ reflects an aspiration for high-end items with advanced features, such as professional lenses, which are added to favorite for potential future use when the photography enthusiast’s skills improve. Unlike domain-specific user preferences that only exist in their corresponding domain, domain-independent user preferences exist in both domains but have different meanings in each domain.
 \end{itemize}
 
Based on the above notations, the problem of Causal Deconfounding for Dual-Target CDR is defined as follows.

\noindent \textbf{Causal Deconfounding for Dual-Target CDR.}
\emph{Given the domain-specific and comprehensive user preferences (i.e., ${\mathbf{Z}_{spe}}$ and $\mathbf{E}_u^*$) in each domain, the goal of causal deconfounding for dual-target CDR is to decouple observed single-domain confounders $\mathbf{C}_{sd}$ and cross-domain confounders $\mathbf{C}_{cd}$, eliminate such observed confounders' negative effects to obtain debiased comprehensive user preferences, and incorporate these confounders' positive effects into such debiased preferences to achieve a comprehensive understanding of user-item interactions, thus enhancing the recommendation accuracy in both domains.}

\subsection{Causal Graph}
A causal graph, i.e., a directed acyclic graph (DAG), where edges represent causal relationships between variables. Taking cross-domain confounders $\mathbf{C}_{cd}$ as an example, as illustrated in Fig. \ref{causal_graph}, they affect predicted interactions $\hat y$ via two types of paths: ${\mathbf{C}_{cd}} \to \hat y$ and ${\mathbf{C}_{cd}} \to {\mathbf{Z}_{sha}} \to \mathbf{E}_u^* \to \hat y$. The first type of path reveals that $\mathbf{C}_{cd}$, even if not the primary cause, i.e., users' true preferences, still have a direct positive impact on predicted interactions. The second type of path indicates that the negative impact of $\mathbf{C}_{cd}$ on domain-shared user preferences ${\mathbf{Z}_{sha}}$ induces confounding bias in both domains. Such confounding bias, in turn, skews comprehensive user preferences $\mathbf{E}_u^*$, because ${\mathbf{Z}_{sha}}$ is an essential component for capturing $\mathbf{E}_u^*$ \cite{zhu2023domain}. If the backdoor path ${\mathbf{C}_{cd}} \to {\mathbf{Z}_{sha}}$ is not blocked, $\mathbf{C}_{cd}$ will result in capturing biased comprehensive user preferences, thus yielding inaccurate recommendation results \cite{zhan2022deconfounding}. Similarly, single-domain confounders $\mathbf{C}_{sd}$ also have both positive and negative effects on predicted interactions and user preferences, respectively, thus the backdoor path ${\mathbf{C}_{sd}} \to {\mathbf{Z}_{spe}}$ in each domain should be blocked too.

Overall, the causal graph in Fig. \ref{causal_graph} provides a detailed view of how user preferences, observed confounders, and user-item interactions are causally related in dual-target CDR. In this study, we focus on addressing the confounding bias introduced by observed confounders in cross-domain settings. Firstly, we effectively decouple observed single-domain and cross-domain confounders. Secondly, we perform backdoor adjustment to preserve the positive direct effects of such observed confounders on predicted interactions and eliminate their negative effects on capturing comprehensive user preferences. These steps mitigate confounding biases to a large extent, enable a comprehensive understanding of user-item interactions, and thus improve the recommendation performance in both domains.

\begin{figure}[ht]
\centering
\includegraphics[scale=0.42]{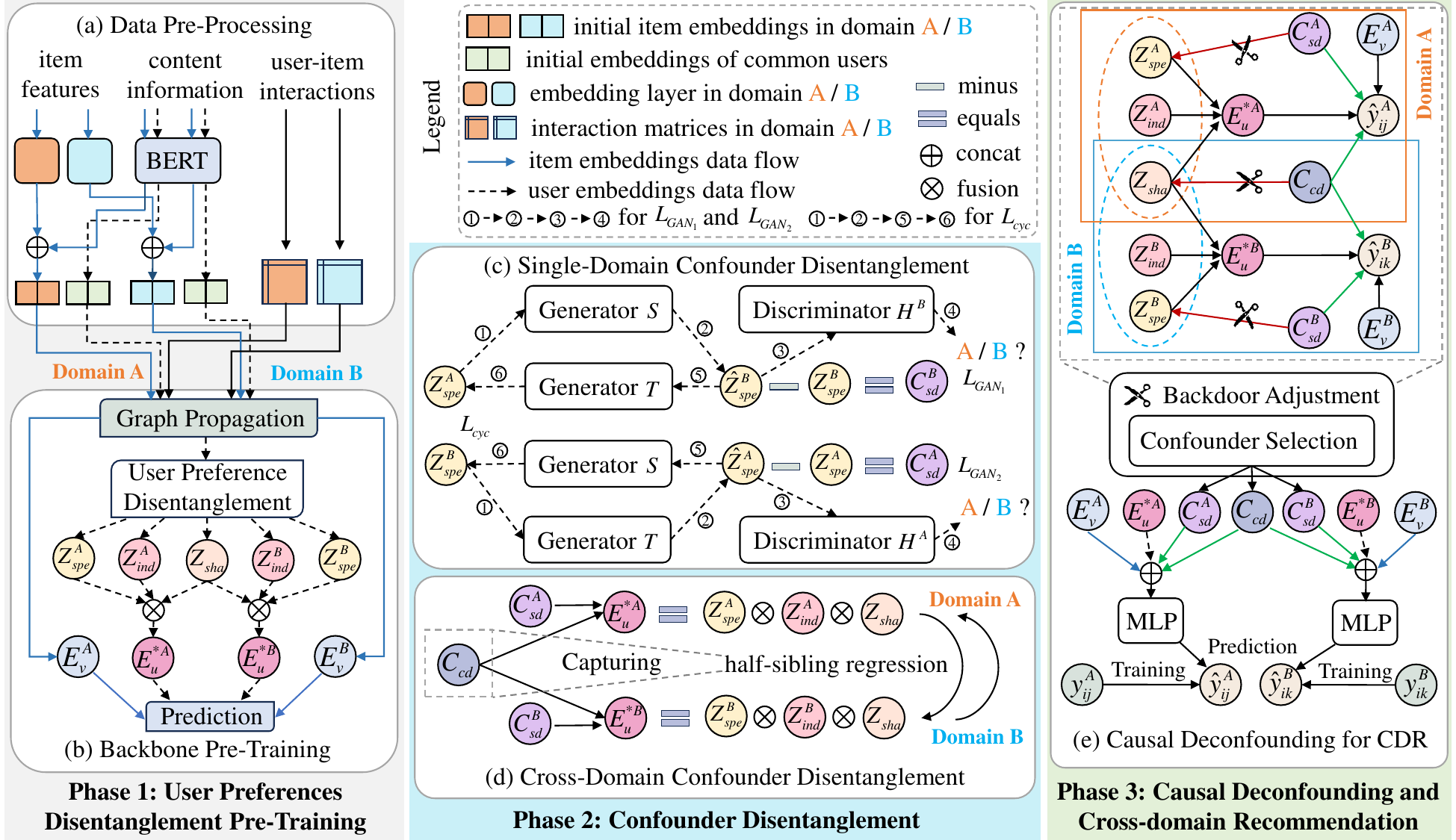} 
\vspace{-0.15in}
\caption{The structure of our CD2CDR Framework. The symbols and arrows not shown in the legend are defined in Table \ref{tab:important_notation} and Fig. \ref{causal_graph}.}
\label{flowchart}
\vspace{-0.2in}
\end{figure}

\subsection{Connection with Existing Works and Our Novel Insights}
Our proposed CD2CDR is the first work to explicitly decouple observed cross-domain confounders and incorporate the observed confounders' positive impacts into debiased comprehensive user preferences for dual-target CDR. Our study indicates that the proposed framework not only builds on existing works, but also provides several novel insights \cite{ding2025few}. Below, we analyze the remaining gaps in existing works, and explain how CD2CDR addresses these gaps and introduces novel insights:

\begin{itemize}[leftmargin=*]
\item \textbf{Neglecting observed confounders in user preference modeling:} Existing approaches \cite{cao2022disencdr,zhang2023disentangled} often emphasize decoupling essential components of user preferences to capture comprehensive user preferences, overlooking the impact of observed confounders on predicted interactions, which can lead to biased user preferences. In contrast, our CD2CDR addresses this gap by disentangling observed confounders and modeling the intricate causal relationships among such confounders, user preferences, and user-item interactions.

\item \textbf{Suboptimal deconfounding due to lack of explicit decoupling of cross-domain confounders:} Existing approaches either use off-the-shelf features as confounders \cite{liu2022causal,zhang2022learning} or neglect the need for decoupling cross-domain confounders \cite{li2021debiasing,zhang2023connecting}, resulting in suboptimal deconfounding performance. By contrast, our CD2CDR focuses on the explicit decoupling of both single-domain and cross-domain confounders, introducing a new perspective on deconfounding in cross-domain settings.

\item \textbf{Overlooking the positive impacts of observed confounders:} Existing approaches \cite{wang2021deconfounded,zhang2021causal} generally focus on intervening in the causal relationships that lead to bias and apply backdoor adjustment to remove these negative impacts, aiming to obtain debiased comprehensive user preferences. However, they often overlook the positive impacts of such confounders on predicted interactions. In contrast, although our model also utilizes backdoor adjustment, which is one of the widely used causal intervention techniques, we incorporate these positive impacts into the debiased comprehensive user preferences. It is worth noting that while backdoor adjustment itself is not our contribution, our novelty lies in the way we leverage it to selectively preserve the positive impacts of the observed confounders and eliminate their negative impacts, thereby achieving a more comprehensive understanding of user-item interactions.
\end{itemize}

\section{The Proposed Model}

\subsection{Overview of CD2CDR}
To enhance the recommendation accuracy in both domains, we propose a novel \textbf{C}ausal \textbf{D}econfounding framework via \textbf{C}onfounder \textbf{D}isentanglement for dual-target \textbf{C}ross-\textbf{D}omain \textbf{R}ecommendation, called CD2CDR. As depicted in Fig. \ref{flowchart}, the framework can be divided into three phases, i.e., \textbf{Phase 1:} User Preference Disentanglement Pre-Training, \textbf{Phase 2:} Confounder Disentanglement, and \textbf{Phase 3:} Causal Deconfounding and Cross-Domain Recommendation. In \textbf{Phase 1}, we obtain disentangled domain-independent and domain-specific user preferences in each domain and domain-shared user preferences by pre-training the backbone introduced in \cite{zhu2023domain}. In \textbf{Phase 2}, we first extract the SDCs in each domain by bidirectionally transforming domain-specific user preferences decoupled in Phase 1. Then, we distill confounding factors that jointly influence comprehensive user preferences in each of both domains as CDCs by adopting half-sibling regression \cite{scholkopf2016modeling}. In \textbf{Phase 3}, we utilize the backdoor adjustment to deconfound the observed confounders disentangled in \textbf{Phase 2}. Specifically, we design a confounder selection function to mitigate negative effects of such confounders on user preferences, thus recovering debiased comprehensive user preferences. We then incorporate the observed confounders' positive effects into such debiased preferences to predict user-item interactions via a multi-layer perceptron (MLP) in each of both domains. 

\subsection{Phase 1: User Preference Disentanglement Pre-Training}
Accurate disentanglement of user preferences is vital to ensure the robustness of subsequent confounder disentanglement process. Since the method introduced in \cite{zhu2023domain} excels at decoupling three essential components of user preferences for modeling comprehensive user preferences, our CD2CDR adopts it as the backbone for user preference disentanglement. To extract more accurate disentangled user preferences, we consider multi-source content information of users and items, e.g., user reviews and item details. Taking domain $A$ as an example, for each categorical feature field of an item (e.g., brand and category), we distill a set of unique features, which are then encoded into vectors using either one-hot or multi-hot encoding. Next, these encoded vectors are concatenated to form the raw feature vector for each item. We then transform the raw feature vectors of items $\mathbf{E}_{vr}^A \in \mathbb{R}^{d^A}$ into the dense embeddings $\mathbf{E}_{vd}^A \in \mathbb{R}^{d}$ as follows:
\begin{equation}
\mathbf{E}_{vd}^A = \mathbf{W}_{rd}^A\mathbf{E}_{vr}^A,
\end{equation}
where $\mathbf{W}_{rd}^A \in {\mathbb{R}^{{d^A} \times {d_d}}}$ is a trainable mapping matrix. ${d_d}$ denotes the dimensionality of dense embeddings. Then, for a user $u_i$, we collect all the user's reviews into a user text document. For an item $v_j$, we collect its title and all reviews on the item into an item text document. Next, we adopt a pre-trained BERT \cite{kenton2019bert} to map the documents of all users and items in the training set into user text embeddings $\mathbf{E}_{ut}^A$ and item text embeddings $\mathbf{E}_{vt}^A$, respectively. Finally, we concatenate $\mathbf{E}_{vd}^A$ and $\mathbf{E}_{vt}^A$ to form combined item embeddings $\mathbf{E}_{vc}^A$. We then transform $\mathbf{E}_{ut}^A$, $\mathbf{E}_{vc}^A$ into fixed-size initial user embeddings $\mathbf{E}_{ui}^A$ and initial item embedding $\mathbf{E}_{vi}^A$ in domain $A$ as follows:
\begin{equation}
\mathbf{E}_{ui}^A = \mathbf{\delta} _u^A\mathbf{E}_{ut}^A, \quad \mathbf{E}_{vi}^A = \mathbf{\delta} _v^A\mathbf{E}_{vc}^A,
\end{equation}
where $\mathbf{\delta} _u^A$ and $\mathbf{\delta} _v^A$ are the mapping functions of MLP layers. Similarly, we can obtain initial user embeddings $\mathbf{E}_{ui}^B$ and initial item embeddings $\mathbf{E}_{vi}^B$ in domain $B$. We then leverage such initial embeddings and the interaction matrices as inputs to pre-train the backbone. Specifically, we aggregate interaction data within each domain to build two heterogeneous graphs, which allow us to learn coarse user and item embeddings for each domain. Next, we apply linear interpolation to the user embeddings of both domains to generate augmented user representations, augmenting the sparser domain. With these coarse user embeddings and augmented user representations, we then employ a user preference disentanglement module, guided by a domain classifier, to decouple domain-independent, domain-specific, and domain-shared user preferences, namely, ${\mathbf{Z}_{ind}}$, ${\mathbf{Z}_{spe}}$, and ${\mathbf{Z}_{sha}}$ (for more information, please refer to \cite{zhu2023domain}). By incorporating the above three components of user preferences using attention mechanism in accordance with DIDA-CDR \cite{zhu2023domain}, we can obtain comprehensive user preferences $\mathbf{E}_u^*$.

\subsection{Phase 2: Confounder Disentanglement}
Since user-item interactions are also influenced by observed confounders apart from comprehensive user preferences, we propose to decouple SDCs and CDCs, as detailed in the following subsections.

\subsubsection{\textbf{Single-Domain Confounder Disentanglement}}
\label{sec:4.3.1}
To explore the SDCs, we utilize bi-directional domain transformation to decouple them from previously obtained domain-specific user preferences. If such SDCs are not identified and explicitly decoupled, their negative effects on domain-specific user preferences can hardly be eliminated. By contrast, if they are well disentangled, the causal deconfounding module can utilize backdoor adjustment to remove the confounding bias, thus obtaining debiased domain-specific user preferences. Inspired by CycleGAN \cite{zhu2017unpaired}, we devise a dual adversarial structure, which consists of two domain transformation generators and two discriminators, to disentangle SDCs in each domain. Specifically, we aim to learn two generators, i.e., $S( \cdot ):{D^A} \to {D^B}$ and $T( \cdot ):{D^B} \to {D^A}$, to transform domain-specific user preferences in each domain. 

Taking domain $B$ as an example, the generator $S(\cdot)$ takes the domain-specific user preferences ${\mathbf{Z}}_{spe}^A$ of common users in $D^A$ as inputs to generate $\hat{\mathbf{Z}}_{spe}^B = S(\mathbf{Z}_{spe}^A)$ that look similar to domain-specific user preferences in domain $B$, i.e., $\mathbf{Z}_{spe}^B$. However, if there are still differences between the simulated preferences $\hat{\mathbf{Z}}_{spe}^B$ and the original ones $\mathbf{Z}_{spe}^B$, such differences are not characteristics of domain-specific user preferences in domain $B$, but should be considered as SDCs \cite{zhang2023video}. To ensure that the generator $S(\cdot)$ is proficient at domain-specific preference simulation, we introduce a discriminator ${H^B}( \cdot )$ to recognize which domain the domain-specific user preferences come from. In the adversarial learning paradigm, the discriminator is expected to improve the ability to differentiate domain-specific user preferences in each domain to achieve better discriminative performance, while the generator is supposed to generate indistinguishable simulations of these domain-specific preferences to confuse such discriminator \cite{su2023cross}. For training the generator $S(\cdot)$ and the corresponding discriminator ${H^B}( \cdot )$, we adopt the adversarial loss \cite{goodfellow2014generative}, which can be expressed as follows:
\begin{equation}
\begin{split}
 &{\mathcal L_{GAN_1}(S,H^B,{D^A},{D^B})} = {\mathbb{E}_{\mathbf{Z}_{spe}^B \sim {\mathbb{P}^B}}}[\log H^B(\mathbf{Z}_{spe}^B)] \\
 &+ {\mathbb{E}_{\mathbf{Z}_{spe}^A \sim {\mathbb{P}^A}}}[\log (1 - H^B(S(\mathbf{Z}_{spe}^A)))],
\end{split}
\end{equation}
where $\mathbb{E}$ is the expectation, and $\mathbb{P}^A$, $\mathbb{P}^B$ denote the feature distribution of domain $A$ and domain $B$, respectively. Similarly, for training the generator $T(\cdot)$ and the corresponding discriminator ${H^A}( \cdot )$, we adopt the adversarial loss ${\mathcal L_{GAN_2}(T,H^A,{D^B},{D^A})}$. However, relying solely on adversarial loss is insufficient to ensure that a user's domain-specific preferences remain aligned with the user's preferences after transformation. If transformed domain-specific user preferences no longer reflect the user's preferences, then such transformation becomes meaningless, serving merely to confuse the discriminator. Hence, the generators should maintain cycle consistency, i.e., $\mathbf{Z}_{spe}^A \to S(\mathbf{Z}_{spe}^A) \to T(S(\mathbf{Z}_{spe}^A)) \approx \mathbf{Z}_{spe}^A$ and $\mathbf{Z}_{spe}^B \to T(\mathbf{Z}_{spe}^B) \to S(T(\mathbf{Z}_{spe}^B)) \approx \mathbf{Z}_{spe}^B$ during the training process (see $\mathcal L_{cyc}$ in Fig. \ref{flowchart}(c)). To this end, we apply a cycle consistency loss, which is represented as follows:
\begin{equation}
\begin{split}
{\mathcal L_{cyc}}(S,T) &= {\mathbb{E}_{\mathbf{Z}_{spe}^A \sim {\mathbb{P}^A}}}[||T(S(\mathbf{Z}_{spe}^A)) - Z_{spe}^A|{|_1}]\\
 &+ {\mathbb{E}_{\mathbf{Z}_{spe}^B \sim {\mathbb{P}^B}}}[||S(T(\mathbf{Z}_{spe}^B)) - \mathbf{Z}_{spe}^B|{|_1}].
\end{split}
\end{equation}

Moreover, the total objective function for training the generators and discriminators can be formulated as follows:
\begin{equation}
\begin{split}
 &{\mathcal L_{sd}}(S,T,H^A,H^B) = {\mathcal L_{GAN_1}}(S,H^B,{D^A},{D^B})\\
 &+ {\mathcal L_{GAN_2}}(T,H^A,{D^B},{D^A}) + \lambda {\mathcal L_{cyc}}(S,T),
\end{split}
\label{eq5}
\end{equation}
where $\lambda$ controls the importance of cycle consistency loss relative to adversarial losses. Following the method introduced in \cite{zhang2023video}, upon training completion, we calculate the differences between the domain-specific user preferences after transformation and the original ones as candidate SDCs. Even though confounding bias may still exist in the original domain-specific user preferences, the differences calculation helps to decouple candidate SDCs. By performing deconfounding on these decoupled confounders, the impact of such biases can be mitigated to a large extent. The differences are defined as follows:

\begin{equation}
\hat{\mathbf{C}}_{sd}^A = T(\mathbf{Z}_{spe}^B) - \mathbf{Z}_{spe}^A, \quad \hat{\mathbf{C}}_{sd}^B = S(\mathbf{Z}_{spe}^A) - \mathbf{Z}_{spe}^B.
\label{eq6}
\end{equation}

Recall the causal graph in Fig. \ref{causal_graph}(a), the negative effects of single-domain confounders $\mathbf{C}_{sd}$ on domain-specific user preferences $\mathbf{Z}_{spe}$ result in confounding bias, leading to inaccurate estimation of $\mathbf{Z}_{spe}$. For example, as depicted in Fig. \ref{confounder_example}(c), in the ‘purchase’ domain, a data-driven RS improperly perceives the ‘free shipping’ (i.e., an SDC), as Alice's ‘purchase’ domain preference, resulting in biased recommendation. Since our well-trained generator excels at simulating Alice's ‘purchase’ domain preferences based on her ‘add to favorite’ domain preferences, if there are still differences as per Eq. (\ref{eq6}), this indicates such differences are not Alice's ‘purchase’ domain preferences but rather SDCs independent of her preferences. Such SDCs (e.g., ‘free shipping’), previously entangled with Alice's ‘purchase’ domain preferences, are decoupled through our SDC disentanglement process. Note that this process specifically targets biased domain-specific user preferences, because unbiased ones are not entangled with such SDCs. Thus, although this process decouples SDCs from biased domain-specific user preferences, this does not imply a causal relationship $\mathbf{Z}_{spe} \to \mathbf{C}_{sd}$ in the causal graph, because SDCs are not generated by such biased domain-specific user preferences. To distill representative SDCs and reduce redundancy, we apply K-means clustering on candidate single-domain confounders $\hat{\mathbf{C}}_{sd}^A$ (or $\hat{\mathbf{C}}_{sd}^B$) and choose $J_{sd}^A$ (or $J_{sd}^B$) cluster centroids to form the potential SDC subspace $\mathcal{C}_{sd}^A$ (or $\mathcal{C}_{sd}^B$).

\subsubsection{\textbf{Cross-Domain Confounder Disentanglement}}
In addition to SDCs, it is more important to identify confounding factors that simultaneously affect user-item interactions in both domains. Inspired by the method introduced in \cite{zhang2023video}, we employ half-sibling regression to disentangle CDCs from the previously obtained comprehensive user preferences in both domains. Half-sibling regression excels at capturing the influence of confounding factors that simultaneously affect multiple observed variables \cite{scholkopf2016modeling}, and thus it is well suited for decoupling CDCs in dual-target CDR. As illustrated in Fig. \ref{causal_graph}(a), ${\mathbf{C}_{cd}} \to \mathbf{E}_u^{*A}$ and ${\mathbf{C}_{cd}} \to \mathbf{E}_u^{*B}$ indicate that CDCs indirectly influence the comprehensive user preferences in each of both domains via ${\mathbf{C}_{cd}} \to {\mathbf{Z}_{sha}} \to \mathbf{E}_u^{*A}$ and ${\mathbf{C}_{cd}} \to {\mathbf{Z}_{sha}} \to \mathbf{E}_u^{*B}$. The core idea of half-sibling regression is: if $\mathbf{E}_u^{*A}$ and $\mathcal{C}_{sd}^B$ are independent, then predicting $\mathbf{E}_u^{*B}$ based on $\mathbf{E}_u^{*A}$ becomes a method to selectively capture the influence of ${\mathbf{C}_{cd}}$ on $\mathbf{E}_u^{*B}$ (see Fig. \ref{flowchart}(d)). Similarly, predicting $\mathbf{E}_u^{*A}$ based on $\mathbf{E}_u^{*B}$ serves to capture the influence of ${\mathbf{C}_{cd}}$ on $\mathbf{E}_u^{*A}$ (for more information, please refer to \cite{scholkopf2016modeling}). Therefore, we can apply half-sibling regression to decouple $\mathbf{C}_{cd}$ from $\mathbf{E}_u^{*A}$ and $\mathbf{E}_u^{*B}$. Taking the regression from domain $A$ to domain $B$ as an example, our goal is to estimate a transformation matrix ${\mathbf{W}^{A \to B}}$ such that:
\begin{equation}
\mathbf{E}_u^{*B} \approx \mathbf{E}_u^{*A}{\mathbf{W}^{A \to B}},
\end{equation}
using ridge regression, and regression results are expressed as:
\begin{equation}
{\mathbf{W}^{A \to B}} = {[{(\mathbf{E}_u^{*A})^\top}\mathbf{E}_u^{*A} + \alpha \mathbf{I}]^{ - 1}}{(\mathbf{E}_u^{*A})^\top}\mathbf{E}_u^{*B},
\end{equation}
where $\alpha $ denotes the regularization parameter. We assume that $\mathbf{E}_u^{*A}$ and $\mathbf{C}_{sd}^B$ are independent, because $\mathbf{E}_u^{*A}$ are comprehensive user preferences in domain $A$, while $\mathbf{C}_{sd}^B$ are SDCs specific to domain $B$. When we estimate a transformation matrix ${\mathbf{W}^{A \to B}}$ to predict $\mathbf{E}_u^{*B}$ using $\mathbf{E}_u^{*A}$, the influence of $\mathbf{C}_{sd}^B$ on $\mathbf{E}_u^{*B}$ will not be captured. This is because $\mathbf{E}_u^{*A}$ are independent from $\mathbf{C}_{sd}^B$, and as a result, utilizing $\mathbf{E}_u^{*A}$ cannot predict $\mathbf{C}_{sd}^B$ and the influence of $\mathbf{C}_{sd}^B$ on $\mathbf{E}_u^{*B}$. By contrast, the influence of $\mathbf{C}_{cd}$ on $\mathbf{E}_u^{*B}$ will be captured, because $\mathbf{C}_{cd}$ simultaneously affect $\mathbf{E}_u^{*A}$ and $\mathbf{E}_u^{*B}$, which means the regression results will only capture $\mathbf{C}_{cd}$. Hence, the regression results can be identified as candidate cross-domain confounders:
\begin{equation}
{\hat{\mathbf{C}}_{cd}^{A \to B}} = \mathbf{E}_u^{*A}{\mathbf{W}^{A \to B}}.
\end{equation}
Similarly, we can obtain the regression results from domain $B$ to domain $A$, denoted as $\hat{\mathbf{C}}_{cd}^{B \to A}$. For cross-domain confounders, K-means clustering is also employed on the candidate cross-domain confounders $\hat{\mathbf{C}}_{cd}^{A \to B}$ and $\hat{\mathbf{C}}_{cd}^{B \to A}$, with the $J_{cd}$ cluster centroids forming the potential CDC subspace $\mathcal{C}_{cd}$.

\subsection{Phase 3: Causal Deconfounding and Cross-Domain Recommendation}
After the confounder disentanglement, we obtain the potential SDC subspaces $\mathcal{C}_{sd}^A$ and $\mathcal{C}_{sd}^B$, and potential CDC subspace $\mathcal{C}_{cd}$. From a causal perspective, if the backdoor paths (i.e., ${\mathbf{C}_{sd}} \to {\mathbf{Z}_{spe}}$ and ${\mathbf{C}_{cd}} \to {\mathbf{Z}_{sha}}$) are not blocked, the observed confounders $C$ will simultaneously influence user preferences $Z$ and user-item interactions $Y$ (see Fig. \ref{confounder_example}(a)), and thus cause biased estimation of comprehensive user preferences. To this end, we perform the \emph{do}-calculus intervention based on backdoor adjustment \cite{zhu2022deep} to block the backdoor paths $C \to Z$ and enable our model to more accurately estimate the direct effect $Z \to Y$ (also see Fig. \ref{confounder_example}(a)). Formally, the conventional likelihood $P(Y|Z)$ is defined as:
\begin{equation}
P(Y|Z) = \sum\nolimits_c {P(Y|Z,c)P(c|Z)},
\end{equation}
where $c$ denotes a specific confounder selected from the confounder space $\mathcal{C}$. By applying the \emph{do}-calculus, we exclude all influences directed towards the intervened variable (i.e., $Z$), and then we have:
\begin{equation}
\begin{split}
P(Y|do(Z)) &= \sum\nolimits_c {P(Y|do(Z),c)P(c|do(Z))} \\
 &= \sum\nolimits_c {P(Y|Z,c)P(c)}.
\end{split}
\end{equation}
For brevity, the detailed proof of the transformations ${P(Y|do(Z),c) = {P(Y|Z,c)}}$ and ${P(c|do(Z)) = {P(c)}}$ is omitted, which can be found in \cite{pearl2018book}. In fact, transforming $P(c|do(Z))$ into the prior probability of confounders $P(c)$ blocks backdoor paths $C \to Z$. As a result, $P(Y|do(Z))$ mainly focus on modeling the direct effect $Z \to Y$. Specifically, we implement the backdoor adjustment by modeling $P(Y|Z,c)$ with an interaction prediction network, which is expressed as follows:
\begin{equation}
P(Y|do(Z)) = {\mathbb{E}_c}[P(Y|Z,c)] = {\mathbb{E}_c}[f(\mathbf{E}_u^*,{\mathbf{E}_v},\mathbf{c})],
\label{eq12}
\end{equation}
where $f(\cdot)$ denotes a neural network, namely, MLP, to predict the probabilities of user-item interactions \cite{he2017neural}. $\mathbf{E}_u^*$ and $\mathbf{E}_v$ are comprehensive user preferences and pre-trained item embeddings obtained by the backbone in Phase 1, respectively. In other words, based on two subspaces of disentangled observed confounders in domain $A$ and domain $B$, i.e., ${\mathcal{C}^A} = \mathcal{C}_{sd}^A \cup {\mathcal{C}_{cd}}$ and ${\mathcal{C}^B} = \mathcal{C}_{sd}^B \cup {\mathcal{C}_{cd}}$, we apply backdoor adjustment to rectify the biased recommendations in each domain using Eq. (\ref{eq12}). Since the decoupled observed confounders are incorporated as part of the input to MLP, the direct influence of such confounders on user-item interactions $C \to Y$ is also considered. Moreover, inspired by \cite{zhang2023video}, we devise a confounder selection function to effectively control decoupled confounders for more accurate deconfounding.

Taking domain $A$ as an example, the confounder selection function is defined as follows:
\begin{equation}
\begin{split}
\phi (\mathbf{E}_u^{*A},\mathbf{E}_v^A,\mathbf{c}) &= \frac{{\exp (\mathbf{W}_u^A\mathbf{E}_u^{*A} \cdot \mathbf{W}_{uc}^A\mathbf{c})}}{{2\sum\nolimits_{\mathbf{c}'} {\exp (\mathbf{W}_u^A\mathbf{E}_u^{*A} \cdot \mathbf{W}_{uc}^A\mathbf{c}')} }} \\
&+ \frac{{\exp (\mathbf{W}_v^A\mathbf{E}_v^A \cdot \mathbf{W}_{vc}^A\mathbf{c})}}{{2\sum\nolimits_{\mathbf{c}'} {\exp (\mathbf{W}_v^A\mathbf{E}_v^A \cdot \mathbf{W}_{vc}^A\mathbf{c}')} }},
\end{split}
\end{equation}
where $\mathbf{c}'$ denotes any confounder selected from confounder subspace $\mathcal{C}^A$ and $\cdot$ is the dot product. $\mathbf{W}_u^A$, $\mathbf{W}_{uc}^A$, $\mathbf{W}_v^A$, $\mathbf{W}_{vc}^A$ are trainable matrices for embedding transformation. We then formulate the expectation ${\mathbb{E}_c}[f(\mathbf{E}_u^*,{\mathbf{E}_v},\mathbf{c})]$ as follows:
\begin{equation}
\begin{split}
&{\mathbb{E}_c}[f(\mathbf{E}_u^*,{\mathbf{E}_v},\mathbf{c})] = f(\mathbf{Q}_{in}^A)  \\
&= f[{\mathbf{W}_{fc}}(\mathbf{E}_u^{*A}||\mathbf{E}_v^A||\sum\nolimits_c {p(c)\mathbf{c}} \phi (\mathbf{E}_u^{*A},\mathbf{E}_v^A,\mathbf{c})],
\end{split}
\label{eq14}
\end{equation}
where $\mathbf{W}_{fc}$ is the weight matrix of the fully connected (FC) layer and $||$ is the concatenation operator. In practice, we assume a uniform distribution for the prior probability $p(c)$. In addition, $\mathbf{Q}_{in}^A={\mathbf{W}_{fc}}(\mathbf{E}_u^{*A}||\mathbf{E}_v^A||\sum\nolimits_c {p(c)\mathbf{c}} \phi (\mathbf{E}_u^{*A},\mathbf{E}_v^A,\mathbf{c})$ denotes the input for MLP in domain $A$. Moreover, the predicted interaction $\hat y_{ij}^A$ between an user ${u_i}$ and an item ${v_j}$ in domain $A$ is represented as follows:
\begin{equation}
\hat y_{ij}^A = \mathbf{\delta} _{out}^A(\mathbf{\delta} _l^A(...\mathbf{\delta} _2^A(\mathbf{\delta} _1^A(\mathbf{Q}_{in}^A))...)),
\end{equation}
where $\mathbf{\delta} _l^A$ is the mapping function for $l$-th MLP layer, and there are $l$ MLP layers including $\delta _{out}^A$ in domain $A$. Similarly, the predicted interaction $\hat y_{ij}^B$ in domain $B$ can be obtained.

The essence of our causal deconfounding module lies in blocking the backdoor paths $C \to Z$, allowing the model to concentrate on the direct effects of users' true preferences on the predicted interactions $Z \to Y$, and disregard the interference of observed confounders on these preferences. Specifically, the confounder selection function assigns different weights to the potential observed confounders, mitigates the effects of those irrelevant or harmful confounders to the prediction task, and enhances the direct effects of beneficial confounders on predicted interactions. Therefore, this module enables the model to eliminate the negative effects of such observed confounders to learn debiased comprehensive user preferences, and preserve the positive effects of these confounders on predicted interactions, thereby achieving a more comprehensive understanding of user-item interactions in both domains. Finally, we employ cross-entropy loss to fine-tune the user preference disentanglement backbone $g(\cdot)$ and the interaction prediction network $f(\cdot)$. To be specific, the final objective function in domain $A$ can be defined as follows:
\begin{equation}
{g^*},{f^*} = \mathop {\arg \min }\limits_{g,f} \sum\limits_{y \in {\mathcal{Y}^{A+}} \cup {\mathcal{Y}^{A-}}} {\ell (\hat y,y)},
\label{eq16}
\end{equation}
where $\hat y$ and $y$ are the predicted interaction and corresponding observed interaction, respectively. ${\ell}(\hat y,y)$ denotes the cross-entropy loss function. ${{\mathcal{Y}^{A+}}}$ denotes the observed interaction set, and ${{\mathcal{Y}^{A-}}}$ corresponds to a specific quantity of negative samples, which are randomly selected from unseen user-item interaction set in domain $A$ to mitigate the over-fitting. During the fine-tuning process, $g(\cdot)$ serves as the backbone, with the original prediction module being replaced by the interaction prediction network $f(\cdot)$. Likewise, we can obtain the objective function and predicted user-item interaction $\hat y_{ik}^B$ in domain $B$.

\subsection{Time Complexity Analysis}
\label{time_complexity}
Our CD2CDR mainly focuses on four modules: (1) Graph Propagation, (2) User Preference Disentanglement, (3) Confounder Disentanglement, and (4) Causal Deconfounding and Cross-domain Recommendation. While the first two modules are part of backbone model \cite{zhu2023domain}, the latter two constitute our novel framework. For simplicity and consistency, we assume that all embedding dimensions are $d$ and the number of layers in each network structure within each module is $L$ \cite{zhao2023cross}. The time complexity for each module can be analyzed as follows:

\noindent \textbf{(1) Graph Propagation:} Assuming the graph has $(m + n)$ nodes, where $m$ and $n$ are the number of users and items, respectively. In addition, assuming the average number of neighboring nodes for each node is ${\bar N}$, the time complexity for the graph propagation process per node is $O({\bar N} d)$. The total time complexity for graph propagation, using a graph convolutional network (GCN) with $L$ layers, is $O(L (m + n) {\bar N} d)$. Given that ${\bar N} \ll (m + n)$, this simplifies to $O(L (m + n) d)$.

\noindent \textbf{(2) User Preference Disentanglement:} Next, we conduct user preference disentanglement using an architecture similar to the VAE encoder. Given that this module is implemented with an MLP consisting of $L$ layers, the time complexity of user preference disentanglement is approximately $O(L m d^2)$. The time complexity of domain classifier can be ignored as it is relatively simple compared to main disentanglement module.

\noindent \textbf{(3) Confounder Disentanglement:} Then, we perform the confounder disentanglement module, which involves SDC and CDC disentanglement. For SDC disentanglement, we implement a structure similar to CycleGAN, using an MLP with $L$ layers to decouple candidate SDCs. The time complexity of SDC disentanglement can be roughly estimated as $O(L d^2 m)$. For CDC disentanglement, ridge regression is used to calculate a transformation matrix ${\mathbf{W}^{A \to B}}$ for obtaining candidate CDCs. The estimated time complexity is $O(m d^2 + d^3)$. Considering $d \ll m$, it simplifies to $O(m d^2)$. To identify representative SDCs and CDCs and eliminate redundancy, we perform K-means clustering on the candidate SDCs and CDCs. Given that the number of cluster centroids is $J$, the time complexity for the K-means clustering is estimated to be $O(m d J)$. Thus, the overall time complexity for the confounder disentanglement module is $O(L d^2 m + m d^2 + m d J)$, which simplifies to $O(m d (L d + J))$.

\noindent \textbf{(4) Causal Deconfounding and Cross-domain Recommendation:} Finally, we utilize the confounder selection function to effectively control the decoupled observed confounders, achieving more accurate deconfounding with a time complexity of approximately $O(m n J d)$. Subsequently, we concatenate the user embeddings, item embeddings, and selected confounders, feeding them into the MLP for prediction. Given that the prediction module consists of $L$ MLP layers, the time complexity can be estimated as $O(L m n d^2)$. Thus, the overall time complexity for the causal deconfounding and cross-domain recommendation module is $O(m n J d + L m n d^2)$, which simplifies to $O(m n d (J + L d))$.

Overall, the time complexity of our CD2CDR can be approximated as $O(m n d^2 (J + L))$, where $J$ is the number of cluster centroids, and $m$ and $n$ are the number of users and items, respectively. This approximation is based on combining the time complexities of all four modules and simplifying by focusing on the dominant terms. The overall time complexity exhibits a non-linear relationship with the number of users, items, observed confounders, and embedding dimensions.

\section{Experiments and Analysis}
Extensive experiments are conducted on seven real-world datasets to answer the following four research questions:

\begin{itemize}[leftmargin=*]
    \item \noindent \textbf{RQ1.} How does our model perform in comparison with state-of-the-art baseline models (see Section \ref{sec:V-B})?
    \item \noindent \textbf{RQ2.} How do different components, namely, confounder disentanglement, causal deconfounding and cycle consistency loss, influence the recommendation accuracy of our model (see Section \ref{sec:V-C})?
    \item \noindent \textbf{RQ3.} How do different backbone models impact the recommendation accuracy of our model (see Section \ref{sec:V-D})?
    \item \noindent \textbf{RQ4.} How do different hyper-parameter settings affect the recommendation accuracy of our model (see Section \ref{sec:V-E})?
\end{itemize}

\subsection{Experimental Settings}
\subsubsection{\textbf{Experimental Datasets}}
Semantic information, such as item titles containing details about free shipping, sales promotion, category, and brand, helps to disentangle user preferences and observed confounders. In e-commerce scenarios, semantic information is easily accessible and crucial for gaining a more comprehensive understanding of user-item interactions. To comprehensively evaluate our CD2CDR model, we conduct experiments in two distinct recommendation scenarios: (1) CDR with fully overlapping user sets and (2) cross-system recommendation (CSR) with only overlapping items and completely non-overlapping users.

For the CDR scenario, we select two real-world e-commerce datasets that provide rich semantic information, ratings, reviews and item metadata, namely, Rec-Tmall\footnote{https://tianchi.aliyun.com/dataset/140281} dataset \cite{he2023dmbin} and Amazon dataset \cite{cao2022disencdr}. For Amazon dataset, we choose two relevant domains, namely, Amazon-Electronics and Amazon-Cloth. Similarly, for Rec-Tmall dataset, we select three relevant behaviors as business domains, namely, Add to Favorite, Purchase, and Add to Cart\footnote{For brevity, we refer to these subsets as Tmall-Favorite, Tmall-Purchase, Tmall-Cart, Amazon-Elec, and Amazon-Cloth, respectively in subsequent discussions.}. In the Tmall-Favorite domain, most users engage in exploration-oriented behaviors, adding items they find appealing to their favorites without an immediate intent to purchase. In contrast, the Tmall-Purchase and Tmall-Cart domains reflect purchase-oriented behaviors, where users are more likely to select items that match their true preferences and may result in actual purchases. By defining these distinct behaviors as domains, we broaden the concept of domains to encompass varying user intents, thereby enhancing the flexibility of our CDR framework for broader application scenarios \cite{li2021debiasing}.

For the CSR scenario, we utilize two widely-used movie recommendation datasets: MovieLens 20M \cite{harper2015movielens} and Douban-Movie \cite{zhu2020graphical}. These datasets contain ratings and side information on common movies from different user communities, creating a scenario where user sets are completely non-overlapping while item sets are partially overlapping. This cross-system setting broadens our experimental scope beyond CDR to a more challenging CSR scenario that better testifies the effectiveness of our CD2CDR.

\begin{table}[t]
\caption{The statistics for three dual-target CDR tasks and a dual-target CSR task.}
\vspace{-0.15in}
\label{dataset_statistics}
\begin{threeparttable}
\begin{tabular}{cccccc}
\hline
Tasks                     & Datasets       & \#Users & \#Items & \#Interactions & Density \\ \hline
\multirow{2}{*}{Task \#1} & Tmall-Favorite & 25,434  & 99,237  & 500,876        & 0.020\% \\
                          & Tmall-Purchase & 25,434  & 28,817  & 55,057         & 0.008\% \\ \hline
\multirow{2}{*}{Task \#2} & Tmall-Favorite & 39,657  & 104,496 & 807,493        & 0.019\% \\
                          & Tmall-Cart     & 39,657  & 44,172  & 354,499        & 0.020\% \\ \hline
\multirow{2}{*}{Task \#3} & Amazon-Elec    & 15,761  & 51,447  & 224,689        & 0.027\% \\
                          & Amazon-Cloth   & 15,761  & 48,781  & 133,609        & 0.017\% \\ \hline
\multirow{2}{*}{Task \#4} & MovieLens      & 10,000  & 9,395$^{\dagger}$   & 1,462,905      & 1.56\%  \\
                          & Douban-Movie   & 2,712   & 34,893$^{\dagger}$  & 1,278,401      & 1.35\%  \\ \hline
\end{tabular}
\vspace{0.05in}
\begin{tablenotes}
\footnotesize
\item[${\dagger}$] There are 4,115 common items between MovieLens and Douban-Movie datasets.
\end{tablenotes}
\end{threeparttable}
\vspace{-0.15in}
\end{table}

\subsubsection{\textbf{Experimental Tasks}}
We construct four experimental tasks: three dual-target CDR tasks using e-commerce datasets and one CSR task using movie datasets. All tasks involve transforming explicit ratings into implicit feedback. For the CDR scenario, we design three tasks with fully overlapping user sets: (1) Tmall-Favorite and Tmall-Purchase, (2) Tmall-Favorite and Tmall-Cart, and (3) Amazon-Elec and Amazon-Cloth. These tasks are chosen to test the model's ability to handle diverse recommendations across different user interactions and preferences in e-commerce settings.

For Task \#1, users and items with fewer than 20 interactions are removed from Tmall-Favorite, and those with fewer than 5 interactions are filtered out from Tmall-Purchase. For the Task \#2 and Task \#3, users and items with fewer than 20 interactions in Task \#2 and those with fewer than 5 interactions in Task \#3 are filtered out. In line with the preprocessing operation taken for the two Amazon subsets in DisenCDR \cite{cao2022disencdr}, we also conduct the same operation on three Rec-Tmall subsets to remove the cold-start item entry for testing.

For the CSR scenario (Task \#4: MovieLens and Douban-Movie), we follow the filtering approach in \cite{ijcai2019p587,zhu2019dtcdr}, retaining users and items with at least 5 interactions in Douban-Movie and extracting a subset of 10,000 users with at least 5 interactions from MovieLens 20M. We then identify common items across the two datasets, enabling knowledge transfer through overlapping items despite having completely non-overlapping users. The statistics are shown in Table \ref{dataset_statistics}.

\subsubsection{\textbf{Parameter Settings}}
The settings of our backbone DIDA-CDR are consistent with those listed in its original paper \cite{zhu2023domain}, including the number of GCN layers, embedding dimension and information fusion approach, etc. In the interaction prediction network, the structure is $e \to 32 \to 16 \to q$, where $e$ is the combined size after the mapping of FC layer in Eq. (\ref{eq14}), and $q$ is the output size, i.e., the dimension of latent factors. We vary $e$ in the range of $\{64,128\}$ and $q$ in the range of $\{8,16\}$, and finally set $e = 128$ and $q = 8$. The initial parameters for all the above layers are set following a Gaussian distribution $X \sim \mathcal N(0,0.01)$. In line with the approach used in GA-DTCDR \cite{zhu2020graphical}, for each observed interaction, we randomly select 7 non-interacted items to serve as negative examples. For a fair comparison, we employ grid search to fine-tune the parameters of all models. Specifically, we select the learning rate in $\{0.01, 0.005, 0.001, 0.0005, 0.0001\}$. Moreover, we adopt the Adam optimizer \cite{kingma2014adam} for all models with a batch size of 1024. In addition, we keep the number of cluster centroids $J_{sd}^A = J_{sd}^B = J_{cd}$ and vary them in $\{2, 5, 10, 20, 50\}$. Furthermore, we investigate the weight of cycle consistency loss $\lambda$ in $\{0.1, 1, 2, 5, 10\}$, and the regularization parameter $\alpha $ in $\{0.1, 1, 10, 20, 50\}$. The influence of the above parameters on our CD2CDR is particularly discussed in Section \ref{sec:V-E}. 

\subsubsection{\textbf{Model Training}}
\label{model_training}
Since our CD2CDR can be divided into three phases, we first pre-train the backbone of our model with 50 epochs\footnote{The number of training epochs for each phase is chosen in $\{10, 20, 30, 40, 50\}$.} to obtain disentangled user preferences and comprehensive user preferences. Next, we train the generators and discriminators in the dual adversarial structure with 30 epochs to decouple SDCs, apply half-sibling regression, a computational method inherently without a training process \cite{scholkopf2016modeling}, to decouple CDCs, and then save cluster centroids of both SDCs and CDCs. Finally, we replace the prediction module in the backbone with the interaction prediction network to fine-tune the overall CD2CDR with 20 epochs.

To enhance the stability of our model training in the dual adversarial structure, inspired by \cite{zhu2017unpaired}, we replace the negative log-likelihood objective with a least-squares loss for the adversarial losses $\mathcal L_{GAN_1}(S,H^B,{D^A},{D^B})$ and $\mathcal L_{GAN_2}(T,H^A,{D^B},{D^A})$. This replacement enables the generator to produce higher-quality outputs and improves training stability. The reasons are as follows. Firstly, the least-squares loss penalizes generated samples far from the decision boundary, guiding the generator to adjust these samples closer to the boundary. This process reduces the discrepancy between generated and real data distributions, improving the quality of generated samples. Secondly, the distance-based penalization produces more gradients to guide the generator’s updates, mitigating the gradient vanishing issue and thereby stabilizing the generator’s learning process. For further details, please refer to \cite{mao2017least}. Taking $\mathcal L_{GAN_1}(S,H^B,{D^A},{D^B})$ as an example, the generator $S(\cdot)$ is trained to minimize ${\mathbb{E}_{\mathbf{Z}_{spe}^A \sim {\mathbb{P}^A}}}[(H^B(S(\mathbf{Z}_{spe}^A))-1)^2]$, while the corresponding discriminator ${H^B}( \cdot )$ is trained to minimize ${\mathbb{E}_{\mathbf{Z}_{spe}^B \sim {\mathbb{P}^B}}}[(H^B(\mathbf{Z}_{spe}^B)-1)^2] + {\mathbb{E}_{\mathbf{Z}_{spe}^A \sim {\mathbb{P}^A}}}[H^B(S(\mathbf{Z}_{spe}^A))^2]$. Likewise, $\mathcal L_{GAN_2}(T,H^A,{D^B},{D^A})$ is optimized in a similar manner. In addition, we adjust the weight of the cycle consistency loss to balance the adversarial process and the cycle consistency constraint, ensuring stable convergence of the dual adversarial training. Detailed results of this weight adjustment can be found in Section \ref{sec:5.3.3} and Section \ref{sec:5.5.2}.

During each epoch, we shuffle and split the training data for both domains into batches. We then iterate through batches, training on domain A and domain B in parallel. This approach allows the model to learn from both domains within the same epoch, ensuring that the model parameters are updated based on information from both domains. This form of joint learning helps improve the generalization performance across domains. Note that Eq. (\ref{eq5}) and Eq. (\ref{eq16}) are not optimized jointly. Since observed confounders are no longer entangled with debiased user preferences after deconfounding, the joint optimization of Eq. (\ref{eq5}) and Eq. (\ref{eq16}) for decoupling these confounders from such preferences becomes redundant. For a fair comparison, other baselines are trained for 100 epochs to confirm their convergence.

\subsubsection{\textbf{Evaluation Metrics}}
Given the widespread use of leave-one-out approach in baselines, e.g., GA-DTCDR \cite{zhu2020graphical} and DisenCDR \cite{cao2022disencdr}, we adopt it as well to validate the recommendation accuracy of our CD2CDR and baselines. Moreover, the test set is created by the final interaction of each user, while the training set is formed by the remaining interaction records of each user. In line with the methods introduced in \cite{cao2022disencdr,cao2022cross}, for every interaction in the test set, we randomly select 999 non-interacted items as negative samples for the test user, and then predict scores for a total of 1000 items to perform ranking. The leave-one-out approach mainly uses Hit Ratio (HR) and Normalized Discounted Cumulative Gain (NDCG), which are commonly adopted in ranking evaluations \cite{zhu2021unified}. In our experiments, these metrics are applied to validate the recommendation accuracy within top-10 rankings, and all experiments are conducted five times with average results reported in this paper.

\subsubsection{\textbf{Comparison Methods}}
We choose seventeen state-of-the-art baseline models to conduct a comparison against the proposed CD2CDR. We then categorize the seventeen baseline models into four groups: (I) Single-Domain Recommendation (SDR), (II) Single-Target CDR, (III) Disentanglement-Based Dual-Target CDR, and (IV) Debiasing Dual-Target CDR. To the best of our knowledge, our CD2CDR is the first Deconfounding Dual-Target CDR model in the literature. Thus, we select three representative Debiasing Dual-Target CDR approaches as alternatives for Deconfounding Dual-Target CDR baselines. Moreover, although there are some methods that identify unobserved domain-specific confounders and even unobserved general confounders, or utilize backdoor adjustment in the single-domain manner, they are not selected as baseline models. This is because they focus on different settings, i.e., domain generalization \cite{zhang2023connecting,lin2024pre}, CDSR \cite{zhao2023sequential,xu2024rethinking,zhang2024transferring}, click-through rate (CTR) prediction \cite{wang2022causalint,menglin2024c2dr} and different item groups \cite{zhang2021causal,wang2021deconfounded}, respectively, from our model. Furthermore, in Table \ref{tab:baseline-comparison}, we present an in-depth analysis of embedding strategies and main ideas of seventeen baselines and our CD2CDR. Detailed descriptions of these baselines are listed as follows.

\begin{table}[]
\caption{The comparison of the baselines and our proposed model.}
\vspace{-0.15in}
\label{tab:baseline-comparison}
\resizebox{\textwidth}{!}{
\begin{tabular}{ccc|c|l}
\hline
\multicolumn{3}{c|}{\textbf{Model}}                                                                                                                                                                                             & \textbf{Embedding Strategy}                   & \multicolumn{1}{c}{\textbf{Main Idea}}                                        \\ \hline
\multicolumn{1}{c|}{\multirow{17}{*}{Baselines}} & \multicolumn{1}{c|}{\multirow{3}{*}{\begin{tabular}[c]{@{}c@{}}Single-Domain \\ Recommendation (SDR)\end{tabular}}}              & \textbf{NGCF} \cite{wang2019neural}       & Graph Embedding                               & Devising an embedding propagation layer to encode the collaborative signal    \\ \cline{3-5} 
\multicolumn{1}{c|}{}                            & \multicolumn{1}{c|}{}                                                                                                            & \textbf{LightGCN} \cite{he2020lightgcn}   & Graph Embedding                               & Designing a simplified GCN using linear message propagation for RSs           \\ \cline{3-5} 
\multicolumn{1}{c|}{}                            & \multicolumn{1}{c|}{}                                                                                                            & \textbf{DCCF} \cite{ren2023disentangled}  & Graph Contrastive Learning \& Disentanglement & Decoupling user intent by adaptively integrating self-supervised augmentation \\ \cline{2-5} 
\multicolumn{1}{c|}{}                            & \multicolumn{1}{c|}{\multirow{3}{*}{\begin{tabular}[c]{@{}c@{}}Single-Target Cross-Domain\\  Recommendation (CDR)\end{tabular}}} & \textbf{BPR\_EMCDR} \cite{man2017cross}   & Linear Matrix Factorization (MF)              & Maping the latent factors of common users/items for knowledge transfer        \\ \cline{3-5} 
\multicolumn{1}{c|}{}                            & \multicolumn{1}{c|}{}                                                                                                            & \textbf{BPR\_DCDCSR} \cite{zhu2018deep}   & Linear MF                                     & Considering the rating sparsity degrees of individual users and items         \\ \cline{3-5} 
\multicolumn{1}{c|}{}                            & \multicolumn{1}{c|}{}                                                                                                            & \textbf{CUT} \cite{li2024aiming}          & Contrastive Learning                          & Filtering users’ collaborative information with user similarity constraints   \\ \cline{2-5} 
\multicolumn{1}{c|}{}                            & \multicolumn{1}{c|}{\multirow{8}{*}{\begin{tabular}[c]{@{}c@{}}Disentanglement-Based \\ Dual-Target CDR\end{tabular}}}           & \textbf{BiTGCF} \cite{liu2020cross}       & Graph Embedding                               & Integrating in-domain feature propagation and inter-domain feature transfer   \\ \cline{3-5} 
\multicolumn{1}{c|}{}                            & \multicolumn{1}{c|}{}                                                                                                            & \textbf{GA-DTCDR} \cite{zhu2020graphical} & Graph Embedding                               & Generating and effectively combining more representative embeddings           \\ \cline{3-5} 
\multicolumn{1}{c|}{}                            & \multicolumn{1}{c|}{}                                                                                                            & \textbf{DisenCDR} \cite{cao2022disencdr}  & Graph Embedding \& VAE                        & Developing two mutual-information-based disentanglement regularizers          \\ \cline{3-5} 
\multicolumn{1}{c|}{}                            & \multicolumn{1}{c|}{}                                                                                                            & \textbf{CausalCDR} \cite{li2024causalcdr} & Causal Embedding \& VAE                       & Using causal embeddings to model the joint distribution of user interactions  \\ \cline{3-5} 
\multicolumn{1}{c|}{}                            & \multicolumn{1}{c|}{}                                                                                                            & \textbf{GDCCDR} \cite{liu2024graph}       & Graph Embedding \& Contrastive Learning       & Using decoulpled graph for feature disentanglement and personalized transfer  \\ \cline{3-5} 
\multicolumn{1}{c|}{}                            & \multicolumn{1}{c|}{}                                                                                                            & \textbf{CrossAug} \cite{mao2024cross}     & Graph Embedding \& Augmentation               & Utilizing cross-domain interactions via intra- and inter-domain augmentation  \\ \cline{3-5} 
\multicolumn{1}{c|}{}                            & \multicolumn{1}{c|}{}                                                                                                            & \textbf{HJID} \cite{du2024joint}          & Graph Embedding \& Disentanglement            & Devising a hierarchical subspace disentanglement method for latent factors    \\ \cline{3-5} 
\multicolumn{1}{c|}{}                            & \multicolumn{1}{c|}{}                                                                                                            & \textbf{DIDA-CDR} \cite{zhu2023domain}    & Graph Embedding \& Disentanglement            & Decoupling three components to capture comprehensive user preferences         \\ \cline{2-5} 
\multicolumn{1}{c|}{}                            & \multicolumn{1}{c|}{\multirow{3}{*}{Debiasing Dual-Target CDR}}                                                                  & \textbf{SCDGN} \cite{li2022debiasing}     & Graph Embedding \& Semantic Clustering        & Distilling unbiased graph knowledge and learning debiasing vectors            \\ \cline{3-5} 
\multicolumn{1}{c|}{}                            & \multicolumn{1}{c|}{}                                                                                                            & \textbf{CDRIB} \cite{cao2022cross}        & Graph Embedding \& Information Bottleneck     & Designing two information bottleneck regularizers for debiasing               \\ \cline{3-5} 
\multicolumn{1}{c|}{}                            & \multicolumn{1}{c|}{}                                                                                                            & \textbf{IPSCDR} \cite{li2021debiasing}    & In accordance with the backbone (DIDA-CDR)    & Eliminating selection bias and transferring debiased user preferences         \\ \hline
\multicolumn{1}{c|}{Our model}                   & \multicolumn{1}{c|}{Deconfounding Dual-Target CDR}                                                                               & \textbf{CD2CDR}                           & In accordance with the backbone (DIDA-CDR)    & Disentangling fine-grained observed confounders and deconfounding             \\ \hline
\end{tabular}
}
\vspace{-0.3in}
\end{table}

\begin{itemize}[leftmargin=*]
\item \textbf{NGCF} \cite{wang2019neural} (I) is a classic recommendation framework based on a graph neural network that encodes collaborative signals through high-order connectivities using stacked embedding propagation layers.
\item \textbf{LightGCN} \cite{he2020lightgcn} (I) is a representative recommendation model that simplifies the GCN by using linear message propagation to learn user and item embeddings through neighborhood aggregation.
\item \textbf{DCCF} \cite{ren2023disentangled} (I) is a state-of-the-art framework for disentangling user intent in collaborative filtering by adaptively integrating self-supervised augmentation, leveraging global context and cross-view contrastive learning to enhance generalization and robustness in graph-based recommendation.
\item \textbf{BPR\_EMCDR} \cite{man2017cross} (II) utilizes Bayesian Personalized Ranking model (BPR) \cite{rendle2012bpr} as its matrix factorization model and maps the latent factors of common users/items across different domains for effective knowledge transfer.
\item \textbf{BPR\_DCDCSR} \cite{zhu2018deep} (II) integrates the latent factors from both domains by considering the sparsity degrees of individual users/items in each domain, creating more accurate benchmark factors to guide the deep neural network to map the latent factors across domains.
\item \textbf{CUT} \cite{li2024aiming} (II) is a state-of-the-art single-target CDR model, which uses the user similarity in the target domain as a filter for collaborative information from the source domain. Through a user transformation layer and contrastive loss, CUT constrains user representations to preserve user relationships in the target domain during information transfer.
\item \textbf{BiTGCF} \cite{liu2020cross} (III) combines high-order feature propagation on user-item graphs with a novel knowledge transfer mechanism, effectively balancing users' common features with domain-specific features across domains.
\item \textbf{GA-DTCDR} \cite{zhu2020graphical} (III) generates more representative user/item embeddings by constructing heterogeneous graphs from two domains and applies an element-wise attention mechanism to combine the embeddings of common users to enhance the recommendation accuracy in both domains.
\item \textbf{DisenCDR} \cite{cao2022disencdr} (III) uses two mutual-information-based regularizers to decouple domain-shared and domain-specific information, transferring only domain-shared information across domains to improve recommendation performance.
\item \textbf{CausalCDR} \cite{li2024causalcdr} (III) incorporates causality into CDR by using causal embeddings to model the joint distribution of interactions and utilizes an adversarial domain classifier to decouple the domain-specific and domain-shared features.
\item \textbf{GDCCDR} \cite{liu2024graph} (III) leverages two distinct contrastive learning-based constraints for feature disentanglement: one preserves domain-invariant features across domains, and the other disentangles domain-specific features via mutual information, with meta-networks supporting the personalized transfer of domain-invariant features.
\item \textbf{CrossAug} \cite{mao2024cross} (III) uses intra- and inter-domain data augmentation based on cross-reconstructed representations, while utilizing Householder transformations for domain-shared center alignment to mitigate the domain shift.
\item \textbf{HJID} \cite{du2024joint} (III) uses a hierarchical subspace disentanglement method to split user representations into generic shallow and domain-specific deep subspaces, utilizing a causal data generation graph to decouple domain-shared and domain-specific latent factors, thus enhancing robustness against distribution shifts across domains.
\item \textbf{DIDA-CDR} \cite{zhu2023domain} (III) is a state-of-the-art disentanglement-based dual-target CDR model that uniquely decouples domain-independent user preferences, as well as domain-shared and domain-specific user preferences, to capture more comprehensive user preferences for recommendation.
\item \textbf{SCDGN} \cite{li2022debiasing} (IV) builds a cross-domain user-cluster graph and employs a debiasing graph convolutional layer to extract and transfer unbiased graph knowledge between domains.
\item \textbf{CDRIB} \cite{cao2022cross} (IV) devises two information bottleneck regularizers to simultaneously model user-item interactions within and across domains, aiming to debias the user and item representations.
\item \textbf{IPSCDR} \cite{li2021debiasing} (IV) employs a generalized IPS estimator to mitigate selection bias in cross-domain contexts and devises three types of restrictions to learn propensity scores in the presence of unobserved domain-specific confounders. Since it is model-agnostic, for a fair comparison, we implement IPSCDR using the same state-of-the-art backbone (i.e., DIDA-CDR) as employed in our proposed CD2CDR.
\end{itemize}

Overall, our baselines cover both single-domain and cross-domain recommendation models. In the experiments, we use our CD2CDR framework to extend all the above Disentanglement-Based Dual-Target CDR baselines. Experimental results (see Section \ref{sec:V-D}) demonstrate that CD2CDR is highly extendable and compatible with most off-the-shelf disentanglement-based dual-target CDR backbones, making it suitable for a wide range of recommendation scenarios.

\subsection{Performance Comparison (for RQ1)}
\label{sec:V-B}
Table \ref{baseline_comparison} displays a comparative analysis of the performance\footnote{We only display experimental results when the embedding dimension $d=64$ in Table \ref{baseline_comparison} due to space limitation. For other values of $d$ that are not shown, similarly, our CD2CDR also significantly outperforms other baselines.} of different methods across all four tasks using HR@10 and NDCG@10 as evaluation metrics. It is worth mentioning that the Single-Target CDR baseline models are trained in both domains, but only their results in the data-sparser domain are reported, because they are designed to enhance the recommendation accuracy in the data-sparser domain. We can observe from Table \ref{baseline_comparison}: 

\begin{itemize}[leftmargin=*]
\item[(1)] Our CD2CDR improves Disentanglement-Based Dual-Target CDR baselines by an average of 16.73\% and 18.40\% w.r.t. HR@10 and NDCG@10, respectively. Among this type of baselines, BiTGCF \cite{liu2020cross} performs well on the Rec-Tmall dataset, outperforming GA-DTCDR \cite{zhu2020graphical} and DisenCDR \cite{cao2022disencdr}, but still falls short compared to CrossAug \cite{mao2024cross}. CrossAug, which utilizes cross-domain data augmentation and domain-shared center alignment, achieves competitive performance comparable to Debiasing Dual-Target CDR baselines. However, our CD2CDR still outperforms CrossAug by 11.95\% and 13.17\% w.r.t. HR@10 and NDCG@10, respectively. This is because, in addition to the user preference disentanglement, we adopt the confounder disentanglement, which effectively decouples observed SDCs and CDCs. By decoupling these confounders, we account for the fact that user interactions are not solely driven by their true preferences but also by observed confounders. Such confounders' positive influences can be secondary causes for user-item interactions, while their negative influences will result in capturing biased comprehensive user preferences. Effectively decoupling such observed confounders allows us to consider a more comprehensive range of factors affecting user-item interactions, thereby achieving better recommendation performance in both domains;

\item[(2)] Our CD2CDR improves Debiasing Dual-Target CDR baselines by an average of 13.03\% and 14.85\% w.r.t. HR@10 and NDCG@10, respectively. This demonstrates that deconfounding the observed confounders in each of both domains effectively benefits the prediction of user-item interactions in dual-target CDR;

\item[(3)] Our CD2CDR improves the best-performing baseline, i.e., IPSCDR \cite{li2021debiasing}, which is implemented with the same backbone as our model. Specifically, our CD2CDR outperforms IPSCDR with an average increase of 6.17\% and 8.23\% w.r.t. HR@10 and NDCG@10, respectively. This is because our CD2CDR particularly takes observed CDCs into consideration and our causal deconfounding module can not only eliminate observed confounders' negative effects on user preferences, but also preserve their positive effects on predicted interactions, thus gaining a more comprehensive understanding of user-item interactions;
\item[(4)] In the challenging CSR scenario (Task \#4) where user sets are completely non-overlapping, our CD2CDR still shows strong performance, outperforming the best-performing baseline model by an average of 4.75\% and 6.33\% w.r.t. HR@10 and NDCG@10, respectively. This demonstrates that our model adapts effectively to item-wise knowledge transfer through common items across different systems, and can extract item embeddings that are not entangled with observed confounders, enabling more accurate matching with comprehensive user preferences despite the absence of user overlap. This evaluation in the cross-system context extends our experimental scope beyond CDR scenarios, further validating the robustness and effectiveness of our model in more challenging CSR settings.
 \end{itemize}

\begin{table}[ht]
\caption{Comparative performance analysis (\%) of different methods in all four tasks using HR@10 and NDCG@10 as evaluation metrics \protect\cite{zhu2020graphical}. For experimental results, the best results are highlighted in bold and the results of best-performing baseline model are underlined (* denotes $p < 0.05$ in the paired t-test between the best-performing baseline model and CD2CDR) \protect\cite{zhu2023domain}.}
\vspace{-0.1in}
\label{baseline_comparison}
\centering
\resizebox{\textwidth}{!}{
\begin{tabular}{c|cccccc|cccccc}
\hline
\multirow{3}{*}{\textbf{Datasets}} & \multicolumn{6}{c|}{\textbf{SDR Baselines}}                                                         & \multicolumn{6}{c}{\textbf{Single-Target CDR Baselines}}                                        \\ \cline{2-13} 
                                   & \multicolumn{2}{c|}{NGCF}          & \multicolumn{2}{c|}{LightGCN}      & \multicolumn{2}{c|}{DCCF} & \multicolumn{2}{c|}{BPR\_EMCDR}   & \multicolumn{2}{c|}{BPR\_DCDCSR}  & \multicolumn{2}{c}{CUT} \\ \cline{2-13} 
                                   & HR    & \multicolumn{1}{c|}{NDCG}  & HR    & \multicolumn{1}{c|}{NDCG}  & HR          & NDCG        & HR    & \multicolumn{1}{c|}{NDCG} & HR    & \multicolumn{1}{c|}{NDCG} & HR          & NDCG      \\ \hline
Tmall-Favorite                     & 12.39 & \multicolumn{1}{c|}{6.27}  & 12.62 & \multicolumn{1}{c|}{6.35}  & 12.54       & 6.32        & -     & \multicolumn{1}{c|}{-}    & -     & \multicolumn{1}{c|}{-}    & -           & -         \\
Tmall-Purchase                     & 6.46  & \multicolumn{1}{c|}{4.01}  & 6.74  & \multicolumn{1}{c|}{4.06}  & 6.88        & 4.21        & 5.31  & \multicolumn{1}{c|}{3.56} & 5.88  & \multicolumn{1}{c|}{3.73} & 7.25        & 4.34      \\ \hline
Tmall-Favorite                     & 12.48 & \multicolumn{1}{c|}{6.29}  & 12.81 & \multicolumn{1}{c|}{6.43}  & 12.65       & 6.37        & -     & \multicolumn{1}{c|}{-}    & -     & \multicolumn{1}{c|}{-}    & -           & -         \\
Tmall-Cart                         & 10.04 & \multicolumn{1}{c|}{5.23}  & 10.85 & \multicolumn{1}{c|}{5.58}  & 10.98       & 5.8         & 8.79  & \multicolumn{1}{c|}{4.87} & 9.46  & \multicolumn{1}{c|}{5.11} & 11.48       & 6.16      \\ \hline
Amazon-Elec                        & 21.85 & \multicolumn{1}{c|}{12.36} & 21.73 & \multicolumn{1}{c|}{11.61} & 21.57       & 11.14       & -     & \multicolumn{1}{c|}{-}    & -     & \multicolumn{1}{c|}{-}    & -           & -         \\
Amazon-Cloth                       & 11.62 & \multicolumn{1}{c|}{6.18}  & 12.04 & \multicolumn{1}{c|}{6.22}  & 12.23       & 6.25        & 10.69 & \multicolumn{1}{c|}{5.47} & 11.44 & \multicolumn{1}{c|}{6.15} & 13.52       & 7.1       \\ \hline
MovieLens                          & 13.17 & \multicolumn{1}{c|}{6.85}  & 13.46 & \multicolumn{1}{c|}{7.09}  & 13.35       & 7.04        & -     & \multicolumn{1}{c|}{-}    & -     & \multicolumn{1}{c|}{-}    & -           & -         \\
\multicolumn{1}{l|}{Douban-Movie}  & 10.48 & \multicolumn{1}{c|}{5.38}  & 10.71 & \multicolumn{1}{c|}{5.48}  & 10.83       & 5.56        & 9.72  & \multicolumn{1}{c|}{5.19} & 10.36 & \multicolumn{1}{c|}{5.28} & 11.25       & 6.07      \\ \hline
\end{tabular}
}
\vspace{-0.05in}
\end{table}

\begin{table}[ht]
\vspace{-0.15in}
\begin{tabular}{c|cccccccc}
\hline
\multirow{3}{*}{\textbf{Datasets}} & \multicolumn{8}{c}{\textbf{Disentanglement-Based Dual-Target CDR Baselines}}                                                                 \\ \cline{2-9} 
                                   & \multicolumn{2}{c|}{BiTGCF}        & \multicolumn{2}{c|}{GA-DTCDR}      & \multicolumn{2}{c|}{DisenCDR}      & \multicolumn{2}{c}{CausalCDR} \\ \cline{2-9} 
                                   & HR    & \multicolumn{1}{c|}{NDCG}  & HR    & \multicolumn{1}{c|}{NDCG}  & HR    & \multicolumn{1}{c|}{NDCG}  & HR            & NDCG          \\ \hline
Tmall-Favorite                     & 15.68 & \multicolumn{1}{c|}{8.63}  & 14.87 & \multicolumn{1}{c|}{7.82}  & 15.34 & \multicolumn{1}{c|}{8.25}  & 15.96         & 8.69          \\
Tmall-Purchase                     & 9.24  & \multicolumn{1}{c|}{5.06}  & 8.44  & \multicolumn{1}{c|}{4.53}  & 9.01  & \multicolumn{1}{c|}{4.94}  & 9.17          & 5.03          \\ \hline
Tmall-Favorite                     & 15.80 & \multicolumn{1}{c|}{8.65}  & 14.91 & \multicolumn{1}{c|}{8.05}  & 15.39 & \multicolumn{1}{c|}{8.53}  & 16.25         & 9.10          \\
Tmall-Cart                         & 13.45 & \multicolumn{1}{c|}{7.13}  & 12.66 & \multicolumn{1}{c|}{6.38}  & 13.13 & \multicolumn{1}{c|}{6.81}  & 13.56         & 7.18          \\ \hline
Amazon-Elec                        & 23.42 & \multicolumn{1}{c|}{13.65} & 24.79 & \multicolumn{1}{c|}{13.87} & 24.53 & \multicolumn{1}{c|}{14.02} & 25.14         & 14.47         \\
Amazon-Cloth                       & 14.31 & \multicolumn{1}{c|}{7.59}  & 14.58 & \multicolumn{1}{c|}{7.64}  & 15.81 & \multicolumn{1}{c|}{8.56}  & 15.93         & 8.65          \\ \hline
MovieLens                          & 15.28 & \multicolumn{1}{c|}{8.25}  & 15.53 & \multicolumn{1}{c|}{8.54}  & 19.28 & \multicolumn{1}{c|}{10.53} & 19.87         & 10.89         \\
\multicolumn{1}{l|}{Douban-Movie}  & 11.79 & \multicolumn{1}{c|}{6.19}  & 11.96 & \multicolumn{1}{c|}{6.21}  & 14.72 & \multicolumn{1}{c|}{7.66}  & 15.29         & 8.27          \\ \hline
\end{tabular}
\vspace{-0.05in}
\end{table}

\begin{table}[ht]
\vspace{-0.15in}
\begin{tabular}{c|cccccccc}
\hline
\multirow{3}{*}{\textbf{Datasets}} & \multicolumn{8}{c}{\textbf{Disentanglement-Based Dual-Target CDR Baselines}}                                                                \\ \cline{2-9} 
                                   & \multicolumn{2}{c|}{GDCCDR}        & \multicolumn{2}{c|}{CrossAug}      & \multicolumn{2}{c|}{HJID}          & \multicolumn{2}{c}{DIDA-CDR} \\ \cline{2-9} 
                                   & HR    & \multicolumn{1}{c|}{NDCG}  & HR    & \multicolumn{1}{c|}{NDCG}  & HR    & \multicolumn{1}{c|}{NDCG}  & HR            & NDCG         \\ \hline
Tmall-Favorite                     & 16.24 & \multicolumn{1}{c|}{9.05}  & 16.31 & \multicolumn{1}{c|}{9.07}  & 16.45 & \multicolumn{1}{c|}{9.12}  & 16.60         & 9.14         \\
Tmall-Purchase                     & 9.65  & \multicolumn{1}{c|}{5.19}  & 9.74  & \multicolumn{1}{c|}{5.20}  & 9.88  & \multicolumn{1}{c|}{5.22}  & 10.03         & 5.28         \\ \hline
Tmall-Favorite                     & 16.53 & \multicolumn{1}{c|}{9.13}  & 16.66 & \multicolumn{1}{c|}{9.15}  & 16.79 & \multicolumn{1}{c|}{9.19}  & 17.02         & 9.26         \\
Tmall-Cart                         & 14.21 & \multicolumn{1}{c|}{7.56}  & 14.38 & \multicolumn{1}{c|}{7.59}  & 14.43 & \multicolumn{1}{c|}{7.60}  & 14.56         & 7.71         \\ \hline
Amazon-Elec                        & 25.69 & \multicolumn{1}{c|}{14.58} & 25.73 & \multicolumn{1}{c|}{14.62} & 25.94 & \multicolumn{1}{c|}{14.71} & 26.12         & 14.83        \\
Amazon-Cloth                       & 16.72 & \multicolumn{1}{c|}{9.16}  & 16.85 & \multicolumn{1}{c|}{9.24}  & 17.27 & \multicolumn{1}{c|}{9.42}  & 17.75         & 9.70         \\ \hline
MovieLens                          & 20.14 & \multicolumn{1}{c|}{10.92} & 20.43 & \multicolumn{1}{c|}{10.96} & 20.64 & \multicolumn{1}{c|}{11.01} & 21.08         & 11.09        \\
\multicolumn{1}{l|}{Douban-Movie}  & 15.58 & \multicolumn{1}{c|}{8.63}  & 15.82 & \multicolumn{1}{c|}{8.67}  & 16.11 & \multicolumn{1}{c|}{8.85}  & 16.34         & 9.06         \\ \hline
\end{tabular}
\vspace{-0.05in}
\end{table}

\begin{table}[ht]
\vspace{-0.15in}
\begin{tabular}{c|cccccc|cc|cc}
\hline
\multirow{3}{*}{\textbf{Datasets}} & \multicolumn{6}{c|}{\textbf{Debiasing Dual-Target CDR Baselines}}                                     & \multicolumn{2}{c|}{\textbf{\begin{tabular}[c]{@{}c@{}}Our Model\end{tabular}}} & \multicolumn{2}{c}{\textbf{Improvement}}                                                            \\ \cline{2-11} 
                                   & \multicolumn{2}{c|}{SCDGN}         & \multicolumn{2}{c|}{CDRIB}         & \multicolumn{2}{c|}{IPSCDR} & \multicolumn{2}{c|}{CD2CDR}                                                                                  & \multicolumn{2}{c}{\textbf{\begin{tabular}[c]{@{}c@{}}(CD2CDR vs.\\  best baselines)\end{tabular}}} \\ \cline{2-11} 
                                   & HR    & \multicolumn{1}{c|}{NDCG}  & HR    & \multicolumn{1}{c|}{NDCG}  & HR           & NDCG         & HR                                                    & NDCG                                                 & HR                                               & NDCG                                             \\ \hline
Tmall-Favorite                     & 15.07 & \multicolumn{1}{c|}{8.16}  & 15.9  & \multicolumn{1}{c|}{8.62}  & {\ul 17.14}  & {\ul 9.38}   & \textbf{18.01*}                                       & \textbf{9.87*}                                       & 5.08\%                                           & 5.22\%                                           \\
Tmall-Purchase                     & 8.73  & \multicolumn{1}{c|}{4.75}  & 9.56  & \multicolumn{1}{c|}{5.17}  & {\ul 10.51}  & {\ul 5.39}   & \textbf{11.38*}                                       & \textbf{6.12*}                                       & 8.28\%                                           & 13.54\%                                          \\ \hline
Tmall-Favorite                     & 15.14 & \multicolumn{1}{c|}{8.21}  & 16.22 & \multicolumn{1}{c|}{9.04}  & {\ul 17.43}  & {\ul 9.47}   & \textbf{18.35*}                                       & \textbf{9.98*}                                       & 5.28\%                                           & 5.39\%                                           \\
Tmall-Cart                         & 12.82 & \multicolumn{1}{c|}{6.48}  & 13.98 & \multicolumn{1}{c|}{7.51}  & {\ul 15.05}  & {\ul 8.02}   & \textbf{16.37*}                                       & \textbf{9.06*}                                       & 8.77\%                                           & 12.97\%                                          \\ \hline
Amazon-Elec                        & 24.69 & \multicolumn{1}{c|}{13.85} & 25.06 & \multicolumn{1}{c|}{14.34} & {\ul 26.78}  & {\ul 15.19}  & \textbf{28.11*}                                       & \textbf{16.24*}                                      & 4.97\%                                           & 6.91\%                                           \\
Amazon-Cloth                       & 15.19 & \multicolumn{1}{c|}{8.24}  & 16.67 & \multicolumn{1}{c|}{9.18}  & {\ul 18.26}  & {\ul 9.93}   & \textbf{19.62*}                                       & \textbf{10.87*}                                      & 7.45\%                                           & 9.47\%                                           \\ \hline
MovieLens                          & 20.81 & \multicolumn{1}{c|}{11.05} & 19.36 & \multicolumn{1}{c|}{10.55} & {\ul 21.74}        & {\ul 11.67}        & \textbf{22.73*}                                       & \textbf{12.56*}                                      & 4.55\%                                           & 7.63\%                                           \\
\multicolumn{1}{l|}{Douban-Movie}  & 15.88 & \multicolumn{1}{c|}{8.67}  & 15.13 & \multicolumn{1}{c|}{8.17}  & {\ul 16.75}        & {\ul 9.18}         & \textbf{17.59*}                                       & \textbf{9.61*}                                       & 5.01\%                                           & 4.68\%                                           \\ \hline
\end{tabular}
\vspace{-0.05in}
\end{table}

\begin{table}[ht]
\caption{Ablation study of different components in our CD2CDR across three dual-target CDR tasks and a dual-target CSR task. The best results are highlighted in bold.}
\label{ablation_study}
\vspace{-0.1in}
\begin{tabular}{c|cc|cc|cc|cc|cc}
\hline
\multirow{2}{*}{Datasets}         & \multicolumn{2}{c|}{CD2CDR\_Cross} & \multicolumn{2}{c|}{CD2CDR\_Single} & \multicolumn{2}{c|}{CD2CDR\_Coarse} & \multicolumn{2}{c|}{CD2CDR\_Cycle} & \multicolumn{2}{c}{CD2CDR}      \\ \cline{2-11} 
                                  & HR               & NDCG            & HR               & NDCG             & HR               & NDCG             & HR               & NDCG            & HR             & NDCG           \\ \hline
Tmall-Favorite                    & 17.25            & 9.41            & 17.48            & 9.56             & 15.03            & 8.14             & 17.86            & 9.70            & \textbf{18.01} & \textbf{9.87}  \\
Tmall-Purchase                    & 10.67            & 5.40            & 10.82            & 5.53             & 8.68             & 4.69             & 11.22            & 6.04            & \textbf{11.38} & \textbf{6.12}  \\ \hline
Tmall-Favorite                    & 17.64            & 9.59            & 17.85            & 9.69             & 15.11            & 8.16             & 18.23            & 9.89            & \textbf{18.35} & \textbf{9.98}  \\
Tmall-Cart                        & 15.21            & 8.25            & 15.56            & 8.61             & 12.75            & 6.43             & 16.18            & 9.01            & \textbf{16.37} & \textbf{9.06}  \\ \hline
Amazon-Elec                       & 26.83            & 15.23           & 26.97            & 15.32            & 24.48            & 13.79            & 27.94            & 16.15           & \textbf{28.11} & \textbf{16.24} \\
Amazon-Cloth                      & 18.32            & 9.96            & 18.44            & 10.01            & 14.76            & 7.68             & 19.37            & 10.58           & \textbf{19.62} & \textbf{10.87} \\ \hline
MovieLens                         & 21.89            & 12.38           & 22.03            & 12.45            & 20.23            & 10.93            & 22.55            & 12.49           & \textbf{22.73} & \textbf{12.56} \\
\multicolumn{1}{l|}{Douban-Movie} & 16.91            & 9.22            & 17.16            & 9.40             & 15.46            & 8.54             & 17.38            & 9.51            & \textbf{17.59} & \textbf{9.61}  \\ \hline
\end{tabular}
\vspace{-0.1in}
\end{table}

\subsection{Ablation Study (for RQ2)}
\label{sec:V-C}
To highlight the significance of each component in enhancing the recommendation accuracy of our model, we reconstruct our CD2CDR into four variants and perform an ablation study for all four tasks.

\subsubsection{\textbf{Impact of Confounder Disentanglement}}
We modify our proposed CD2CDR to form two variants, namely \textbf{CD2CDR\_Cross} and \textbf{CD2CDR\_Single}, by removing the SDC disentanglement and CDC disentanglement, respectively. From Table \ref{ablation_study}, we can observe that with SDC disentanglement module, our proposed CD2CDR outperforms \textbf{CD2CDR\_Cross} with an average improvement of 5.49  \%. This shows that the dual adversarial structure can effectively disentangle observed SDCs, and SDCs play an important role in predicting user-item interactions in each domain. In addition, our proposed CD2CDR improves \textbf{CD2CDR\_Single} by an average increase of 4.23\%. This indicates that half-sibling regression is well suited for decoupling observed CDCs, which are essential factors for achieving a comprehensive understanding of user-item interactions in both domains. Overall, our confounder disentanglement module can explicitly decouple more accurate observed confounders, especially the CDCs, thus enable our model to obtain better recommendation performance via accurate causal deconfounding.

\subsubsection{\textbf{Impact of Causal Deconfounding}}
Moreover, another variant, namely \textbf{CD2CDR\_Coarse}, directly incorporates decoupled observed confounders with biased comprehensive user preferences in each domain and does not include the causal deconfounding module. From Table \ref{ablation_study}, we can observe that without the causal deconfounding module, the recommendation accuracy of \textbf{CD2CDR\_Coarse} drops by 21.35\% on average, making it less effective compared to the Debiasing Dual-Target CDR baselines. This shows that the causal deconfounding module indeed helps the model control the negative effects of SDCs and CDCs on user preferences. By recovering debiased comprehensive user preferences and then incorporating the positive effects of SDCs and CDCs into such preferences, the module enables the model to obtain the better recommendation accuracy in both domains.

\subsubsection{\textbf{Impact of Cycle Consistency Loss}}
\label{sec:5.3.3}
In addition, we construct another variant, namely \textbf{CD2CDR\_Cycle}, by removing the cycle consistency loss in the SDC disentanglement module. From Table \ref{ablation_study}, we can observe that our CD2CDR improves \textbf{CD2CDR\_Cycle} by an average of 1.01\%. This demonstrates that the cycle consistency loss effectively preserves users' domain-specific preferences during the transformation process, ensuring the transformed preferences accurately reflect the original user preferences rather than merely confusing the discriminator. By incorporating the cycle consistency loss to stabilize the adversarial loss, our model can more accurately disentangle SDCs, providing strong support for explicitly considering the impact of observed confounders on user preferences and user-item interactions.

Overall, our ablation study demonstrates the importance of each component in our CD2CDR model. Similar trends are observed in the CSR scenario (Task \#4) where knowledge transfer relies on common items, further confirming the effectiveness of these components across different systems.

\begin{figure}[!t]
\centering
\includegraphics[scale=0.5]{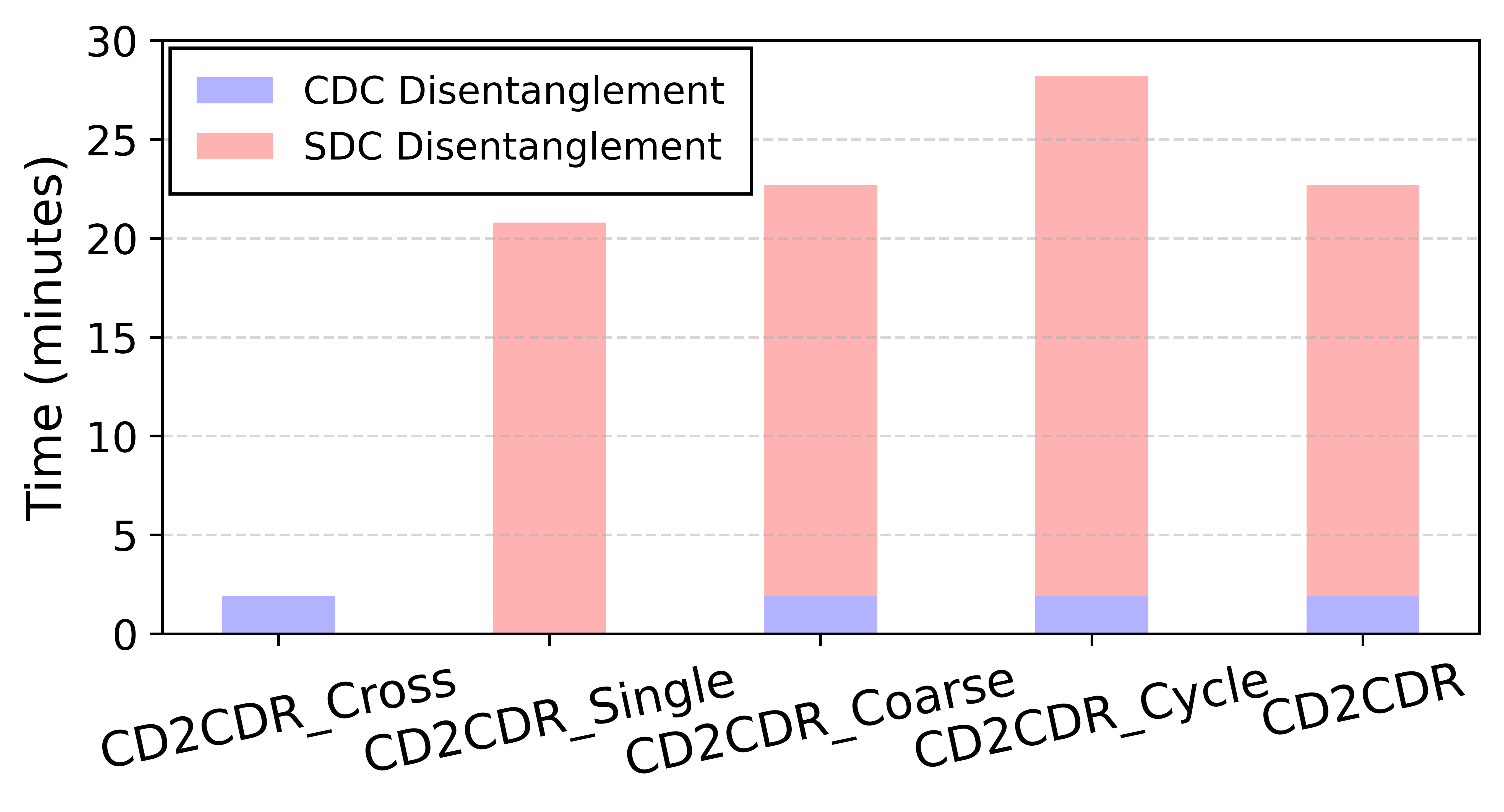} 
\vspace{-0.2in}
\caption{Average time comparison of confounder disentanglement phase across four tasks for CD2CDR and its variants.}
\label{confounder_disentanglement_time}
\vspace{-0.2in}
\end{figure}

\subsubsection{\textbf{Empirical Analysis of Time Complexity}}
To comprehensively evaluate the trade-off between effectiveness and efficiency of our model and its variants, we further conduct an empirical analysis of their time complexity. Fig. \ref{confounder_disentanglement_time} illustrates the average time consumption of the confounder disentanglement phase across all four tasks.  In Fig. \ref{confounder_disentanglement_time}, the time of confounder disentanglement phrase is divided into CDC disentanglement time (implemented via half-sibling regression) and SDC disentanglement time (implemented through dual adversarial training).

As shown in Fig. \ref{confounder_disentanglement_time}, \textbf{CD2CDR\_Cross} consumes significantly less time compared to other variants since it only employs half-sibling regression, which inherently requires no iterative training as mentioned in Section \ref{model_training}. However, this computational efficiency comes with a performance degradation of 5.49\% in recommendation metrics. In contrast, \textbf{CD2CDR\_Single} requires approximately 10 times more computational time since it relies solely on the dual adversarial structure, which demands multiple training epochs to converge, yet still underperforms our CD2CDR by 4.23\% in recommendation metrics. In addition, \textbf{CD2CDR\_Coarse} shows comparable time consumption to CD2CDR as it uses identical confounder disentanglement processes, despite suffering a substantial 21.35\% performance drop. Notably, \textbf{CD2CDR\_Cycle} consumes more time than our CD2CDR despite removing the cycle consistency loss, while also showing 1.01\% lower performance. The increased time consumption occurs because without the stabilizing effect of cycle consistency loss, the adversarial training requires more iterations to reach convergence. Moreover, the absence of cycle consistency loss leads to less accurate SDC disentanglement, which explains the observed performance degradation.

It is worth noting that the confounder disentanglement phase represents only a fraction of the overall computational cost in the entire training pipeline (as discussed in Section \ref{time_complexity}). Despite the differences in how CD2CDR and its variants implement confounder disentanglement and causal deconfounding, the major computational cost for both CD2CDR and its variants typically comes from the shared pretraining phase and final recommendation phase. As a result, the time differences observed in confounder disentanglement have a relatively limited impact on total training time. Based on the above analysis, CD2CDR achieves a good balance between effectiveness and efficiency, providing superior recommendation performance with reasonable computational requirements.

\subsection{Impact of Different Backbones (for RQ3)}
\label{sec:V-D}

\begin{figure}[ht]
\centering
\subfigure[]{
\begin{tikzpicture}[scale=0.55]
\pgfplotsset{
    width=0.65\textwidth,
    height=0.41\textwidth
}
\begin{axis}[
    ybar,
    bar width=10pt,
    ylabel={HR@10},
    ylabel style ={font = \huge, yshift=0.2cm},
    xlabel style ={font = \huge},
    enlarge x limits={abs=2cm},
    scaled ticks=false,
    tick label style={/pgf/number format/fixed, font=\huge},
    ymin=0, ymax=0.21,
    symbolic x coords={Tmall-Favorite, Tmall-Purchase},
    xtick=data,
    ytick={0,0.04,0.08,0.12,0.16,0.20},
]
\addplot coordinates {
(Tmall-Favorite,0.1710) (Tmall-Purchase,0.1025)};
\addplot coordinates {
(Tmall-Favorite,0.1608) (Tmall-Purchase,0.0912)};
\addplot coordinates {
(Tmall-Favorite,0.1664) (Tmall-Purchase,0.0973)};
\addplot coordinates {
(Tmall-Favorite,0.1723) (Tmall-Purchase,0.1016)};
\addplot coordinates {
(Tmall-Favorite,0.1749) (Tmall-Purchase,0.1064)};
\addplot[color=6,fill=2] coordinates {
(Tmall-Favorite,0.1756) (Tmall-Purchase,0.1078)};
\addplot[color=5,fill=11] coordinates {
(Tmall-Favorite,0.1776) (Tmall-Purchase,0.1102)};
\addplot[color=8,fill=4] coordinates {
(Tmall-Favorite,0.1801) (Tmall-Purchase,0.1138)};
\end{axis}
\end{tikzpicture}}
\subfigure[]{
\begin{tikzpicture}[scale=0.55]
\pgfplotsset{
    width=0.65\textwidth,
    height=0.41\textwidth
}
\begin{axis}[
    ybar,
    bar width=10pt,
    ylabel={NGCG@10},
    ylabel style ={font = \huge, yshift=0.2cm},
    xlabel style ={font = \huge},
    enlarge x limits={abs=2cm},
    scaled ticks=false,
    tick label style={/pgf/number format/fixed, font=\huge},
    ymin=0, ymax=0.11,
    symbolic x coords={Tmall-Favorite, Tmall-Purchase},
    xtick=data,
    ytick={0,0.02,0.04,0.06,0.08,0.10},
    legend style={at={(1.35,0.98)}, anchor=north,legend columns=1, column sep=0.2cm, row sep=3.5pt, draw=black, font=\Large},
]
\addplot coordinates {
(Tmall-Favorite,0.0932) (Tmall-Purchase,0.0536)};
\addplot coordinates {
(Tmall-Favorite,0.0867) (Tmall-Purchase,0.0496)};
\addplot coordinates {
(Tmall-Favorite,0.0915) (Tmall-Purchase,0.0521)};
\addplot coordinates {
(Tmall-Favorite,0.0938) (Tmall-Purchase,0.0535)};
\addplot coordinates {
(Tmall-Favorite,0.0957) (Tmall-Purchase,0.0548)};
\addplot[color=6,fill=2] coordinates {
(Tmall-Favorite,0.0959) (Tmall-Purchase,0.0551)};
\addplot[color=5,fill=11] coordinates {
(Tmall-Favorite,0.0971) (Tmall-Purchase,0.0583)};
\addplot[color=8,fill=4] coordinates {
(Tmall-Favorite,0.0987) (Tmall-Purchase,0.0612)};
\legend{BiTGCF\_CD2CDR, GA-DTCDR\_CD2CDR, DisenCDR\_CD2CDR, CausalCDR\_CD2CDR, GDCCDR\_CD2CDR, CrossAug\_CD2CDR, HJID\_CD2CDR, Our CD2CDR}
\end{axis}
\end{tikzpicture}}
\vspace{-0.25in}
\caption{(a)-(b): Comparative performance analysis between CD2CDR and its seven variants with different backbones.}
\label{tab:different_backbone}
\vspace{-0.2in}
\end{figure}

Since our CD2CDR can be easily combined with disentanglement-based dual-target CDR backbone models, in addition to DIDA-CDR \cite{zhu2023domain}, we select all the other representative and state-of-the-art models from this group as backbones to form the following seven variants, namely, \textbf{BiTGCF\_CD2CDR}, \textbf{GA-DTCDR\_CD2CDR}, \textbf{DisenCDR\_CD2CDR}, \textbf{CausalCDR\_CD2CDR}, \textbf{GDCCDR\_CD2CDR}, \textbf{CrossAug\_CD2CDR} and \textbf{HJID\_CD2CDR}. Our aim is to demonstrate the flexibility and effectiveness of our CD2CDR by integrating it with various representative and state-of-the-art disentanglement-based dual-target CDR backbone models, thereby highlighting its generalizability and extendability across diverse CDR scenarios. The performance comparison of our CD2CDR and its seven variants\footnote{Owing to constraints in space, Fig. \ref{tab:different_backbone} and Fig. \ref{different_parameter} solely present the experimental results for Task \#1, with similar trends observed in other unshown tasks.} with corresponding backbones is shown in Fig. \ref{tab:different_backbone}. We find that when our model employs DIDA-CDR as the backbone, it improves the above seven variants, namely, \textbf{BiTGCF\_CD2CDR}, \textbf{GA-DTCDR\_CD2CDR}, \textbf{DisenCDR\_CD2CDR}, \textbf{CausalCDR\_CD2CDR}, \textbf{GDCCDR\_CD2CDR}, \textbf{CrossAug\_CD2CDR} and \textbf{HJID\_CD2CDR} by an average of 7.97\%, 16.87\%, 11.42\%, 7.74\%, 5.09\%, 4.47\% and 2.34\%, respectively. This improvement can be attributed to the ability of DIDA-CDR to effectively decouple three components of user preferences for modeling more accurate comprehensive user preferences. Notably, the ability of DIDA-CDR aligns well with the requirements of our CD2CDR, which relies on this precise disentanglement to accurately decouple observed confounders. In addition, our model, when combined with various backbones, consistently outperforms these backbones in their original form with an average improvement of 9.05\% and 8.04\% w.r.t. HR@10 and NDCG@10, respectively. This not only shows the superior efficacy of CD2CDR in improving recommendation performance in both domains, but also shows its generalizability to various CDR models.

\subsection{Parameter Sensitivity (for RQ4)}
\label{sec:V-E}
\subsubsection{\textbf{Impact of the number of cluster centroids $J$}}
To explore the impact of number of cluster centroids $J$ on the efficacy of our proposed CD2CDR, we keep $J_{sd}^A = J_{sd}^B = J_{cd}$ and vary them in $\{2, 5, 10, 20, 50\}$. The corresponding experimental results are depicted in Figs. \ref{different_parameter}(a)-\ref{different_parameter}(b). We can observe that as $J$ increases, the recommendation performance initially improves but gradually plateaus beyond 10. This suggests that there is a threshold for $J$, which may vary in different datasets. Beyond this threshold, additional cluster centroids do not significantly improve the recommendation performance. In other words, once $J$ reaches this threshold, the potential confounders represented by these cluster centroids are comprehensive enough for effective deconfounding. With the aim of achieving a balance between model complexity and recommendation accuracy, we finally set $J_{sd}^A = J_{sd}^B = J_{cd} = 10$ in all three tasks. In particular, the comprehensive confounder disentanglement significantly contributes to more accurate estimation of Eq. (\ref{eq12}). More importantly, the experimental results show that our confounder disentanglement module can form effective confounder spaces, where even basic clustering techniques can easily identify key confounders, thereby yielding promising deconfounding results.

\begin{figure}[t]
 \setlength{\belowcaptionskip}{-0.12in}
 \centering
 \small
  \subfigure[]{
  \begin{tikzpicture}
  \begin{axis}[
  width=4.7cm,
  height=3.8cm,
  xmin=0.9, xmax=5.1,
  ymin=0.098, ymax=0.19,
  ylabel={HR@10},
  ylabel style={yshift=-0.5cm, font=\small},
  xlabel={$J$},
  xlabel style={yshift=0.2cm},
  xtick={1,2,3,4,5},
  xticklabels={2,5,10,20,50},
  xticklabel style={font=\small},
  yticklabel style={font=\small, /pgf/number format/.cd, fixed, precision=3},
  ytick={0,0.10,0.14,0.18},
  scaled ticks=false,
  legend style={at={(0.49,0.60)}, font=\footnotesize, anchor=north,legend columns=1, draw=none, fill=none,},
  ymajorgrids=true,
  grid style=dashed,
  ]
  \addplot[color=red3,
  mark=*,
  mark options={solid},
  line width=1pt,mark size=1.5pt,
  smooth] coordinates {
   (1,0.1705)
   (2,0.1743)
   (3,0.1801)
   (4,0.1806)
   (5,0.1809)
   };
  \addplot[ color=red1,
  mark=square,
  mark options={solid},
  line width=1pt,mark size=1.5pt,
  smooth] coordinates {
   (1,0.1041)
   (2,0.1076)
   (3,0.1138)
   (4,0.1141)
   (5,0.1143)
   };
  \legend{Tmall-Favorite, Tmall-Purchase}
  \end{axis}
  \end{tikzpicture}}
  \hspace{0in}
  \vspace{-0.1in}
 \subfigure[]{
  \begin{tikzpicture}
  \begin{axis}[
  width=4.7cm,
  height=3.8cm,
  xmin=0.9, xmax=5.1,
  ymin=0.048, ymax=0.105,
  ylabel={NDCG@10},
  ylabel style={yshift=-0.36cm, font=\small},
  xlabel={$J$},
  xlabel style={yshift=0.2cm},
  xtick={1,2,3,4,5},
  xticklabels={2,5,10,20,50},
  xticklabel style={font=\small},
  yticklabel style={font=\small, /pgf/number format/.cd, fixed, precision=3},
  ytick={0,0.05,0.075,0.10},
  scaled ticks=false,
  legend style={at={(0.49,0.60)}, font=\footnotesize, anchor=north,legend columns=1, draw=none, fill=none,},
  ymajorgrids=true,
  grid style=dashed,
  ]
  \addplot [color=red3,
  mark=*,
  mark options={solid},
  line width=1pt,mark size=1.5pt,
  smooth]coordinates {
   (1,0.0934)
   (2,0.0952)
   (3,0.0987)
   (4,0.0989)
   (5,0.0990)
  };
  \addplot [color=red1,
  mark=square,
  mark options={solid},
  line width=1pt,mark size=1.5pt,
  smooth]coordinates {
   (1,0.0530)
   (2,0.0549)
   (3,0.0612)
   (4,0.0613)
   (5,0.0613)
  };
  \legend{Tmall-Favorite, Tmall-Purchase}
  \end{axis}
  \end{tikzpicture}}
  \hspace{0in}
  \subfigure[]{
  \begin{tikzpicture}
  \begin{axis}[
  width=4.7cm,
  height=3.8cm,
  xmin=0.9, xmax=5.1,
  ymin=0.098, ymax=0.19,
  ylabel={HR@10},
  ylabel style={yshift=-0.5cm, font=\small},
  xlabel={$\lambda$},
  xlabel style={yshift=0.2cm},
  xtick={1,2,3,4,5},
  xticklabels={0.1, 1, 2, 5, 10},
  xticklabel style={font=\small},
  yticklabel style={font=\small, /pgf/number format/.cd, fixed, precision=3},
  ytick={0,0.10,0.14,0.18},
  scaled ticks=false,
  legend style={at={(0.49,0.60)}, font=\footnotesize, anchor=north,legend columns=1, draw=none, fill=none,},
  ymajorgrids=true,
  grid style=dashed,
  ]
  \addplot[color=red3,
  mark=*,
  mark options={solid},
  line width=1pt,mark size=1.5pt,
  smooth] coordinates {
   (1,0.1791)
   (2,0.1801)
   (3,0.1795)
   (4,0.1782)
   (5,0.1768)
   };
  \addplot[ color=red1,
  mark=square,
  mark options={solid},
  line width=1pt,mark size=1.5pt,
  smooth] coordinates {
   (1,0.1129)
   (2,0.1138)
   (3,0.1115)
   (4,0.1092)
   (5,0.1079)
   };
  \legend{Tmall-Favorite, Tmall-Purchase}
  \end{axis}
  \end{tikzpicture}}
  \hspace{0in}
 \subfigure[]{
  \begin{tikzpicture}
  \begin{axis}[
  width=4.7cm,
  height=3.8cm,
  xmin=0.9, xmax=5.1,
  ymin=0.048, ymax=0.105,
  ylabel={NDCG@10},
  ylabel style={yshift=-0.36cm, font=\small},
  xlabel={$\lambda$},
  xlabel style={yshift=0.2cm},
  xtick={1,2,3,4,5},
  xticklabels={0.1, 1, 2, 5, 10},
  xticklabel style={font=\small},
  yticklabel style={font=\small, /pgf/number format/.cd, fixed, precision=3},
  ytick={0,0.05,0.075,0.10},
  scaled ticks=false,
  legend style={at={(0.49,0.60)}, font=\footnotesize, anchor=north,legend columns=1, draw=none, fill=none,},
  ymajorgrids=true,
  grid style=dashed,
  ]
  \addplot [color=red3,
  mark=*,
  mark options={solid},
  line width=1pt,mark size=1.5pt,
  smooth]coordinates {
   (1,0.0974)
   (2,0.0987)
   (3,0.0976)
   (4,0.0969)
   (5,0.0963)
  };
  \addplot [color=red1,
  mark=square,
  mark options={solid},
  line width=1pt,mark size=1.5pt,
  smooth]coordinates {
   (1,0.0608)
   (2,0.0612)
   (3,0.0610)
   (4,0.0605)
   (5,0.0584)
  };
  \legend{Tmall-Favorite, Tmall-Purchase}
  \end{axis}
  \end{tikzpicture}}
  \hspace{0in}
  \subfigure[]{
  \begin{tikzpicture}
  \begin{axis}[
  width=4.7cm,
  height=3.8cm,
  xmin=0.9, xmax=5.1,
  ymin=0.098, ymax=0.19,
  ylabel={HR@10},
  ylabel style={yshift=-0.5cm, font=\small},
  xlabel={$\alpha $},
  xlabel style={yshift=0.2cm},
  xtick={1,2,3,4,5},
  xticklabels={0.1, 1, 10, 20, 50},
  xticklabel style={font=\small},
  yticklabel style={font=\small, /pgf/number format/.cd, fixed, precision=3},
  ytick={0,0.10,0.14,0.18},
  scaled ticks=false,
  legend style={at={(0.49,0.60)}, font=\footnotesize, anchor=north,legend columns=1, draw=none, fill=none,},
  ymajorgrids=true,
  grid style=dashed,
  ]
  \addplot[color=red3,
  mark=*,
  mark options={solid},
  line width=1pt,mark size=1.5pt,
  smooth] coordinates {
   (1,0.1797)
   (2,0.1801)
   (3,0.1790)
   (4,0.1774)
   (5,0.1755)
   };
  \addplot[ color=red1,
  mark=square,
  mark options={solid},
  line width=1pt,mark size=1.5pt,
  smooth] coordinates {
   (1,0.1132)
   (2,0.1138)
   (3,0.1115)
   (4,0.1092)
   (5,0.1079)
   };
  \legend{Tmall-Favorite, Tmall-Purchase}
  \end{axis}
  \end{tikzpicture}}
  \hspace{0in}
 \subfigure[]{
  \begin{tikzpicture}
  \begin{axis}[
  width=4.7cm,
  height=3.8cm,
  xmin=0.9, xmax=5.1,
  ymin=0.048, ymax=0.105,
  ylabel={NDCG@10},
  ylabel style={yshift=-0.36cm, font=\small},
  xlabel={$\alpha $},
  xlabel style={yshift=0.2cm},
  xtick={1,2,3,4,5},
  xticklabels={0.1,1,10,20,50},
  xticklabel style={font=\small},
  yticklabel style={font=\small, /pgf/number format/.cd, fixed, precision=3},
  ytick={0,0.05,0.075,0.10},
  scaled ticks=false,
  legend style={at={(0.49,0.60)}, font=\footnotesize, anchor=north,legend columns=1, draw=none, fill=none,},
  ymajorgrids=true,
  grid style=dashed,
  ]
  \addplot [color=red3,
  mark=*,
  mark options={solid},
  line width=1pt,mark size=1.5pt,
  smooth]coordinates {
   (1,0.0980)
   (2,0.0987)
   (3,0.0973)
   (4,0.0967)
   (5,0.0958)
  };
  \addplot [color=red1,
  mark=square,
  mark options={solid},
  line width=1pt,mark size=1.5pt,
  smooth]coordinates {
   (1,0.0609)
   (2,0.0612)
   (3,0.0587)
   (4,0.0566)
   (5,0.0552)
  };
  \legend{Tmall-Favorite, Tmall-Purchase}
  \end{axis}
  \end{tikzpicture}}
 \vspace{-0.2in}
 \caption{(a)-(b): Impact of the number of cluster centroids $J$. (c)-(d): Impact of the weight of cycle consistency loss $\lambda$. (e)-(f): Impact of the regularization parameter $\alpha $.}
 \label{different_parameter}
\vspace{-0.1in}
\end{figure}

\subsubsection{\textbf{Impact of the weight of cycle consistency loss $\lambda$}}
\label{sec:5.5.2}
To examine the impact of the weight of cycle consistency loss $\lambda$ on the recommendation performance of our model, we test $\lambda$ with values from $\{0.1, 1, 2, 5, 10\}$. The experimental results are depicted in Figs. \ref{different_parameter}(c)-(d). We can observe that smaller values of $\lambda$ (e.g., 0.1 or 1) allow the adversarial losses to play a dominant role during training, driving the generator to better align the distributions of domain-specific user preferences across domains. Meanwhile, the cycle consistency loss still enforces moderate consistency, ensuring the transformation remains meaningful. In contrast, when $\lambda$ is set to larger values (e.g., 5 or 10), the cycle consistency loss becomes overly influential. As a result, the generator focuses primarily on minimizing cycle consistency loss, prioritizing outputs that closely resemble the inputs, rather than using feedback from the discriminator to refine cross-domain transformations. This weakens the discriminator's ability to guide the generator towards producing domain-specific user preferences that are indistinguishable from those in the target domain. Therefore, we select $\lambda = 1$ to ensure a proper balance between the optimization of adversarial losses and cycle consistency loss.

\subsubsection{\textbf{Impact of the regularization parameter $\alpha $}}
To analyze the impact of the regularization parameter $\alpha$ on decoupling candidate CDCs, we test $\alpha$ with values from $\{0.1, 1, 10, 20, 50\}$. The results are shown in Figs. \ref{different_parameter}(e)-\ref{different_parameter}(f). We can observe that small values of $\alpha$ (e.g., 0.1) lead to insufficient regularization, making the regression overly sensitive to noise. By contrast, large $\alpha$ values (e.g., 20 or 50) place too much emphasis on the regularization term, leading the model to underfit important features related to the CDCs. This weakens the model's ability to effectively decouple candidate CDCs. As shown in Figs. \ref{different_parameter}(e)-\ref{different_parameter}(f), when $\alpha = 1$, our CD2CDR achieves the best trade-off between stabilizing numerical computations and maintaining the accuracy of decoupling candidate CDCs, thereby enhancing the effectiveness of subsequent deconfounding in cross-domain settings.

\subsection{Discussion}
\subsubsection{\textbf{Analysis of experimental results}}
\label{results_discussion}
The experimental results indicate that (1) Our CD2CDR consistently improves seventeen baselines. The improvements are primarily due to our proposed confounder disentanglement module and causal deconfounding module. The confounder disentanglement module, designed with the dual adversarial structure and half-sibling regression, efficiently decouples observed SDCs and CDCs. The causal deconfounding module uses backdoor adjustment to deconfound such decoupled observed confounders and incorporates their positive effects into debiased comprehensive user preferences, enhancing the recommendation accuracy in both domains. By contrast, the baseline models either ignore CDCs or only consider the negative effects of observed confounders, limiting their ability to achieve a comprehensive understanding of user-item interactions and thus leading to degraded recommendation performance; (2) Our CD2CDR is tailored for dual-target CDR to deconfound observed SDCs and CDCs from a causal perspective. Through comprehensive ablation studies, we highlight the impact of each module in accurately modeling the causal relationships among observed confounders, user preferences, and user-item interactions; (3) Our CD2CDR is a robust dual-target CDR framework capable of integrating with various state-of-the-art disentanglement-based backbone models, thereby highlighting its generalizability and extendability across diverse CDR scenarios. As demonstrated in Section \ref{sec:V-D}, we combined our CD2CDR framework with several state-of-the-art disentanglement-based dual-target CDR backbones, resulting in an average improvement of 9.05\% and 8.04\% w.r.t HR@10 and NDCG@10, respectively. This indicates that with our framework, existing disentanglement-based dual-target CDR models can be extended for effective deconfounding, which is influential in developing more advanced CDR models; (4) Our CD2CDR demonstrates adaptability to the more challenging CSR scenario. In this scenario, our model also achieves significant improvements over baselines. This adaptability stems from our model's ability to extract more accurate item representations by handling observed confounders such as item popularity. For example, when an item is frequently interacted with, a data-driven RS improperly perceives the item's popularity as its intrinsic properties, creating biased item representations. In contrast to the above negative effect, item popularity can also serve as a secondary cause for predicting user-item interactions, which represents a positive effect. By effectively disentangling and then deconfounding observed confounders to preserve their positive effects while mitigating negative effects, our model enhances the recommendation accuracy in CSR scenarios where only item-wise knowledge transfer is possible.

\subsubsection{\textbf{Analysis of disentangled confounders}}
Beyond the quantitative evaluation, we conduct an in-depth analysis of how disentangled confounders affect recommendation results in dual-target CDR. The fundamental distinction between SDCs and CDCs lies in their scope of influence: SDCs exclusively affect user preferences and user-item interactions within a specific domain, while CDCs simultaneously influence both domains. Through our dual adversarial structure and half-sibling regression approach, we effectively decouple these confounders from user preferences for subsequent deconfounding steps.

Furthermore, these disentangled confounders reveal important insights about confounding biases in dual-target CDR. Both SDCs and CDCs significantly affect user preference modeling and recommendation accuracy in different ways. Specifically, SDCs are mistakenly incorporated as part of domain-specific user preferences, distorting the preference representation. For example, ‘free shipping’ (an SDC) might be incorrectly identified as a user's true preference in the ‘purchase’ domain, leading to biased recommendations. Similarly, CDCs are incorrectly treated as domain-shared user preferences, creating confounding biases that affect both domains. Since domain-shared preferences represent user interests common to both domains, these biases distort the essential knowledge transfer across domains, resulting in inaccurate recommendations on both domains.

In addition, these disentangled confounders also carry valuable information that directly influences user-item interactions through causal paths. For instance, while ‘sales promotion’ (a CDC) should not be misinterpreted as a user's true preference, it acts as a secondary cause that genuinely affects user behavior on both domains. The quality of confounder disentanglement directly determines the effectiveness of subsequent deconfounding steps. With well-disentangled confounders, our backdoor adjustment effectively eliminates their negative influence on user preference modeling, resulting in debiased comprehensive user preferences. Meanwhile, our confounder selection function preserves their positive direct effects on predicting user-item interactions. Our causal deconfounding approach, which removes negative effects while retaining positive ones, significantly enhances recommendation accuracy. This is achieved by providing a more comprehensive understanding of user-item interactions both within and across domains, particularly in scenarios where confounding effects are prominent. Similar principles of confounder disentanglement and causal deconfounding apply in the CSR scenario where item-wise knowledge transfer is crucial. In this scenario, observed confounders like item popularity similarly distort item representations, requiring proper disentangling and then deconfounding as we have analyzed in Section \ref{results_discussion}.

\subsubsection{\textbf{Limitations and future work}}
A limitation of this work is its focus on a dual-target CDR scenario, which does not fully address the multi-domain causal effects presented in real-world applications. In practice, users often interact across numerous domains simultaneously, creating intricate causal effect networks where user preferences in one domain may influence behaviors across multiple other domains through various direct and indirect pathways. Our current dual-target CDR framework may not be able to capture these complex multi-domain interaction patterns, potentially overlooking intricate multi-domain causal effects that exist in multi-target CDR.

In addition, as briefly discussed in Section \ref{sec:4.3.1}, our model faces challenges in eliminating residual confounding bias embedded in the original domain-specific user preferences. While our model mitigates these biases to a large extent, some deeply embedded confounders may still persist and affect deconfounding effectiveness.

Moreover, our model focuses exclusively on observed confounders that can be explicitly decoupled from observed signals. However, in real-world recommendation scenarios, there exist unobserved confounders that cannot be directly disentangled from available data, such as environmental factors. These unobserved confounders may introduce additional bias into the recommendation process, posing a significant challenge for accurate recommendations. To address this challenge, we need to exploit instrumental variables \cite{yuan2023instrumental} or learn substitutes for unobserved confounders \cite{jin2024unbiased}. We will leave the exploration of unobserved confounders and corresponding deconfounding strategies as one of our future research directions.

In the future, we intend to extend our framework to account for multi-domain causal impacts to better reflect real-world scenarios, conduct real-world deployment validation, and enhance scalability to large-scale datasets. Meanwhile, we plan to explore unobserved confounders, and develop more sophisticated causal inference techniques to mitigate the residual confounding bias and other types of biases. Moreover, we intend to explore replacing our current BERT-based text encoding with more advanced large language models (LLMs) \cite{sun2024large}. LLMs can generate higher-quality content representations from user reviews and item details, potentially leading to more accurate user, item, and confounder representations, thereby enhancing both the confounder disentanglement process and overall recommendation performance. Furthermore, we plan to explore simplified frameworks for causal deconfounding in CDR through, for example, knowledge distillation techniques \cite{yang2023multimodal,an2024ddcdr}, where a simplified student model is trained to replicate the causal reasoning capabilities of more complex model. This approach could help achieve similar or better performance with fewer modules and hyperparameters, thereby reducing model complexity and implementation difficulty.

\section{Conclusion}
In this paper, we have proposed a novel Causal Deconfounding framework via Confounder Disentanglement for dual-target CDR, called CD2CDR. CD2CDR not only effectively decouples observed single-domain and cross-domain confounders, but also preserves the positive direct effects of such observed confounders on predicted interactions and eliminates their negative effects on capturing comprehensive user preferences, thereby enhancing the recommendation accuracy in both domains simultaneously. Moreover, we have conducted extensive experiments on seven real-world datasets, which demonstrates that our CD2CDR significantly outperforms the state-of-the-art methods. 

\begin{acks}
This work is supported by ARC Discovery Project DP230100676.
\end{acks}

\bibliographystyle{ACM-Reference-Format}
\bibliography{reference}


\begin{thebibliography}{102}


\ifx \showCODEN    \undefined \def \showCODEN     #1{\unskip}     \fi
\ifx \showDOI      \undefined \def \showDOI       #1{#1}\fi
\ifx \showISBNx    \undefined \def \showISBNx     #1{\unskip}     \fi
\ifx \showISBNxiii \undefined \def \showISBNxiii  #1{\unskip}     \fi
\ifx \showISSN     \undefined \def \showISSN      #1{\unskip}     \fi
\ifx \showLCCN     \undefined \def \showLCCN      #1{\unskip}     \fi
\ifx \shownote     \undefined \def \shownote      #1{#1}          \fi
\ifx \showarticletitle \undefined \def \showarticletitle #1{#1}   \fi
\ifx \showURL      \undefined \def \showURL       {\relax}        \fi
\providecommand\bibfield[2]{#2}
\providecommand\bibinfo[2]{#2}
\providecommand\natexlab[1]{#1}
\providecommand\showeprint[2][]{arXiv:#2}

\bibitem[An et~al\mbox{.}(2024)]%
        {an2024ddcdr}
\bibfield{author}{\bibinfo{person}{Zhicheng An}, \bibinfo{person}{Zhexu Gu}, \bibinfo{person}{Li Yu}, \bibinfo{person}{Ke Tu}, \bibinfo{person}{Zhengwei Wu}, \bibinfo{person}{Binbin Hu}, \bibinfo{person}{Zhiqiang Zhang}, \bibinfo{person}{Lihong Gu}, {and} \bibinfo{person}{Jinjie Gu}.} \bibinfo{year}{2024}\natexlab{}.
\newblock \showarticletitle{{DDCDR}: A Disentangle-based Distillation Framework for Cross-Domain Recommendation}. In \bibinfo{booktitle}{\emph{KDD}}. \bibinfo{pages}{4764--4773}.
\newblock


\bibitem[Cao et~al\mbox{.}(2022a)]%
        {cao2022disencdr}
\bibfield{author}{\bibinfo{person}{Jiangxia Cao}, \bibinfo{person}{Xixun Lin}, \bibinfo{person}{Xin Cong}, \bibinfo{person}{Jing Ya}, \bibinfo{person}{Tingwen Liu}, {and} \bibinfo{person}{Bin Wang}.} \bibinfo{year}{2022}\natexlab{a}.
\newblock \showarticletitle{Disen{CDR}: Learning Disentangled Representations for Cross-Domain Recommendation}. In \bibinfo{booktitle}{\emph{SIGIR}}. \bibinfo{pages}{267--277}.
\newblock


\bibitem[Cao et~al\mbox{.}(2022b)]%
        {cao2022cross}
\bibfield{author}{\bibinfo{person}{Jiangxia Cao}, \bibinfo{person}{Jiawei Sheng}, \bibinfo{person}{Xin Cong}, \bibinfo{person}{Tingwen Liu}, {and} \bibinfo{person}{Bin Wang}.} \bibinfo{year}{2022}\natexlab{b}.
\newblock \showarticletitle{Cross-Domain Recommendation to Cold-Start Users via Variational Information Bottleneck}. In \bibinfo{booktitle}{\emph{ICDE}}. \bibinfo{pages}{2209--2223}.
\newblock


\bibitem[Chen et~al\mbox{.}(2021)]%
        {chen2021multi}
\bibfield{author}{\bibinfo{person}{Wanyu Chen}, \bibinfo{person}{Pengjie Ren}, \bibinfo{person}{Fei Cai}, \bibinfo{person}{Fei Sun}, {and} \bibinfo{person}{Maarten De~Rijke}.} \bibinfo{year}{2021}\natexlab{}.
\newblock \showarticletitle{Multi-interest Diversification for End-to-end Sequential Recommendation}.
\newblock \bibinfo{journal}{\emph{ACM TOIS}} \bibinfo{volume}{40}, \bibinfo{number}{1} (\bibinfo{year}{2021}), \bibinfo{pages}{1--30}.
\newblock


\bibitem[Chen et~al\mbox{.}(2023)]%
        {chen2023toward}
\bibfield{author}{\bibinfo{person}{Xu Chen}, \bibinfo{person}{Ya Zhang}, \bibinfo{person}{Ivor~W Tsang}, \bibinfo{person}{Yuangang Pan}, {and} \bibinfo{person}{Jingchao Su}.} \bibinfo{year}{2023}\natexlab{}.
\newblock \showarticletitle{Toward Equivalent Transformation of User Preferences in Cross Domain Recommendation}.
\newblock \bibinfo{journal}{\emph{ACM TOIS}} \bibinfo{volume}{41}, \bibinfo{number}{1} (\bibinfo{year}{2023}), \bibinfo{pages}{1--31}.
\newblock


\bibitem[Devlin et~al\mbox{.}(2019)]%
        {kenton2019bert}
\bibfield{author}{\bibinfo{person}{Jacob Devlin}, \bibinfo{person}{Ming{-}Wei Chang}, \bibinfo{person}{Kenton Lee}, {and} \bibinfo{person}{Kristina Toutanova}.} \bibinfo{year}{2019}\natexlab{}.
\newblock \showarticletitle{{BERT:} Pre-training of Deep Bidirectional Transformers for Language Understanding}. In \bibinfo{booktitle}{\emph{NAACL-HLT}}. \bibinfo{pages}{4171--4186}.
\newblock


\bibitem[Ding et~al\mbox{.}(2025)]%
        {ding2025few}
\bibfield{author}{\bibinfo{person}{Pengfei Ding}, \bibinfo{person}{Yan Wang}, \bibinfo{person}{Guanfeng Liu}, \bibinfo{person}{Nan Wang}, {and} \bibinfo{person}{Xiaofang Zhou}.} \bibinfo{year}{2025}\natexlab{}.
\newblock \showarticletitle{Few-Shot Causal Representation Learning for Out-of-Distribution Generalization on Heterogeneous Graphs}.
\newblock \bibinfo{journal}{\emph{IEEE TKDE}} \bibinfo{volume}{37}, \bibinfo{number}{4} (\bibinfo{year}{2025}), \bibinfo{pages}{1804--1818}.
\newblock


\bibitem[Du et~al\mbox{.}(2024c)]%
        {du2024towards}
\bibfield{author}{\bibinfo{person}{Jing Du}, \bibinfo{person}{Zesheng Ye}, \bibinfo{person}{Bin Guo}, \bibinfo{person}{Zhiwen Yu}, \bibinfo{person}{Jia Wu}, \bibinfo{person}{Jian Yang}, \bibinfo{person}{Michael Sheng}, {and} \bibinfo{person}{Lina Yao}.} \bibinfo{year}{2024}\natexlab{c}.
\newblock \showarticletitle{Towards Robust Cross-Domain Recommendation with Joint Identifiability of User Preference}.
\newblock \bibinfo{journal}{\emph{arXiv preprint arXiv:2411.17361}} (\bibinfo{year}{2024}).
\newblock


\bibitem[Du et~al\mbox{.}(2024b)]%
        {du2024joint}
\bibfield{author}{\bibinfo{person}{Jing Du}, \bibinfo{person}{Zesheng Ye}, \bibinfo{person}{Bin Guo}, \bibinfo{person}{Zhiwen Yu}, {and} \bibinfo{person}{Lina Yao}.} \bibinfo{year}{2024}\natexlab{b}.
\newblock \showarticletitle{Identifiability of Cross-Domain Recommendation via Causal Subspace Disentanglement}. In \bibinfo{booktitle}{\emph{SIGIR}}. \bibinfo{pages}{2091–2101}.
\newblock


\bibitem[Du et~al\mbox{.}(2022)]%
        {du2022invariant}
\bibfield{author}{\bibinfo{person}{Xiaoyu Du}, \bibinfo{person}{Zike Wu}, \bibinfo{person}{Fuli Feng}, \bibinfo{person}{Xiangnan He}, {and} \bibinfo{person}{Jinhui Tang}.} \bibinfo{year}{2022}\natexlab{}.
\newblock \showarticletitle{Invariant Representation Learning for Multimedia Recommendation}. In \bibinfo{booktitle}{\emph{MM}}. \bibinfo{pages}{619--628}.
\newblock


\bibitem[Du et~al\mbox{.}(2024a)]%
        {du2024disentangled}
\bibfield{author}{\bibinfo{person}{Yingpeng Du}, \bibinfo{person}{Ziyan Wang}, \bibinfo{person}{Zhu Sun}, \bibinfo{person}{Yining Ma}, \bibinfo{person}{Hongzhi Liu}, {and} \bibinfo{person}{Jie Zhang}.} \bibinfo{year}{2024}\natexlab{a}.
\newblock \showarticletitle{Disentangled Multi-interest Representation Learning for Sequential Recommendation}. In \bibinfo{booktitle}{\emph{KDD}}. \bibinfo{pages}{677--688}.
\newblock


\bibitem[Du et~al\mbox{.}(2019)]%
        {du2019sequential}
\bibfield{author}{\bibinfo{person}{Zhengxiao Du}, \bibinfo{person}{Xiaowei Wang}, \bibinfo{person}{Hongxia Yang}, \bibinfo{person}{Jingren Zhou}, {and} \bibinfo{person}{Jie Tang}.} \bibinfo{year}{2019}\natexlab{}.
\newblock \showarticletitle{Sequential Scenario-Specific Meta Learner for Online Recommendation}. In \bibinfo{booktitle}{\emph{KDD}}. \bibinfo{pages}{2895--2904}.
\newblock


\bibitem[Fu et~al\mbox{.}(2019)]%
        {fu2019deeply}
\bibfield{author}{\bibinfo{person}{Wenjing Fu}, \bibinfo{person}{Zhaohui Peng}, \bibinfo{person}{Senzhang Wang}, \bibinfo{person}{Yang Xu}, {and} \bibinfo{person}{Jin Li}.} \bibinfo{year}{2019}\natexlab{}.
\newblock \showarticletitle{Deeply Fusing Reviews and Contents for Cold Start Users in Cross-Domain Recommendation Systems}. In \bibinfo{booktitle}{\emph{AAAI}}. \bibinfo{pages}{94--101}.
\newblock


\bibitem[Gao et~al\mbox{.}(2022)]%
        {gao2022causal}
\bibfield{author}{\bibinfo{person}{Chen Gao}, \bibinfo{person}{Yu Zheng}, \bibinfo{person}{Wenjie Wang}, \bibinfo{person}{Fuli Feng}, \bibinfo{person}{Xiangnan He}, {and} \bibinfo{person}{Yong Li}.} \bibinfo{year}{2022}\natexlab{}.
\newblock \showarticletitle{Causal Inference in Recommender Systems: A Survey and Future Directions}.
\newblock \bibinfo{journal}{\emph{ACM TOIS}} \bibinfo{volume}{42}, \bibinfo{number}{4} (\bibinfo{year}{2022}), \bibinfo{pages}{1--32}.
\newblock


\bibitem[Goodfellow et~al\mbox{.}(2014)]%
        {goodfellow2014generative}
\bibfield{author}{\bibinfo{person}{Ian Goodfellow}, \bibinfo{person}{Jean Pouget-Abadie}, \bibinfo{person}{Mehdi Mirza}, \bibinfo{person}{Bing Xu}, \bibinfo{person}{David Warde-Farley}, \bibinfo{person}{Sherjil Ozair}, \bibinfo{person}{Aaron Courville}, {and} \bibinfo{person}{Yoshua Bengio}.} \bibinfo{year}{2014}\natexlab{}.
\newblock \showarticletitle{Generative Adversarial Nets}. In \bibinfo{booktitle}{\emph{NIPS}}. \bibinfo{pages}{2672--2680}.
\newblock


\bibitem[Gowda et~al\mbox{.}(2021)]%
        {gowda2021pulling}
\bibfield{author}{\bibinfo{person}{Sindhu~CM Gowda}, \bibinfo{person}{Shalmali Joshi}, \bibinfo{person}{Haoran Zhang}, {and} \bibinfo{person}{Marzyeh Ghassemi}.} \bibinfo{year}{2021}\natexlab{}.
\newblock \showarticletitle{Pulling Up by the Causal Bootstraps: Causal Data Augmentation for Pre-training Debiasing}. In \bibinfo{booktitle}{\emph{CIKM}}. \bibinfo{pages}{606--616}.
\newblock


\bibitem[Guo et~al\mbox{.}(2023)]%
        {guo2023disentangled}
\bibfield{author}{\bibinfo{person}{Xiaobo Guo}, \bibinfo{person}{Shaoshuai Li}, \bibinfo{person}{Naicheng Guo}, \bibinfo{person}{Jiangxia Cao}, \bibinfo{person}{Xiaolei Liu}, \bibinfo{person}{Qiongxu Ma}, \bibinfo{person}{Runsheng Gan}, {and} \bibinfo{person}{Yunan Zhao}.} \bibinfo{year}{2023}\natexlab{}.
\newblock \showarticletitle{Disentangled Representations Learning for Multi-target Cross-domain Recommendation}.
\newblock \bibinfo{journal}{\emph{ACM TOIS}} \bibinfo{volume}{41}, \bibinfo{number}{4} (\bibinfo{year}{2023}), \bibinfo{pages}{1--27}.
\newblock


\bibitem[Harper and Konstan(2015)]%
        {harper2015movielens}
\bibfield{author}{\bibinfo{person}{F~Maxwell Harper} {and} \bibinfo{person}{Joseph~A Konstan}.} \bibinfo{year}{2015}\natexlab{}.
\newblock \showarticletitle{The MovieLens Datasets: History and Context}.
\newblock \bibinfo{journal}{\emph{ACM TIIS}} \bibinfo{volume}{5}, \bibinfo{number}{4} (\bibinfo{year}{2015}), \bibinfo{pages}{1--19}.
\newblock


\bibitem[He et~al\mbox{.}(2023a)]%
        {he2023dmbin}
\bibfield{author}{\bibinfo{person}{Tianqi He}, \bibinfo{person}{Kaiyuan Li}, \bibinfo{person}{Shan Chen}, \bibinfo{person}{Haitao Wang}, \bibinfo{person}{Qiang Liu}, \bibinfo{person}{Xingxing Wang}, {and} \bibinfo{person}{Dong Wang}.} \bibinfo{year}{2023}\natexlab{a}.
\newblock \showarticletitle{{DMBIN}: A Dual Multi-behavior Interest Network for Click-Through Rate Prediction via Contrastive Learning}. In \bibinfo{booktitle}{\emph{SIGIR}}. \bibinfo{pages}{1366--1375}.
\newblock


\bibitem[He et~al\mbox{.}(2020)]%
        {he2020lightgcn}
\bibfield{author}{\bibinfo{person}{Xiangnan He}, \bibinfo{person}{Kuan Deng}, \bibinfo{person}{Xiang Wang}, \bibinfo{person}{Yan Li}, \bibinfo{person}{Yongdong Zhang}, {and} \bibinfo{person}{Meng Wang}.} \bibinfo{year}{2020}\natexlab{}.
\newblock \showarticletitle{Light{GCN}: Simplifying and Powering Graph Convolution Network for Recommendation}. In \bibinfo{booktitle}{\emph{SIGIR}}. \bibinfo{pages}{639--648}.
\newblock


\bibitem[He et~al\mbox{.}(2017)]%
        {he2017neural}
\bibfield{author}{\bibinfo{person}{Xiangnan He}, \bibinfo{person}{Lizi Liao}, \bibinfo{person}{Hanwang Zhang}, \bibinfo{person}{Liqiang Nie}, \bibinfo{person}{Xia Hu}, {and} \bibinfo{person}{Tat-Seng Chua}.} \bibinfo{year}{2017}\natexlab{}.
\newblock \showarticletitle{Neural Collaborative Filtering}. In \bibinfo{booktitle}{\emph{WWW}}. \bibinfo{pages}{173--182}.
\newblock


\bibitem[He et~al\mbox{.}(2023b)]%
        {he2023addressing}
\bibfield{author}{\bibinfo{person}{Xiangnan He}, \bibinfo{person}{Yang Zhang}, \bibinfo{person}{Fuli Feng}, \bibinfo{person}{Chonggang Song}, \bibinfo{person}{Lingling Yi}, \bibinfo{person}{Guohui Ling}, {and} \bibinfo{person}{Yongdong Zhang}.} \bibinfo{year}{2023}\natexlab{b}.
\newblock \showarticletitle{Addressing Confounding Feature Issue for Causal Recommendation}.
\newblock \bibinfo{journal}{\emph{ACM TOIS}} \bibinfo{volume}{41}, \bibinfo{number}{3} (\bibinfo{year}{2023}), \bibinfo{pages}{1--23}.
\newblock


\bibitem[Hu et~al\mbox{.}(2019)]%
        {hu2019transfer}
\bibfield{author}{\bibinfo{person}{Guangneng Hu}, \bibinfo{person}{Yu Zhang}, {and} \bibinfo{person}{Qiang Yang}.} \bibinfo{year}{2019}\natexlab{}.
\newblock \showarticletitle{Transfer Meets Hybrid: A Synthetic Approach for Cross-Domain Collaborative Filtering with Text}. In \bibinfo{booktitle}{\emph{TheWebConf}}. \bibinfo{pages}{2822--2829}.
\newblock


\bibitem[Huang et~al\mbox{.}(2024)]%
        {huang2024multi}
\bibfield{author}{\bibinfo{person}{Zhirong Huang}, \bibinfo{person}{Shichao Zhang}, \bibinfo{person}{Debo Cheng}, \bibinfo{person}{Jiuyong Li}, \bibinfo{person}{Lin Liu}, {and} \bibinfo{person}{Guixian Zhang}.} \bibinfo{year}{2024}\natexlab{}.
\newblock \showarticletitle{Multi-Cause Deconfounding for Recommender Systems with Latent Confounders}.
\newblock \bibinfo{journal}{\emph{arXiv preprint arXiv:2410.12366}} (\bibinfo{year}{2024}).
\newblock


\bibitem[Jin et~al\mbox{.}(2024)]%
        {jin2024unbiased}
\bibfield{author}{\bibinfo{person}{Xuanyu Jin}, \bibinfo{person}{Ni Li}, \bibinfo{person}{Wanzeng Kong}, \bibinfo{person}{Jiajia Tang}, {and} \bibinfo{person}{Bing Yang}.} \bibinfo{year}{2024}\natexlab{}.
\newblock \showarticletitle{Unbiased Semantic Representation Learning Based on Causal Disentanglement for Domain Generalization}.
\newblock \bibinfo{journal}{\emph{ACM TOMM}} \bibinfo{volume}{20}, \bibinfo{number}{8} (\bibinfo{year}{2024}), \bibinfo{pages}{1551--6857}.
\newblock


\bibitem[Kanagawa et~al\mbox{.}(2019)]%
        {kanagawa2019cross}
\bibfield{author}{\bibinfo{person}{Heishiro Kanagawa}, \bibinfo{person}{Hayato Kobayashi}, \bibinfo{person}{Nobuyuki Shimizu}, \bibinfo{person}{Yukihiro Tagami}, {and} \bibinfo{person}{Taiji Suzuki}.} \bibinfo{year}{2019}\natexlab{}.
\newblock \showarticletitle{Cross-domain Recommendation via Deep Domain Adaptation}. In \bibinfo{booktitle}{\emph{ECIR}}. \bibinfo{pages}{20--29}.
\newblock


\bibitem[Kingma and Ba(2015)]%
        {kingma2014adam}
\bibfield{author}{\bibinfo{person}{Diederik~P Kingma} {and} \bibinfo{person}{Jimmy Ba}.} \bibinfo{year}{2015}\natexlab{}.
\newblock \showarticletitle{Adam: A Method for Stochastic Optimization}. In \bibinfo{booktitle}{\emph{ICLR}}.
\newblock


\bibitem[Li et~al\mbox{.}(2024a)]%
        {li2024causalcdr}
\bibfield{author}{\bibinfo{person}{Fengxin Li}, \bibinfo{person}{Hongyan Liu}, \bibinfo{person}{Jun He}, {and} \bibinfo{person}{Xiaoyong Du}.} \bibinfo{year}{2024}\natexlab{a}.
\newblock \showarticletitle{Causal{CDR}: Causal Embedding Learning for Cross-domain Recommendation}. In \bibinfo{booktitle}{\emph{SDM}}. \bibinfo{pages}{553--561}.
\newblock


\bibitem[Li et~al\mbox{.}(2024b)]%
        {li2024aiming}
\bibfield{author}{\bibinfo{person}{Hanyu Li}, \bibinfo{person}{Weizhi Ma}, \bibinfo{person}{Peijie Sun}, \bibinfo{person}{Jiayu Li}, \bibinfo{person}{Cunxiang Yin}, \bibinfo{person}{Yancheng He}, \bibinfo{person}{Guoqiang Xu}, \bibinfo{person}{Min Zhang}, {and} \bibinfo{person}{Shaoping Ma}.} \bibinfo{year}{2024}\natexlab{b}.
\newblock \showarticletitle{Aiming at the Target: Filter Collaborative Information for Cross-Domain Recommendation}. In \bibinfo{booktitle}{\emph{SIGIR}}. \bibinfo{pages}{2081--2090}.
\newblock


\bibitem[Li and Tuzhilin(2020)]%
        {li2020ddtcdr}
\bibfield{author}{\bibinfo{person}{Pan Li} {and} \bibinfo{person}{Alexander Tuzhilin}.} \bibinfo{year}{2020}\natexlab{}.
\newblock \showarticletitle{DDTCDR: Deep Dual Transfer Cross Domain Recommendation}. In \bibinfo{booktitle}{\emph{WSDM}}. \bibinfo{pages}{331--339}.
\newblock


\bibitem[Li et~al\mbox{.}(2023)]%
        {li2023causal}
\bibfield{author}{\bibinfo{person}{Qian Li}, \bibinfo{person}{Xiangmeng Wang}, \bibinfo{person}{Zhichao Wang}, {and} \bibinfo{person}{Guandong Xu}.} \bibinfo{year}{2023}\natexlab{}.
\newblock \showarticletitle{Be Causal: De-Biasing Social Network Confounding in Recommendation}.
\newblock \bibinfo{journal}{\emph{ACM TKDD}} \bibinfo{volume}{17}, \bibinfo{number}{1} (\bibinfo{year}{2023}), \bibinfo{pages}{1--23}.
\newblock


\bibitem[Li et~al\mbox{.}(2021)]%
        {li2021debiasing}
\bibfield{author}{\bibinfo{person}{Siqing Li}, \bibinfo{person}{Liuyi Yao}, \bibinfo{person}{Shanlei Mu}, \bibinfo{person}{Wayne~Xin Zhao}, \bibinfo{person}{Yaliang Li}, \bibinfo{person}{Tonglei Guo}, \bibinfo{person}{Bolin Ding}, {and} \bibinfo{person}{Ji-Rong Wen}.} \bibinfo{year}{2021}\natexlab{}.
\newblock \showarticletitle{Debiasing Learning based Cross-domain Recommendation}. In \bibinfo{booktitle}{\emph{KDD}}. \bibinfo{pages}{3190--3199}.
\newblock


\bibitem[Li et~al\mbox{.}(2022)]%
        {li2022debiasing}
\bibfield{author}{\bibinfo{person}{Zhi Li}, \bibinfo{person}{Daichi Amagata}, \bibinfo{person}{Yihong Zhang}, \bibinfo{person}{Takahiro Hara}, \bibinfo{person}{Shuichiro Haruta}, \bibinfo{person}{Kei Yonekawa}, {and} \bibinfo{person}{Mori Kurokawa}.} \bibinfo{year}{2022}\natexlab{}.
\newblock \showarticletitle{Debiasing Graph Transfer Learning via Item Semantic Clustering for Cross-Domain Recommendations}. In \bibinfo{booktitle}{\emph{BigData}}. \bibinfo{pages}{762--769}.
\newblock


\bibitem[Liang et~al\mbox{.}(2024)]%
        {liang2024deconfounding}
\bibfield{author}{\bibinfo{person}{Yuliang Liang}, \bibinfo{person}{Enneng Yang}, \bibinfo{person}{Guibing Guo}, \bibinfo{person}{Wei Cai}, \bibinfo{person}{Linying Jiang}, {and} \bibinfo{person}{Xingwei Wang}.} \bibinfo{year}{2024}\natexlab{}.
\newblock \showarticletitle{Deconfounding User Preference in Recommendation Systems through Implicit and Explicit Feedback}.
\newblock \bibinfo{journal}{\emph{ACM TKDD}} \bibinfo{volume}{18}, \bibinfo{number}{8} (\bibinfo{year}{2024}), \bibinfo{pages}{1--18}.
\newblock


\bibitem[Lin et~al\mbox{.}(2024)]%
        {lin2024pre}
\bibfield{author}{\bibinfo{person}{Ziqian Lin}, \bibinfo{person}{Hao Ding}, \bibinfo{person}{Nghia~Trong Hoang}, \bibinfo{person}{Branislav Kveton}, \bibinfo{person}{Anoop Deoras}, {and} \bibinfo{person}{Hao Wang}.} \bibinfo{year}{2024}\natexlab{}.
\newblock \showarticletitle{Pre-trained Recommender Systems: A Causal Debiasing Perspective}. In \bibinfo{booktitle}{\emph{WSDM}}. \bibinfo{pages}{424--433}.
\newblock


\bibitem[Liu et~al\mbox{.}(2021)]%
        {liu2021mitigating}
\bibfield{author}{\bibinfo{person}{Dugang Liu}, \bibinfo{person}{Pengxiang Cheng}, \bibinfo{person}{Hong Zhu}, \bibinfo{person}{Zhenhua Dong}, \bibinfo{person}{Xiuqiang He}, \bibinfo{person}{Weike Pan}, {and} \bibinfo{person}{Zhong Ming}.} \bibinfo{year}{2021}\natexlab{}.
\newblock \showarticletitle{Mitigating Confounding Bias in Recommendation via Information Bottleneck}. In \bibinfo{booktitle}{\emph{RecSys}}. \bibinfo{pages}{351--360}.
\newblock


\bibitem[Liu et~al\mbox{.}(2023)]%
        {liu2023debiased}
\bibfield{author}{\bibinfo{person}{Dugang Liu}, \bibinfo{person}{Pengxiang Cheng}, \bibinfo{person}{Hong Zhu}, \bibinfo{person}{Zhenhua Dong}, \bibinfo{person}{Xiuqiang He}, \bibinfo{person}{Weike Pan}, {and} \bibinfo{person}{Zhong Ming}.} \bibinfo{year}{2023}\natexlab{}.
\newblock \showarticletitle{Debiased Representation Learning in Recommendation via Information Bottleneck}.
\newblock \bibinfo{journal}{\emph{ACM TORS}} \bibinfo{volume}{1}, \bibinfo{number}{1} (\bibinfo{year}{2023}), \bibinfo{pages}{1--27}.
\newblock


\bibitem[Liu et~al\mbox{.}(2024a)]%
        {liu2024graph}
\bibfield{author}{\bibinfo{person}{Jing Liu}, \bibinfo{person}{Lele Sun}, \bibinfo{person}{Weizhi Nie}, \bibinfo{person}{Peiguang Jing}, {and} \bibinfo{person}{Yuting Su}.} \bibinfo{year}{2024}\natexlab{a}.
\newblock \showarticletitle{Graph Disentangled Contrastive Learning with Personalized Transfer for Cross-Domain Recommendation}. In \bibinfo{booktitle}{\emph{AAAI}}. \bibinfo{pages}{8769--8777}.
\newblock


\bibitem[Liu et~al\mbox{.}(2020)]%
        {liu2020cross}
\bibfield{author}{\bibinfo{person}{Meng Liu}, \bibinfo{person}{Jianjun Li}, \bibinfo{person}{Guohui Li}, {and} \bibinfo{person}{Peng Pan}.} \bibinfo{year}{2020}\natexlab{}.
\newblock \showarticletitle{Cross Domain Recommendation via Bi-directional Transfer Graph Collaborative Filtering Networks}. In \bibinfo{booktitle}{\emph{CIKM}}. \bibinfo{pages}{885--894}.
\newblock


\bibitem[Liu et~al\mbox{.}(2024b)]%
        {liu2024interact}
\bibfield{author}{\bibinfo{person}{Xu Liu}, \bibinfo{person}{Tong Yu}, \bibinfo{person}{Kaige Xie}, \bibinfo{person}{Junda Wu}, {and} \bibinfo{person}{Shuai Li}.} \bibinfo{year}{2024}\natexlab{b}.
\newblock \showarticletitle{Interact with the Explanations: Causal Debiased Explainable Recommendation System}. In \bibinfo{booktitle}{\emph{WSDM}}. \bibinfo{pages}{472--481}.
\newblock


\bibitem[Liu et~al\mbox{.}(2022)]%
        {liu2022causal}
\bibfield{author}{\bibinfo{person}{Yuchen Liu}, \bibinfo{person}{Yabo Chen}, \bibinfo{person}{Wenrui Dai}, \bibinfo{person}{Chenglin Li}, \bibinfo{person}{Junni Zou}, {and} \bibinfo{person}{Hongkai Xiong}.} \bibinfo{year}{2022}\natexlab{}.
\newblock \showarticletitle{Causal Intervention for Generalizable Face Anti-Spoofing}. In \bibinfo{booktitle}{\emph{ICME}}. \bibinfo{pages}{01--06}.
\newblock


\bibitem[Luo et~al\mbox{.}(2024)]%
        {luo2024survey}
\bibfield{author}{\bibinfo{person}{Huishi Luo}, \bibinfo{person}{Fuzhen Zhuang}, \bibinfo{person}{Ruobing Xie}, \bibinfo{person}{Hengshu Zhu}, \bibinfo{person}{Deqing Wang}, \bibinfo{person}{Zhulin An}, {and} \bibinfo{person}{Yongjun Xu}.} \bibinfo{year}{2024}\natexlab{}.
\newblock \showarticletitle{A Survey on Causal Inference for Recommendation}.
\newblock \bibinfo{journal}{\emph{The Innovation}} \bibinfo{volume}{5}, \bibinfo{number}{2} (\bibinfo{year}{2024}), \bibinfo{pages}{100590}.
\newblock


\bibitem[Luo et~al\mbox{.}(2023)]%
        {luo2023mamdr}
\bibfield{author}{\bibinfo{person}{Linhao Luo}, \bibinfo{person}{Yumeng Li}, \bibinfo{person}{Buyu Gao}, \bibinfo{person}{Shuai Tang}, \bibinfo{person}{Sinan Wang}, \bibinfo{person}{Jiancheng Li}, \bibinfo{person}{Tanchao Zhu}, \bibinfo{person}{Jiancai Liu}, \bibinfo{person}{Zhao Li}, {and} \bibinfo{person}{Shirui Pan}.} \bibinfo{year}{2023}\natexlab{}.
\newblock \showarticletitle{{MAMDR}: A Model Agnostic Learning Framework for Multi-Domain Recommendation}. In \bibinfo{booktitle}{\emph{ICDE}}. \bibinfo{pages}{3079--3092}.
\newblock


\bibitem[Ma et~al\mbox{.}(2019)]%
        {ma2019learning}
\bibfield{author}{\bibinfo{person}{Jianxin Ma}, \bibinfo{person}{Chang Zhou}, \bibinfo{person}{Peng Cui}, \bibinfo{person}{Hongxia Yang}, {and} \bibinfo{person}{Wenwu Zhu}.} \bibinfo{year}{2019}\natexlab{}.
\newblock \showarticletitle{Learning Disentangled Representations for Recommendation}. In \bibinfo{booktitle}{\emph{NeurIPS}}. \bibinfo{pages}{5711--5722}.
\newblock


\bibitem[Ma et~al\mbox{.}(2020)]%
        {ma2020disentangled}
\bibfield{author}{\bibinfo{person}{Jianxin Ma}, \bibinfo{person}{Chang Zhou}, \bibinfo{person}{Hongxia Yang}, \bibinfo{person}{Peng Cui}, \bibinfo{person}{Xin Wang}, {and} \bibinfo{person}{Wenwu Zhu}.} \bibinfo{year}{2020}\natexlab{}.
\newblock \showarticletitle{Disentangled Self-Supervision in Sequential Recommenders}. In \bibinfo{booktitle}{\emph{KDD}}. \bibinfo{pages}{483--491}.
\newblock


\bibitem[Man et~al\mbox{.}(2017)]%
        {man2017cross}
\bibfield{author}{\bibinfo{person}{Tong Man}, \bibinfo{person}{Huawei Shen}, \bibinfo{person}{Xiaolong Jin}, {and} \bibinfo{person}{Xueqi Cheng}.} \bibinfo{year}{2017}\natexlab{}.
\newblock \showarticletitle{Cross-Domain Recommendation: An Embedding and Mapping Approach}. In \bibinfo{booktitle}{\emph{IJCAI}}. \bibinfo{pages}{2464--2470}.
\newblock


\bibitem[Mao et~al\mbox{.}(2024)]%
        {mao2024cross}
\bibfield{author}{\bibinfo{person}{Qingyang Mao}, \bibinfo{person}{Qi Liu}, \bibinfo{person}{Zhi Li}, \bibinfo{person}{Likang Wu}, \bibinfo{person}{Bing Lv}, {and} \bibinfo{person}{Zheng Zhang}.} \bibinfo{year}{2024}\natexlab{}.
\newblock \showarticletitle{Cross-reconstructed Augmentation for Dual-target Cross-domain Recommendation}. In \bibinfo{booktitle}{\emph{SIGIR}}. \bibinfo{pages}{2352--2356}.
\newblock


\bibitem[Mao et~al\mbox{.}(2017)]%
        {mao2017least}
\bibfield{author}{\bibinfo{person}{Xudong Mao}, \bibinfo{person}{Qing Li}, \bibinfo{person}{Haoran Xie}, \bibinfo{person}{Raymond~YK Lau}, \bibinfo{person}{Zhen Wang}, {and} \bibinfo{person}{Stephen Paul~Smolley}.} \bibinfo{year}{2017}\natexlab{}.
\newblock \showarticletitle{Least Squares Generative Adversarial Networks}. In \bibinfo{booktitle}{\emph{ICCV}}. \bibinfo{pages}{2794--2802}.
\newblock


\bibitem[Menglin et~al\mbox{.}(2024)]%
        {menglin2024c2dr}
\bibfield{author}{\bibinfo{person}{Kong Menglin}, \bibinfo{person}{Jia Wang}, \bibinfo{person}{Yushan Pan}, \bibinfo{person}{Haiyang Zhang}, {and} \bibinfo{person}{Muzhou Hou}.} \bibinfo{year}{2024}\natexlab{}.
\newblock \showarticletitle{C$^2${DR}: Robust Cross-Domain Recommendation based on Causal Disentanglement}. In \bibinfo{booktitle}{\emph{WSDM}}. \bibinfo{pages}{341--349}.
\newblock


\bibitem[Ouyang et~al\mbox{.}(2022)]%
        {ouyang2022causality}
\bibfield{author}{\bibinfo{person}{Cheng Ouyang}, \bibinfo{person}{Chen Chen}, \bibinfo{person}{Surui Li}, \bibinfo{person}{Zeju Li}, \bibinfo{person}{Chen Qin}, \bibinfo{person}{Wenjia Bai}, {and} \bibinfo{person}{Daniel Rueckert}.} \bibinfo{year}{2022}\natexlab{}.
\newblock \showarticletitle{Causality-Inspired Single-Source Domain Generalization for Medical Image Segmentation}.
\newblock \bibinfo{journal}{\emph{IEEE TMI}} \bibinfo{volume}{42}, \bibinfo{number}{4} (\bibinfo{year}{2022}), \bibinfo{pages}{1095--1106}.
\newblock


\bibitem[Pearl and Mackenzie(2018)]%
        {pearl2018book}
\bibfield{author}{\bibinfo{person}{Judea Pearl} {and} \bibinfo{person}{Dana Mackenzie}.} \bibinfo{year}{2018}\natexlab{}.
\newblock \bibinfo{booktitle}{\emph{The Book of Why: The New Science of Cause and Effect}}.
\newblock \bibinfo{publisher}{Basic books}.
\newblock


\bibitem[Ren et~al\mbox{.}(2023)]%
        {ren2023disentangled}
\bibfield{author}{\bibinfo{person}{Xubin Ren}, \bibinfo{person}{Lianghao Xia}, \bibinfo{person}{Jiashu Zhao}, \bibinfo{person}{Dawei Yin}, {and} \bibinfo{person}{Chao Huang}.} \bibinfo{year}{2023}\natexlab{}.
\newblock \showarticletitle{Disentangled Contrastive Collaborative Filtering}. In \bibinfo{booktitle}{\emph{SIGIR}}. \bibinfo{pages}{1137--1146}.
\newblock


\bibitem[Rendle et~al\mbox{.}(2009)]%
        {rendle2012bpr}
\bibfield{author}{\bibinfo{person}{Steffen Rendle}, \bibinfo{person}{Christoph Freudenthaler}, \bibinfo{person}{Zeno Gantner}, {and} \bibinfo{person}{Lars Schmidt-Thieme}.} \bibinfo{year}{2009}\natexlab{}.
\newblock \showarticletitle{{BPR}: Bayesian Personalized Ranking from Implicit Feedback}. In \bibinfo{booktitle}{\emph{UAI}}. \bibinfo{pages}{452–461}.
\newblock


\bibitem[Sato et~al\mbox{.}(2020)]%
        {sato2020unbiased}
\bibfield{author}{\bibinfo{person}{Masahiro Sato}, \bibinfo{person}{Sho Takemori}, \bibinfo{person}{Janmajay Singh}, {and} \bibinfo{person}{Tomoko Ohkuma}.} \bibinfo{year}{2020}\natexlab{}.
\newblock \showarticletitle{Unbiased Learning for the Causal Effect of Recommendation}. In \bibinfo{booktitle}{\emph{RecSys}}. \bibinfo{pages}{378--387}.
\newblock


\bibitem[Sch{\"o}lkopf et~al\mbox{.}(2016)]%
        {scholkopf2016modeling}
\bibfield{author}{\bibinfo{person}{Bernhard Sch{\"o}lkopf}, \bibinfo{person}{David~W Hogg}, \bibinfo{person}{Dun Wang}, \bibinfo{person}{Daniel Foreman-Mackey}, \bibinfo{person}{Dominik Janzing}, \bibinfo{person}{Carl-Johann Simon-Gabriel}, {and} \bibinfo{person}{Jonas Peters}.} \bibinfo{year}{2016}\natexlab{}.
\newblock \showarticletitle{Modeling Confounding by Half-Sibling Regression}.
\newblock \bibinfo{journal}{\emph{PNAS}} \bibinfo{volume}{113}, \bibinfo{number}{27} (\bibinfo{year}{2016}), \bibinfo{pages}{7391--7398}.
\newblock


\bibitem[Sheth et~al\mbox{.}(2022)]%
        {sheth2022domain}
\bibfield{author}{\bibinfo{person}{Paras Sheth}, \bibinfo{person}{Raha Moraffah}, \bibinfo{person}{K~Sel{\c{c}}uk Candan}, \bibinfo{person}{Adrienne Raglin}, {and} \bibinfo{person}{Huan Liu}.} \bibinfo{year}{2022}\natexlab{}.
\newblock \showarticletitle{Domain Generalization--A Causal Perspective}.
\newblock \bibinfo{journal}{\emph{arXiv preprint arXiv:2209.15177}} (\bibinfo{year}{2022}).
\newblock


\bibitem[Song et~al\mbox{.}(2024)]%
        {song2024mitigating}
\bibfield{author}{\bibinfo{person}{Zijian Song}, \bibinfo{person}{Wenhan Zhang}, \bibinfo{person}{Lifang Deng}, \bibinfo{person}{Jiandong Zhang}, \bibinfo{person}{Zhihua Wu}, \bibinfo{person}{Kaigui Bian}, {and} \bibinfo{person}{Bin Cui}.} \bibinfo{year}{2024}\natexlab{}.
\newblock \showarticletitle{Mitigating Negative Transfer in Cross-Domain Recommendation via Knowledge Transferability Enhancement}. In \bibinfo{booktitle}{\emph{KDD}}. \bibinfo{pages}{2745--2754}.
\newblock


\bibitem[Su et~al\mbox{.}(2023)]%
        {su2023cross}
\bibfield{author}{\bibinfo{person}{Hongzu Su}, \bibinfo{person}{Jingjing Li}, \bibinfo{person}{Zhekai Du}, \bibinfo{person}{Lei Zhu}, \bibinfo{person}{Ke Lu}, {and} \bibinfo{person}{Heng~Tao Shen}.} \bibinfo{year}{2023}\natexlab{}.
\newblock \showarticletitle{Cross-domain Recommendation via Dual Adversarial Adaptation}.
\newblock \bibinfo{journal}{\emph{ACM TOIS}} \bibinfo{volume}{42}, \bibinfo{number}{3} (\bibinfo{year}{2023}), \bibinfo{pages}{1--26}.
\newblock


\bibitem[Sun et~al\mbox{.}(2024)]%
        {sun2024large}
\bibfield{author}{\bibinfo{person}{Zhu Sun}, \bibinfo{person}{Hongyang Liu}, \bibinfo{person}{Xinghua Qu}, \bibinfo{person}{Kaidong Feng}, \bibinfo{person}{Yan Wang}, {and} \bibinfo{person}{Yew~Soon Ong}.} \bibinfo{year}{2024}\natexlab{}.
\newblock \showarticletitle{Large Language Models for Intent-Driven Session Recommendations}. In \bibinfo{booktitle}{\emph{SIGIR}}. \bibinfo{pages}{324--334}.
\newblock


\bibitem[Tran and Lauw(2023)]%
        {tran2023multi}
\bibfield{author}{\bibinfo{person}{Nhu-Thuat Tran} {and} \bibinfo{person}{Hady~W Lauw}.} \bibinfo{year}{2023}\natexlab{}.
\newblock \showarticletitle{Multi-Representation Variational Autoencoder via Iterative Latent Attention and Implicit Differentiation}. In \bibinfo{booktitle}{\emph{CIKM}}. \bibinfo{pages}{2462--2471}.
\newblock


\bibitem[Wang et~al\mbox{.}(2023)]%
        {wang2023deconfounding}
\bibfield{author}{\bibinfo{person}{Junyan Wang}, \bibinfo{person}{Yiqi Jiang}, \bibinfo{person}{Yang Long}, \bibinfo{person}{Xiuyu Sun}, \bibinfo{person}{Maurice Pagnucco}, {and} \bibinfo{person}{Yang Song}.} \bibinfo{year}{2023}\natexlab{}.
\newblock \showarticletitle{Deconfounding Causal Inference for Zero-Shot Action Recognition}.
\newblock \bibinfo{journal}{\emph{IEEE TMM}}  \bibinfo{volume}{26} (\bibinfo{year}{2023}), \bibinfo{pages}{3976--3986}.
\newblock


\bibitem[Wang et~al\mbox{.}(2022b)]%
        {wang2022generalizing}
\bibfield{author}{\bibinfo{person}{Jindong Wang}, \bibinfo{person}{Cuiling Lan}, \bibinfo{person}{Chang Liu}, \bibinfo{person}{Yidong Ouyang}, \bibinfo{person}{Tao Qin}, \bibinfo{person}{Wang Lu}, \bibinfo{person}{Yiqiang Chen}, \bibinfo{person}{Wenjun Zeng}, {and} \bibinfo{person}{S~Yu Philip}.} \bibinfo{year}{2022}\natexlab{b}.
\newblock \showarticletitle{Generalizing to Unseen Domains: A Survey on Domain Generalization}.
\newblock \bibinfo{journal}{\emph{IEEE TKDE}} \bibinfo{volume}{35}, \bibinfo{number}{8} (\bibinfo{year}{2022}), \bibinfo{pages}{8052--8072}.
\newblock


\bibitem[Wang et~al\mbox{.}(2021)]%
        {wang2021deconfounded}
\bibfield{author}{\bibinfo{person}{Wenjie Wang}, \bibinfo{person}{Fuli Feng}, \bibinfo{person}{Xiangnan He}, \bibinfo{person}{Xiang Wang}, {and} \bibinfo{person}{Tat-Seng Chua}.} \bibinfo{year}{2021}\natexlab{}.
\newblock \showarticletitle{Deconfounded Recommendation for Alleviating Bias Amplification}. In \bibinfo{booktitle}{\emph{KDD}}. \bibinfo{pages}{1717--1725}.
\newblock


\bibitem[Wang et~al\mbox{.}(2019)]%
        {wang2019neural}
\bibfield{author}{\bibinfo{person}{Xiang Wang}, \bibinfo{person}{Xiangnan He}, \bibinfo{person}{Meng Wang}, \bibinfo{person}{Fuli Feng}, {and} \bibinfo{person}{Tat-Seng Chua}.} \bibinfo{year}{2019}\natexlab{}.
\newblock \showarticletitle{Neural Graph Collaborative Filtering}. In \bibinfo{booktitle}{\emph{SIGIR}}. \bibinfo{pages}{165--174}.
\newblock


\bibitem[Wang et~al\mbox{.}(2022c)]%
        {wang2022causal}
\bibfield{author}{\bibinfo{person}{Xiangmeng Wang}, \bibinfo{person}{Qian Li}, \bibinfo{person}{Dianer Yu}, \bibinfo{person}{Peng Cui}, \bibinfo{person}{Zhichao Wang}, {and} \bibinfo{person}{Guandong Xu}.} \bibinfo{year}{2022}\natexlab{c}.
\newblock \showarticletitle{Causal Disentanglement for Semantics-Aware Intent Learning in Recommendation}.
\newblock \bibinfo{journal}{\emph{IEEE TKDE}} \bibinfo{volume}{35}, \bibinfo{number}{10} (\bibinfo{year}{2022}), \bibinfo{pages}{9836--9849}.
\newblock


\bibitem[Wang et~al\mbox{.}(2022a)]%
        {wang2022causalint}
\bibfield{author}{\bibinfo{person}{Yichao Wang}, \bibinfo{person}{Huifeng Guo}, \bibinfo{person}{Bo Chen}, \bibinfo{person}{Weiwen Liu}, \bibinfo{person}{Zhirong Liu}, \bibinfo{person}{Qi Zhang}, \bibinfo{person}{Zhicheng He}, \bibinfo{person}{Hongkun Zheng}, \bibinfo{person}{Weiwei Yao}, \bibinfo{person}{Muyu Zhang}, {et~al\mbox{.}}} \bibinfo{year}{2022}\natexlab{a}.
\newblock \showarticletitle{Causal{I}nt: Causal Inspired Intervention for Multi-Scenario Recommendation}. In \bibinfo{booktitle}{\emph{KDD}}. \bibinfo{pages}{4090--4099}.
\newblock


\bibitem[Wang et~al\mbox{.}(2020)]%
        {wang2020causal}
\bibfield{author}{\bibinfo{person}{Yixin Wang}, \bibinfo{person}{Dawen Liang}, \bibinfo{person}{Laurent Charlin}, {and} \bibinfo{person}{David~M Blei}.} \bibinfo{year}{2020}\natexlab{}.
\newblock \showarticletitle{Causal Inference for Recommender Systems}. In \bibinfo{booktitle}{\emph{RecSys}}. \bibinfo{pages}{426--431}.
\newblock


\bibitem[Wang et~al\mbox{.}(2022d)]%
        {wang2022unbiased}
\bibfield{author}{\bibinfo{person}{Zhenlei Wang}, \bibinfo{person}{Shiqi Shen}, \bibinfo{person}{Zhipeng Wang}, \bibinfo{person}{Bo Chen}, \bibinfo{person}{Xu Chen}, {and} \bibinfo{person}{Ji-Rong Wen}.} \bibinfo{year}{2022}\natexlab{d}.
\newblock \showarticletitle{Unbiased Sequential Recommendation with Latent Confounders}. In \bibinfo{booktitle}{\emph{TheWebConf}}. \bibinfo{pages}{2195--2204}.
\newblock


\bibitem[Wu et~al\mbox{.}(2025)]%
        {wu2025instrumental}
\bibfield{author}{\bibinfo{person}{Anpeng Wu}, \bibinfo{person}{Kun Kuang}, \bibinfo{person}{Ruoxuan Xiong}, {and} \bibinfo{person}{Fei Wu}.} \bibinfo{year}{2025}\natexlab{}.
\newblock \showarticletitle{Instrumental Variables in Causal Inference and Machine Learning: A Survey}.
\newblock \bibinfo{journal}{\emph{Comput. Surveys}} (\bibinfo{year}{2025}).
\newblock


\bibitem[Wu et~al\mbox{.}(2022)]%
        {ijcai2022p787}
\bibfield{author}{\bibinfo{person}{Peng Wu}, \bibinfo{person}{Haoxuan Li}, \bibinfo{person}{Yuhao Deng}, \bibinfo{person}{Wenjie Hu}, \bibinfo{person}{Quanyu Dai}, \bibinfo{person}{Zhenhua Dong}, \bibinfo{person}{Jie Sun}, \bibinfo{person}{Rui Zhang}, {and} \bibinfo{person}{Xiao-Hua Zhou}.} \bibinfo{year}{2022}\natexlab{}.
\newblock \showarticletitle{On the Opportunity of Causal Learning in Recommendation Systems: Foundation, Estimation, Prediction and Challenges}. In \bibinfo{booktitle}{\emph{IJCAI}}. \bibinfo{pages}{5646--5653}.
\newblock


\bibitem[Xie et~al\mbox{.}(2023)]%
        {xie2023rethinking}
\bibfield{author}{\bibinfo{person}{Yueqi Xie}, \bibinfo{person}{Jingqi Gao}, \bibinfo{person}{Peilin Zhou}, \bibinfo{person}{Qichen Ye}, \bibinfo{person}{Yining Hua}, \bibinfo{person}{Jae~Boum Kim}, \bibinfo{person}{Fangzhao Wu}, {and} \bibinfo{person}{Sunghun Kim}.} \bibinfo{year}{2023}\natexlab{}.
\newblock \showarticletitle{Rethinking Multi-Interest Learning for Candidate Matching in Recommender Systems}. In \bibinfo{booktitle}{\emph{RecSys}}. \bibinfo{pages}{283--293}.
\newblock


\bibitem[Xu et~al\mbox{.}(2025)]%
        {xu2025causal}
\bibfield{author}{\bibinfo{person}{Shuyuan Xu}, \bibinfo{person}{Jianchao Ji}, \bibinfo{person}{Yunqi Li}, \bibinfo{person}{Yingqiang Ge}, \bibinfo{person}{Juntao Tan}, {and} \bibinfo{person}{Yongfeng Zhang}.} \bibinfo{year}{2025}\natexlab{}.
\newblock \showarticletitle{Causal Inference for Recommendation: Foundations, Methods and Applications}.
\newblock \bibinfo{journal}{\emph{ACM TIST}} \bibinfo{volume}{16}, \bibinfo{number}{3} (\bibinfo{year}{2025}), \bibinfo{pages}{1--51}.
\newblock


\bibitem[Xu et~al\mbox{.}(2023)]%
        {xu2023deconfounded}
\bibfield{author}{\bibinfo{person}{Shuyuan Xu}, \bibinfo{person}{Juntao Tan}, \bibinfo{person}{Shelby Heinecke}, \bibinfo{person}{Vena~Jia Li}, {and} \bibinfo{person}{Yongfeng Zhang}.} \bibinfo{year}{2023}\natexlab{}.
\newblock \showarticletitle{Deconfounded Causal Collaborative Filtering}.
\newblock \bibinfo{journal}{\emph{ACM TORS}} \bibinfo{volume}{1}, \bibinfo{number}{4} (\bibinfo{year}{2023}), \bibinfo{pages}{1--25}.
\newblock


\bibitem[Xu et~al\mbox{.}(2024)]%
        {xu2024rethinking}
\bibfield{author}{\bibinfo{person}{Wujiang Xu}, \bibinfo{person}{Qitian Wu}, \bibinfo{person}{Runzhong Wang}, \bibinfo{person}{Mingming Ha}, \bibinfo{person}{Qiongxu Ma}, \bibinfo{person}{Linxun Chen}, \bibinfo{person}{Bing Han}, {and} \bibinfo{person}{Junchi Yan}.} \bibinfo{year}{2024}\natexlab{}.
\newblock \showarticletitle{Rethinking Cross-Domain Sequential Recommendation under Open-World Assumptions}. In \bibinfo{booktitle}{\emph{TheWebConf}}. \bibinfo{pages}{3173--3184}.
\newblock


\bibitem[Yang et~al\mbox{.}(2023)]%
        {yang2023multimodal}
\bibfield{author}{\bibinfo{person}{Wei Yang}, \bibinfo{person}{Jie Yang}, {and} \bibinfo{person}{Yuan Liu}.} \bibinfo{year}{2023}\natexlab{}.
\newblock \showarticletitle{Multimodal Optimal Transport Knowledge Distillation for Cross-domain Recommendation}. In \bibinfo{booktitle}{\emph{CIKM}}. \bibinfo{pages}{2959--2968}.
\newblock


\bibitem[Yu et~al\mbox{.}(2023)]%
        {yu2023deconfounded}
\bibfield{author}{\bibinfo{person}{Dianer Yu}, \bibinfo{person}{Qian Li}, \bibinfo{person}{Xiangmeng Wang}, {and} \bibinfo{person}{Guandong Xu}.} \bibinfo{year}{2023}\natexlab{}.
\newblock \showarticletitle{Deconfounded Recommendation via Causal Intervention}.
\newblock \bibinfo{journal}{\emph{Neurocomputing}} (\bibinfo{year}{2023}), \bibinfo{pages}{128--139}.
\newblock


\bibitem[Yuan et~al\mbox{.}(2019)]%
        {ijcai2019p587}
\bibfield{author}{\bibinfo{person}{Feng Yuan}, \bibinfo{person}{Lina Yao}, {and} \bibinfo{person}{Boualem Benatallah}.} \bibinfo{year}{2019}\natexlab{}.
\newblock \showarticletitle{{DAR}ec: Deep Domain Adaptation for Cross-Domain Recommendation via Transferring Rating Patterns}. In \bibinfo{booktitle}{\emph{IJCAI}}. \bibinfo{pages}{4227--4233}.
\newblock


\bibitem[Yuan et~al\mbox{.}(2023)]%
        {yuan2023instrumental}
\bibfield{author}{\bibinfo{person}{Junkun Yuan}, \bibinfo{person}{Xu Ma}, \bibinfo{person}{Ruoxuan Xiong}, \bibinfo{person}{Mingming Gong}, \bibinfo{person}{Xiangyu Liu}, \bibinfo{person}{Fei Wu}, \bibinfo{person}{Lanfen Lin}, {and} \bibinfo{person}{Kun Kuang}.} \bibinfo{year}{2023}\natexlab{}.
\newblock \showarticletitle{Instrumental Variable-Driven Domain Generalization with Unobserved Confounders}.
\newblock \bibinfo{journal}{\emph{ACM TKDD}} \bibinfo{volume}{17}, \bibinfo{number}{8} (\bibinfo{year}{2023}), \bibinfo{pages}{1--21}.
\newblock


\bibitem[Zhan et~al\mbox{.}(2022)]%
        {zhan2022deconfounding}
\bibfield{author}{\bibinfo{person}{Ruohan Zhan}, \bibinfo{person}{Changhua Pei}, \bibinfo{person}{Qiang Su}, \bibinfo{person}{Jianfeng Wen}, \bibinfo{person}{Xueliang Wang}, \bibinfo{person}{Guanyu Mu}, \bibinfo{person}{Dong Zheng}, \bibinfo{person}{Peng Jiang}, {and} \bibinfo{person}{Kun Gai}.} \bibinfo{year}{2022}\natexlab{}.
\newblock \showarticletitle{Deconfounding Duration Bias in Watch-time Prediction for Video Recommendation}. In \bibinfo{booktitle}{\emph{KDD}}. \bibinfo{pages}{4472--4481}.
\newblock


\bibitem[Zhang et~al\mbox{.}(2023e)]%
        {zhang2023debiasing}
\bibfield{author}{\bibinfo{person}{Qing Zhang}, \bibinfo{person}{Xiaoying Zhang}, \bibinfo{person}{Yang Liu}, \bibinfo{person}{Hongning Wang}, \bibinfo{person}{Min Gao}, \bibinfo{person}{Jiheng Zhang}, {and} \bibinfo{person}{Ruocheng Guo}.} \bibinfo{year}{2023}\natexlab{e}.
\newblock \showarticletitle{Debiasing Recommendation by Learning Identifiable Latent Confounders}. In \bibinfo{booktitle}{\emph{KDD}}. \bibinfo{pages}{3353–3363}.
\newblock


\bibitem[Zhang et~al\mbox{.}(2023d)]%
        {zhang2023disentangled}
\bibfield{author}{\bibinfo{person}{Ruohan Zhang}, \bibinfo{person}{Tianzi Zang}, \bibinfo{person}{Yanmin Zhu}, \bibinfo{person}{Chunyang Wang}, \bibinfo{person}{Ke Wang}, {and} \bibinfo{person}{Jiadi Yu}.} \bibinfo{year}{2023}\natexlab{d}.
\newblock \showarticletitle{Disentangled Contrastive Learning for Cross-Domain Recommendation}. In \bibinfo{booktitle}{\emph{DASFAA}}. \bibinfo{pages}{163--178}.
\newblock


\bibitem[Zhang et~al\mbox{.}(2023a)]%
        {zhang2023video}
\bibfield{author}{\bibinfo{person}{Shengyu Zhang}, \bibinfo{person}{Xusheng Feng}, \bibinfo{person}{Wenyan Fan}, \bibinfo{person}{Wenjing Fang}, \bibinfo{person}{Fuli Feng}, \bibinfo{person}{Wei Ji}, \bibinfo{person}{Shuo Li}, \bibinfo{person}{Li Wang}, \bibinfo{person}{Shanshan Zhao}, \bibinfo{person}{Zhou Zhao}, {et~al\mbox{.}}} \bibinfo{year}{2023}\natexlab{a}.
\newblock \showarticletitle{Video-Audio Domain Generalization via Confounder Disentanglement}. In \bibinfo{booktitle}{\emph{AAAI}}. \bibinfo{pages}{15322--15330}.
\newblock


\bibitem[Zhang et~al\mbox{.}(2024)]%
        {zhang2024transferring}
\bibfield{author}{\bibinfo{person}{Shengyu Zhang}, \bibinfo{person}{Qiaowei Miao}, \bibinfo{person}{Ping Nie}, \bibinfo{person}{Mengze Li}, \bibinfo{person}{Zhengyu Chen}, \bibinfo{person}{Fuli Feng}, \bibinfo{person}{Kun Kuang}, {and} \bibinfo{person}{Fei Wu}.} \bibinfo{year}{2024}\natexlab{}.
\newblock \showarticletitle{Transferring Causal Mechanism over Meta-representations for Target-unknown Cross-domain Recommendation}.
\newblock \bibinfo{journal}{\emph{ACM TOIS}} \bibinfo{volume}{42}, \bibinfo{number}{4} (\bibinfo{year}{2024}), \bibinfo{pages}{1--27}.
\newblock


\bibitem[Zhang et~al\mbox{.}(2022a)]%
        {zhang2022re4}
\bibfield{author}{\bibinfo{person}{Shengyu Zhang}, \bibinfo{person}{Lingxiao Yang}, \bibinfo{person}{Dong Yao}, \bibinfo{person}{Yujie Lu}, \bibinfo{person}{Fuli Feng}, \bibinfo{person}{Zhou Zhao}, \bibinfo{person}{Tat-Seng Chua}, {and} \bibinfo{person}{Fei Wu}.} \bibinfo{year}{2022}\natexlab{a}.
\newblock \showarticletitle{Re4: Learning to Re-contrast, Re-attend, Re-construct for Multi-interest Recommendation}. In \bibinfo{booktitle}{\emph{TheWebConf}}. \bibinfo{pages}{2216--2226}.
\newblock


\bibitem[Zhang et~al\mbox{.}(2021)]%
        {zhang2021causal}
\bibfield{author}{\bibinfo{person}{Yang Zhang}, \bibinfo{person}{Fuli Feng}, \bibinfo{person}{Xiangnan He}, \bibinfo{person}{Tianxin Wei}, \bibinfo{person}{Chonggang Song}, \bibinfo{person}{Guohui Ling}, {and} \bibinfo{person}{Yongdong Zhang}.} \bibinfo{year}{2021}\natexlab{}.
\newblock \showarticletitle{Causal Intervention for Leveraging Popularity Bias in Recommendation}. In \bibinfo{booktitle}{\emph{SIGIR}}. \bibinfo{pages}{11--20}.
\newblock


\bibitem[Zhang et~al\mbox{.}(2023c)]%
        {zhang2023connecting}
\bibfield{author}{\bibinfo{person}{Yang Zhang}, \bibinfo{person}{Yue Shen}, \bibinfo{person}{Dong Wang}, \bibinfo{person}{Jinjie Gu}, {and} \bibinfo{person}{Guannan Zhang}.} \bibinfo{year}{2023}\natexlab{c}.
\newblock \showarticletitle{Connecting Unseen Domains: Cross-Domain Invariant Learning in Recommendation}. In \bibinfo{booktitle}{\emph{SIGIR}}. \bibinfo{pages}{1894--1898}.
\newblock


\bibitem[Zhang et~al\mbox{.}(2022b)]%
        {zhang2022learning}
\bibfield{author}{\bibinfo{person}{Yi-Fan Zhang}, \bibinfo{person}{Zhang Zhang}, \bibinfo{person}{Da Li}, \bibinfo{person}{Zhen Jia}, \bibinfo{person}{Liang Wang}, {and} \bibinfo{person}{Tieniu Tan}.} \bibinfo{year}{2022}\natexlab{b}.
\newblock \showarticletitle{Learning Domain Invariant Representations for Generalizable Person Re-Identification}.
\newblock \bibinfo{journal}{\emph{IEEE TIP}}  \bibinfo{volume}{32} (\bibinfo{year}{2022}), \bibinfo{pages}{509--523}.
\newblock


\bibitem[Zhang et~al\mbox{.}(2023b)]%
        {zhang2023hierarchical}
\bibfield{author}{\bibinfo{person}{Zeyu Zhang}, \bibinfo{person}{Heyang Gao}, \bibinfo{person}{Hao Yang}, {and} \bibinfo{person}{Xu Chen}.} \bibinfo{year}{2023}\natexlab{b}.
\newblock \showarticletitle{Hierarchical Invariant Learning for Domain Generalization Recommendation}. In \bibinfo{booktitle}{\emph{KDD}}. \bibinfo{pages}{3470--3479}.
\newblock


\bibitem[Zhao et~al\mbox{.}(2023a)]%
        {zhao2023sequential}
\bibfield{author}{\bibinfo{person}{Chuang Zhao}, \bibinfo{person}{Xinyu Li}, \bibinfo{person}{Ming He}, \bibinfo{person}{Hongke Zhao}, {and} \bibinfo{person}{Jianping Fan}.} \bibinfo{year}{2023}\natexlab{a}.
\newblock \showarticletitle{Sequential Recommendation via an Adaptive Cross-domain Knowledge Decomposition}. In \bibinfo{booktitle}{\emph{CIKM}}. \bibinfo{pages}{3453--3463}.
\newblock


\bibitem[Zhao et~al\mbox{.}(2023b)]%
        {zhao2023cross}
\bibfield{author}{\bibinfo{person}{Chuang Zhao}, \bibinfo{person}{Hongke Zhao}, \bibinfo{person}{Xiaomeng Li}, \bibinfo{person}{Ming He}, \bibinfo{person}{Jiahui Wang}, {and} \bibinfo{person}{Jianping Fan}.} \bibinfo{year}{2023}\natexlab{b}.
\newblock \showarticletitle{Cross-Domain Recommendation via Progressive Structural Alignment}.
\newblock \bibinfo{journal}{\emph{IEEE TKDE}} \bibinfo{volume}{36}, \bibinfo{number}{6} (\bibinfo{year}{2023}), \bibinfo{pages}{2401--2415}.
\newblock


\bibitem[Zheng et~al\mbox{.}(2021)]%
        {zheng2021disentangling}
\bibfield{author}{\bibinfo{person}{Yu Zheng}, \bibinfo{person}{Chen Gao}, \bibinfo{person}{Xiang Li}, \bibinfo{person}{Xiangnan He}, \bibinfo{person}{Yong Li}, {and} \bibinfo{person}{Depeng Jin}.} \bibinfo{year}{2021}\natexlab{}.
\newblock \showarticletitle{Disentangling User Interest and Conformity for Recommendation with Causal Embedding}. In \bibinfo{booktitle}{\emph{TheWebConf}}. \bibinfo{pages}{2980--2991}.
\newblock


\bibitem[Zhou et~al\mbox{.}(2022)]%
        {zhou2022domain}
\bibfield{author}{\bibinfo{person}{Kaiyang Zhou}, \bibinfo{person}{Ziwei Liu}, \bibinfo{person}{Yu Qiao}, \bibinfo{person}{Tao Xiang}, {and} \bibinfo{person}{Chen~Change Loy}.} \bibinfo{year}{2022}\natexlab{}.
\newblock \showarticletitle{Domain Generalization: A Survey}.
\newblock \bibinfo{journal}{\emph{IEEE TPAMI}} \bibinfo{volume}{45}, \bibinfo{number}{4} (\bibinfo{year}{2022}), \bibinfo{pages}{4396--4415}.
\newblock


\bibitem[Zhu et~al\mbox{.}(2019)]%
        {zhu2019dtcdr}
\bibfield{author}{\bibinfo{person}{Feng Zhu}, \bibinfo{person}{Chaochao Chen}, \bibinfo{person}{Yan Wang}, \bibinfo{person}{Guanfeng Liu}, {and} \bibinfo{person}{Xiaolin Zheng}.} \bibinfo{year}{2019}\natexlab{}.
\newblock \showarticletitle{{DTCDR}: A Framework for Dual-Target Cross-Domain Recommendation}. In \bibinfo{booktitle}{\emph{CIKM}}. \bibinfo{pages}{1533--1542}.
\newblock


\bibitem[Zhu et~al\mbox{.}(2018)]%
        {zhu2018deep}
\bibfield{author}{\bibinfo{person}{Feng Zhu}, \bibinfo{person}{Yan Wang}, \bibinfo{person}{Chaochao Chen}, \bibinfo{person}{Guanfeng Liu}, \bibinfo{person}{Mehmet Orgun}, {and} \bibinfo{person}{Jia Wu}.} \bibinfo{year}{2018}\natexlab{}.
\newblock \showarticletitle{A Deep Framework for Cross-Domain and Cross-System Recommendations}. In \bibinfo{booktitle}{\emph{IJCAI}}. \bibinfo{pages}{3711--3717}.
\newblock


\bibitem[Zhu et~al\mbox{.}(2020)]%
        {zhu2020graphical}
\bibfield{author}{\bibinfo{person}{Feng Zhu}, \bibinfo{person}{Yan Wang}, \bibinfo{person}{Chaochao Chen}, \bibinfo{person}{Guanfeng Liu}, {and} \bibinfo{person}{Xiaolin Zheng}.} \bibinfo{year}{2020}\natexlab{}.
\newblock \showarticletitle{A Graphical and Attentional Framework for Dual-Target Cross-Domain Recommendation}. In \bibinfo{booktitle}{\emph{IJCAI}}. \bibinfo{pages}{3001--3008}.
\newblock


\bibitem[Zhu et~al\mbox{.}(2021a)]%
        {ijcai2021p639}
\bibfield{author}{\bibinfo{person}{Feng Zhu}, \bibinfo{person}{Yan Wang}, \bibinfo{person}{Chaochao Chen}, \bibinfo{person}{Jun Zhou}, \bibinfo{person}{Longfei Li}, {and} \bibinfo{person}{Guanfeng Liu}.} \bibinfo{year}{2021}\natexlab{a}.
\newblock \showarticletitle{Cross-Domain Recommendation: Challenges, Progress, and Prospects}. In \bibinfo{booktitle}{\emph{IJCAI}}. \bibinfo{pages}{4721--4728}.
\newblock


\bibitem[Zhu et~al\mbox{.}(2021b)]%
        {zhu2021unified}
\bibfield{author}{\bibinfo{person}{Feng Zhu}, \bibinfo{person}{Yan Wang}, \bibinfo{person}{Jun Zhou}, \bibinfo{person}{Chaochao Chen}, \bibinfo{person}{Longfei Li}, {and} \bibinfo{person}{Guanfeng Liu}.} \bibinfo{year}{2021}\natexlab{b}.
\newblock \showarticletitle{A Unified Framework for Cross-Domain and Cross-System Recommendations}.
\newblock \bibinfo{journal}{\emph{IEEE TKDE}} \bibinfo{volume}{35}, \bibinfo{number}{2} (\bibinfo{year}{2021}), \bibinfo{pages}{1171--1184}.
\newblock


\bibitem[Zhu et~al\mbox{.}(2023b)]%
        {zhu2023domain}
\bibfield{author}{\bibinfo{person}{Jiajie Zhu}, \bibinfo{person}{Yan Wang}, \bibinfo{person}{Feng Zhu}, {and} \bibinfo{person}{Zhu Sun}.} \bibinfo{year}{2023}\natexlab{b}.
\newblock \showarticletitle{Domain Disentanglement with Interpolative Data Augmentation for Dual-Target Cross-Domain Recommendation}. In \bibinfo{booktitle}{\emph{RecSys}}. \bibinfo{pages}{515--527}.
\newblock


\bibitem[Zhu et~al\mbox{.}(2017)]%
        {zhu2017unpaired}
\bibfield{author}{\bibinfo{person}{Jun-Yan Zhu}, \bibinfo{person}{Taesung Park}, \bibinfo{person}{Phillip Isola}, {and} \bibinfo{person}{Alexei~A Efros}.} \bibinfo{year}{2017}\natexlab{}.
\newblock \showarticletitle{Unpaired Image-to-Image Translation using Cycle-Consistent Adversarial Networks}. In \bibinfo{booktitle}{\emph{ICCV}}. \bibinfo{pages}{2223--2232}.
\newblock


\bibitem[Zhu et~al\mbox{.}(2024)]%
        {zhu2024mitigating}
\bibfield{author}{\bibinfo{person}{Xinyuan Zhu}, \bibinfo{person}{Yang Zhang}, \bibinfo{person}{Xun Yang}, \bibinfo{person}{Dingxian Wang}, {and} \bibinfo{person}{Fuli Feng}.} \bibinfo{year}{2024}\natexlab{}.
\newblock \showarticletitle{Mitigating Hidden Confounding Effects for Causal Recommendation}.
\newblock \bibinfo{journal}{\emph{IEEE TKDE}} \bibinfo{volume}{36}, \bibinfo{number}{9} (\bibinfo{year}{2024}), \bibinfo{pages}{4794--4805}.
\newblock


\bibitem[Zhu et~al\mbox{.}(2023a)]%
        {zhu2023causal}
\bibfield{author}{\bibinfo{person}{Yaochen Zhu}, \bibinfo{person}{Jing Ma}, {and} \bibinfo{person}{Jundong Li}.} \bibinfo{year}{2023}\natexlab{a}.
\newblock \showarticletitle{Causal Inference in Recommender Systems: A Survey of Strategies for Bias Mitigation, Explanation, and Generalization}.
\newblock \bibinfo{journal}{\emph{arXiv preprint arXiv:2301.00910}} (\bibinfo{year}{2023}).
\newblock


\bibitem[Zhu et~al\mbox{.}(2022)]%
        {zhu2022deep}
\bibfield{author}{\bibinfo{person}{Yaochen Zhu}, \bibinfo{person}{Xubin Ren}, \bibinfo{person}{Jing Yi}, {and} \bibinfo{person}{Zhenzhong Chen}.} \bibinfo{year}{2022}\natexlab{}.
\newblock \showarticletitle{Deep Deconfounded Content-based Tag Recommendation for {UGC} with Causal Intervention}.
\newblock \bibinfo{journal}{\emph{arXiv preprint arXiv:2205.14380}} (\bibinfo{year}{2022}).
\newblock


\end{thebibliography}

\appendix

\end{document}